\documentclass[aps,pra,showpacs,twoside,twocolumn,10pt,longbibliography,floatfix]{revtex4-2}
\usepackage[colorlinks=true, citecolor=red, urlcolor=blue ]{hyperref}
\usepackage{times,epsfig,amssymb,amsfonts,amsmath,subfigure,mathtools,amsthm,braket,soul,enumitem,color}
\usepackage[normalem]{ulem}

\newcommand{\stkout}[1]{\ifmmode\text{\sout{\ensuremath{#1}}}\else\sout{#1}\fi}
\usepackage{cases}
\usepackage{amsmath}
\usepackage{dsfont}
 \usepackage{relsize}
\usepackage{bm}
\usepackage[dvipsnames]{xcolor}

\makeatletter
\DeclareRobustCommand*{\bfseries}{%
  \not@math@alphabet\bfseries\mathbf
  \fontseries\bfdefault\selectfont
  \boldmath
}
\makeatother

\makeatletter
\newenvironment{smal}[1][\small]
 {\par\nopagebreak\leavevmode\vspace*{-\baselineskip}%
  \skip0=\abovedisplayskip
  #1%
  \def\maketag@@@##1{\hbox{\m@th\normalfont\normalsize##1}}%
  \abovedisplayskip=\skip0
  \align}
 {\endalign\ignorespacesafterend}
\makeatother
	\definecolor{blue-violet}{rgb}{0.54, 0.17, 0.89}

\begin{document}
\raggedbottom

\title{Invariance of success probability in Grover's quantum search under local noise with memory}

\author{Sheikh Parvez Mandal$^{1,2}$, Ahana Ghoshal$^2$, Chirag Srivastava$^2$, and Ujjwal Sen$^2$}

 \affiliation{$^1$Indian Institute of Science Education and Research, Pune 411 008, India\\
 $^2$Harish-Chandra Research Institute,  A CI of Homi Bhabha National Institute, Chhatnag Road, Jhunsi, Prayagraj 211 019, India}

\begin{abstract}

We analyze the robustness of Grover's quantum search algorithm performed by a quantum register under a possibly  time-correlated noise acting locally on the qubits. We model the noise as originating from an arbitrary but fixed unitary evolution, $U$, of some noisy qubits. The noise can occur with some probability in the interval between any pair of consecutive noiseless Grover evolutions. Although each run of the algorithm is a unitary process, the noise model leads to decoherence when all possible runs are considered. We derive a set of unitary $U$'s, called the `good noises,' for which the success probability of the algorithm at any given time remains unchanged with varying the non-trivial total number ($m$) of noisy qubits in the register. The result holds irrespective of the presence of any time-correlations in the noise. We show that only when $U$ is either of the Pauli matrices $\sigma_x$ and $\sigma_z$ (which give rise to $m$-qubit bit-flip and phase-damping channels respectively in the time-correlation-less case), the algorithm's success probability stays unchanged when increasing or decreasing $m$. In contrast, when $U$ is the Pauli matrix $\sigma_y$ (giving rise to $m$-qubit bit-phase flip channel in the time-correlation-less case), the success probability at all times stays unaltered as long as the parity (even or odd) of the total number $m$ remains the same. This asymmetry between the Pauli operators stems from the inherent symmetry-breaking existing within the Grover circuit. We further show that the positions of the noisy sites are irrelevant in case of any of the Pauli noises. The results are illustrated in the cases of time-correlated and time-correlation-less noise. We find that the former case leads to a better performance of the noisy algorithm. We also discuss physical scenarios where our chosen noise model is of relevance.

\end{abstract}

\maketitle

\newcommand*{\bigchi}{\mbox{\large$\chi$}}
\newcommand{\olsi}[1]{\,\overline{\!{#1}}} 
\newcommand{\Mod}[1]{\,\left(\mathrm{mod}\ #1\right)}

\section{Introduction}
\label{Intro}
The last few decades saw the advent and flourishing of the field of quantum information and computation. One of the most important classes of discoveries made in this field has to be that of quantum algorithms which provide or are believed to provide substantial computational advantages over their classical counterparts. The most significant ones include the Deutsch-Jozsa algorithm \cite{Deutsch,Deutsch_1}, Shor’s factoring algorithm~\cite{Shor,Shor1}, the quantum search algorithms~\cite{Grover,Grover_2,Boyer,Lidar,Shenvi, Kempe} and the quantum simulation algorithms~\cite{manin, Feynman,lloyd,bernien,zhan,neill2018}. 
The advantages of these quantum algorithms are assumed to be derived from the efficient use of quantum coherence and entanglement.

After Grover's seminal proposal \cite{Grover,Grover_2} of his eponymous quantum search algorithm, which has been shown to be a special case of the more general \textit{amplitude amplification} algorithm~\cite{Brassard}, an extensive amount of research effort has been directed towards implementing and studying the effects of noise on the efficiency of the algorithm in an actual quantum device. The experimental implementation  of the algorithm was first done using nuclear magnetic resonance techniques~\cite{Chuang_NMR}. Later on, the efficiency of the Grover's algorithm was studied in~\cite{Zalka} and a generalization of the algorithm for an arbitrary amplitude distribution was done in~\cite{Biham2}. For more works on the applications of the quantum search algorithm, see~\cite{abram,GuiLu, Kwiat,Sheng,Long,Biham3,Heinrich,Roland,Xiao} and for some experimental implementations, see~\cite{Jones,Vander,Ermakov,Bhattacharya,Zhang, WaltherP,Brickman,DiCarlo,Figg}.

Even if a quantum algorithm theoretically provides a significantly better efficiency in comparison with its classical counterpart, the efficiency in an implementation of the same undoubtedly depends on the actual fabrication of the relevant quantum circuit. 
Due to possible impurities in  circuit components and their erroneous implementations,
there may arise fluctuations or drifts, which can affect the performance of the quantum algorithm considerably. Therefore, characterizing such deviations from the ideal situation, caused by decoherence and noise, is important to assess the usefulness of an algorithm.
The disturbances may cause a unitary noise on the ideal system, i.e., a small perturbation can arise in the Hamiltonians describing the unitary gates, conserving the hermiticity of the Hamiltonian as well as the unitarity of the quantum gates.  See e.g.~\cite{Bernstein,bassi,Preskill,N&C}.

Studies on the consequences of noisy scenarios in quantum algorithms has started some decades back \cite{Barnes}. The effect of noise on the Grover's search algorithm
was studied in~\cite{Altaba}, which investigated 
the effect of random Gaussian noise on the algorithm's efficiency at each step.
A perturbative method was used 
in~\cite{Azuma} to study decoherence in a noisy Grover algorithm where each qubit suffers phase-flip error independently after each step. 
The effect of a noisy oracle was considered in~\cite{Tu, Bae}.
In~\cite{kwek}, the effect of depolarizing channels on all qubits was examined and it was found that the number of iterations needed to obtain the maximal efficiency of the success probability decreases with increasing decoherence. The effect of the Grover unitary becoming noisy was considered in~\cite{Biham} using a noisy Hadamard gate, with unbiased and isotropic noise, uncorrelated in each iteration of the Grover operators. An upper bound on the strength of the noise parameters up to which the algorithm works efficiently was deduced. 
A comparison of the effects of several completely positive trace preserving maps on the efficiency and computational complexity of the algorithm was described in~\cite{Gawron}. The performance of the algorithm under localized dephasing was studied in~\cite{Reitzner}. For more discussions and further ramifications of noise on the Grover's algorithm, see~\cite{Salas,  Hasegawa, Cohn}. 
Fault-ignorant quantum search was proposed in~\cite{Wolf} where the searched element is reached eventually but with the runtime depending on the noise level. Steane's \cite{Steane2} quantum error correction code was also employed in presence of the depolarizing channel in~\cite{Botsinis}. On the other hand, noise with correlations in time \cite{alicki_,bialczak,bylander,Paladino,Bose,haake,Daffer,Maniscalco_2006,buscemi,Ali_,benedetti2014non,Addis_2016,schwarma} and space \cite{flammia,aliferis,daharonov} have been observed in realistic quantum computing devices and detrimental effects of such noise on quantum error correcting codes have also been reported \cite{Clemens,Klesse,Novais,Cafaro,Clader}. 

In this paper, we study the effects of a noise that originates from probabilistic noisy unitary evolution of some register qubits in between any two Grover operations. In particular, we find a set of noisy qubit unitaries, for which the success probability of the algorithm remains unaffected by the number of noisy qubits. We refer to those special noise unitaries as the ``good noises." We extend our investigation to a type of time-correlated noise considered in \cite{Macchia,Macchia_2,Daems}, and examine its effects on the performance of the algorithm.  

We have organized the paper as follows. After reviewing the noiseless Grover algorithm in Sec.~\ref{noiseless algorithm}, we introduce our noise model and its physical motivation
in Sec.~\ref{noise}. 
Dynamics of the register under the Markovian-correlated noise is analyzed in Sec.~\ref{Example}. The time-correlation-less case of our noise model and its connection to some fundamental decoherence processes are then elucidated in Sec.~\ref{sec:unital}. In Sec.~\ref{Proof}, we give an overview of our analysis for finding the `good noises.' A measure of the algorithm's performance is introduced in Sec.~\ref{performance}. The effect of a memory-less and that of a Markovian-correlated noise on the efficiency of the Grover's algorithm are numerically studied in Sec.~\ref{numerics}. Sec.~\ref{Conclusion} concludes the paper.

\section{Grover Search: The noiseless case and our noise model}
The Grover search algorithm that we consider here
aims to find a single marked element from a search space of finite size. It is known to attain a quadratic speed-up over the best classical search. In our paper, we consider Grover search under a time-correlated local noise. In the succeeding subsections we discuss the ideal Grover algorithm and then introduce our noise model.

\subsection{The noiseless scenario}
\label{noiseless algorithm}
\begin{figure}[t]
    \centering
    \includegraphics[width=\linewidth]{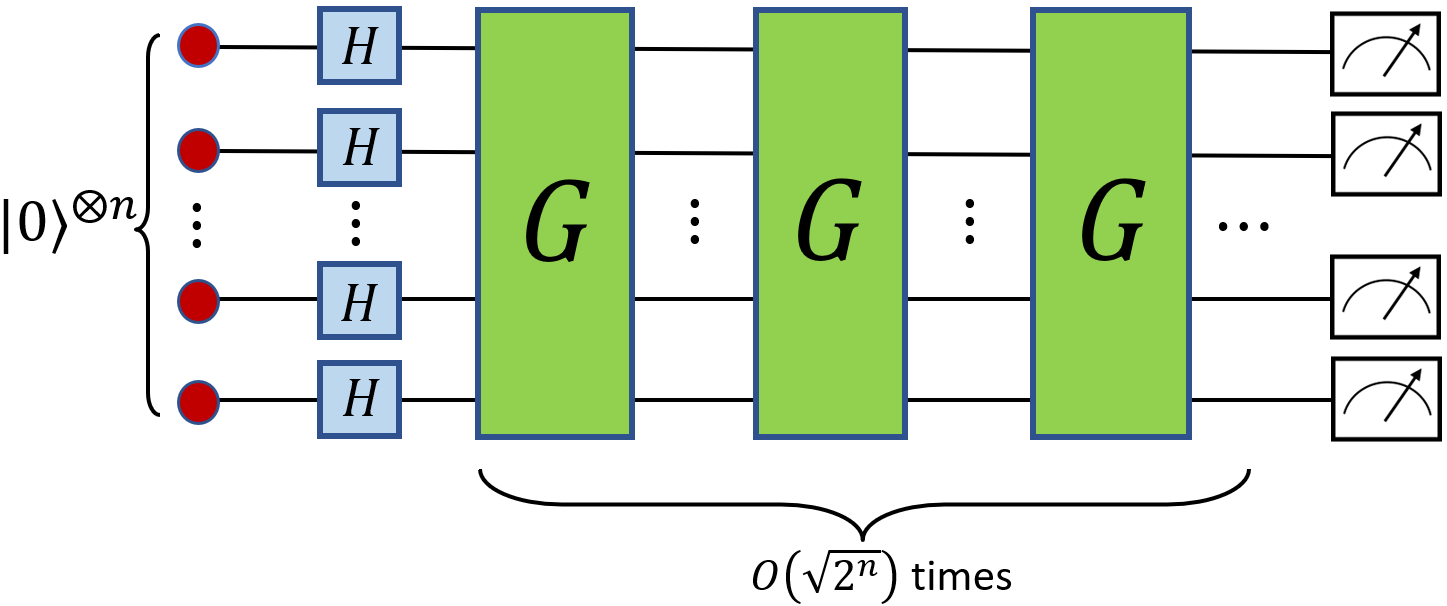}
    \caption{Grover's search algorithm in the noiseless situation. The register containing a string of $n$ qubits, each in the state $\ket{0}$, is each subjected to a Hadamard operation in the first step. The second step is the operation of Grover operator, for $t$ times, which is followed by a measurement on the output state of the register in the computational basis. Time taken to reach the maximal success probability is $O(2^{n/2})$. Further discussions presented in text.}
    \label{fig:my_label}
\end{figure}

The search algorithm is concerned with a search space $\{x\} = \{1,2,\ldots,N\}$ with $N=2^n$ elements. There exists a function $f: \{x\} \rightarrow \{0,1\}$ defined such that 
\begin{equation}
f(x)=
\begin{cases}
1, \quad \text{for} \quad x=w\;\; \text{(`marked element')}\\
0, \quad \text{for} \quad x\neq w.
\end{cases}
\label{eq:fx_1}
\end{equation}
To search for the marked element $w$, a classical computer evaluates $f$ for each of the elements until the value 1, i.e., the marked element is found. This requires $O(N)$ operations. The advantage of Grover's search algorithm over the classical one is that, by using a sequence of unitary operations, it can find the marked element in only $O(\sqrt{N})$ queries to $f$. The steps of the algorithm are described as follows and a schematic demonstration is shown in Fig.~\ref{fig:my_label}.

\begin{figure}
    \centering
    \includegraphics[width=.47\linewidth]{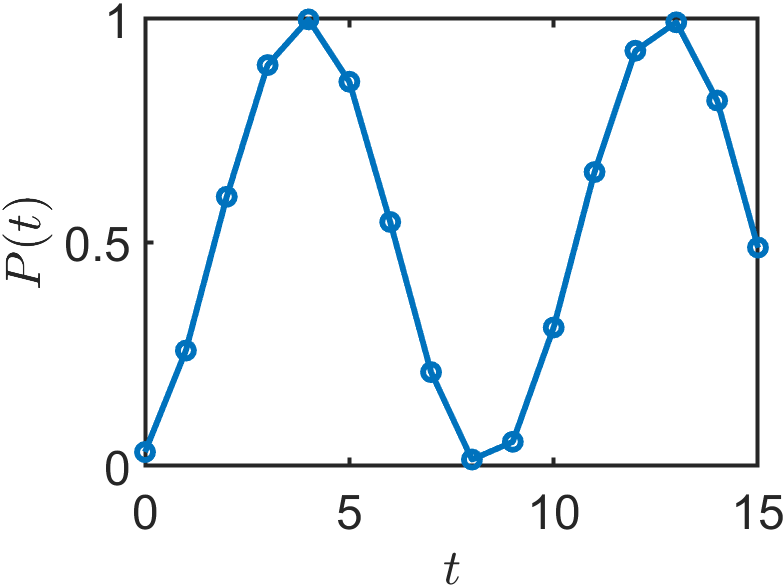}
    \caption{Noiseless Grover algorithm for $n=5$ qubits in the register. The number, $t$, of Grover iterations is along the horizontal axis and the success probability, $P(t)$, for finding the marked state, is plotted on the vertical axis. The smallest \(t\) for which \(P(t)\) is maximal is given by \(t=4\).
    All  quantities plotted are dimensionless.}
    \label{fig:Noiseless grover}
\end{figure}

It starts with all the qubits of an $n$-qubit register in the $|0 \rangle$ state, where $\ket{0}$ is the eigenvector of  Pauli-$z$ operator with eigenvalue 1. The next step is to act on each qubit by the Hadamard operator, $H = \frac{1}{\sqrt{2}}(\sigma_x +\sigma_z)$, where $\sigma_x$ and $\sigma_z$ are Pauli operators. This takes the total register to to an \textit{uniform superposition state},
 \begin{smal}[\small]
    |s \rangle &= \left( \frac{|0\rangle +|1\rangle}{\sqrt{2}} \right)^{\otimes n}  = {\frac{1}{\sqrt{N}}}\left(\sum_{\substack{x = 1 \\ x\neq w}}^{N} |x\rangle + |w\rangle \right),
    \label{eq:1s}
\end{smal}
where $|w\rangle$ is the \textit{marked state}, i.e., the state corresponding to the element we are searching for in the database of $N = 2^n$ elements. The state $|s\rangle$  is then acted on by the \textit{Grover operator} $G$, given by $G = D\ O$, where $D = (2|s\rangle \langle s|- \mathds{1}_{N})$ is called the \textit{Diffuser} and $O = (\mathds{1}_{N}-2|w\rangle \langle w|)$ is the \textit{Oracle}. For a detailed discussion about the construction of the \textit{Diffuser} $D$, \textit{Oracle} $O$ and the Grover $G$ unitaries, see e.g.~\cite{Kitaev,N&C}. The operator $G$ has the form,
\begin{equation}
    G = -\mathds{1}_N + 2|s\rangle \langle s|- \frac{4}{\sqrt{N}}|s\rangle \langle w| + 2 |w\rangle \langle w|. \label{eq:3}
\end{equation}
It acts on successive states until the state of the register, $|\psi(t)\rangle = G^t |s\rangle$
reaches close enough to the marked state $|w\rangle$. Here $t$ stands for the number of times the Grover operator is employed after the first step, i.e., after the Hadamard
operation. The \textit{success probability}, i.e., the probability to find the marked state after $t^{th}$ operation, is given as $P(t) = |\langle w|\psi(t)\rangle|^2$. It can be checked that the marked element is reached after around $t=\lfloor \frac{\pi}{4}\sqrt{N}\rfloor$ Grover iterations. See Fig.~\ref{fig:Noiseless grover}.
\subsection{The noise model}
\label{noise}
For a large database, the number of iterations of the Grover operator will be large, although quadratically smaller than that for the classical algorithm, to reach the first maximal success probability. A high number of applications of Grover operator may result in some noise or fluctuations in the circuit parameters performing the computation -- affecting the efficiency of the algorithm \cite{Preskill2018,murali2019}. In this paper, we consider that in the interval between any two consecutive Grover operations, some $m$ out of the total $n$ qubits evolve under some unitary $U$ that we call the `noise unitary,' at a rate specified by the \textit{noise probability}. Such local single qubit errors in a quantum register due to probabilistic unitary qubit evolutions have been studied previously in numerous settings (see the discussions and references in Secs. \ref{Intro} and \ref{sec:unital}).

We express the effect of this noisy evolution in the form of the total noise unitary $\bigchi_m$ acting on the whole register.
For example, it can be $\bigchi_{m}=\left(U\otimes (\mathds{1}_2)^{\otimes (n-m)} \otimes U^{\otimes (m-1)}\right)$, meaning $m$ noisy qubits evolving under $U$ and $(n-m)$ noiseless qubits acted on by the identity operator $\mathds{1}_2$. We will call the number of noisy qubits $m$ as the `noise strength'. The positions of the $m$ noise sites are allowed to be arbitrary, but fixed during a given run of the algorithm. The noise $\bigchi_m$ occurs with some well-defined probability after every Grover iteration and we can incorporate its effect on the algorithm by defining a new unitary $G'=\bigchi_m\  G$, which we will call the `noisy Grover operator'. Using Eq.~\eqref{eq:3},
\begin{equation}
G' = - \bigchi_m + 2 \left(\bigchi_m|s\rangle\langle s|+\bigchi_m|w\rangle\langle w|\right)- \frac{4}{\sqrt{N}} \bigchi_m|s\rangle \langle w|.
\label{eq:6}
\end{equation}

The probabilistic occurrences of the noise could possibly
even be correlated in time and we assume in this paper
that the noise at each consecutive time steps is Markovian-correlated.

The motivation behind choosing this type of noise model comes from the possibility of spatiotemporally correlated errors~\cite{Kwiatkowski,LViola} and unwanted qubit cross-talks~\cite{gambetta_2012,proctor2019,sarovar2020,murali2020,pengzhao} in the currently available experimental setups for implementing Grover search algorithm. We discuss below one such noisy scenario.

\subsubsection{A physical scenario motivating the noise model}
\label{sec:physical_scenario}
\begin{figure}
    \centering
    \includegraphics[width=\linewidth]{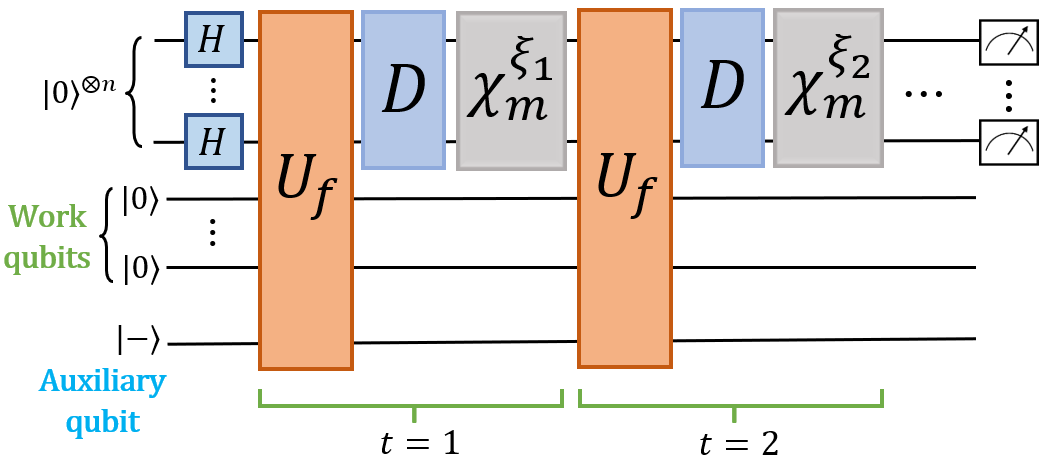}
    \caption{Circuit diagram for Grover algorithm under time-correlated noise. The $n$-qubit register is initialized in the state $|0\rangle^{\otimes n}$. Total noise unitary $\bigchi_m$ acts in between two consecutive perfect Grover iterations. $\xi_t$ is the Markovian process introduced in Sec.~\ref{sec:physical_scenario}. $D$ denotes the diffuser unitary of Sec.~\ref{noiseless algorithm}. $U_f$ implements the Oracle $O$ using the work and auxiliary qubits initialized in $|0\rangle$ and $|-\rangle$ states respectively. $t$ denotes the number of noisy Grover iterations.}
    \label{fig:Noisy_grover}
\end{figure}

Let us first discuss the ideal experimental setup of the implementation of Grover search algorithm. We have already discussed in the previous section that the Grover operator, $G$, contains two parts: one is the Diffuser $D$ and the other is the Oracle $O$. The ideal experimental setup can be seen in Fig.~\ref{fig:Noisy_grover}, if we igonore the noisy evolutions $\bigchi_m^{\xi_i}$ for $i=1,2\ldots$. As shown in Fig.~\ref{fig:Noisy_grover}, the oracle $O$
can be implemented by introducing an auxiliary qubit and an `oracle workspace'~\cite{N&C} to the circuit. The auxiliary qubit in this case needs to be initialized in the state $|-\rangle = \frac{|0\rangle-|1\rangle}{\sqrt{2}}$, and then evolved together with the quantum register, under the unitary $U_f$, that acts on the joint state of the register and auxiliary qubit, $|x\rangle\otimes|q\rangle$, as follows:
\begin{equation}
    U_f(|x\rangle \otimes  |q\rangle) = |x\rangle \otimes |q \oplus f(x)\rangle.
\end{equation}
Here the function $f(x)$ is given in Eq. \eqref{eq:fx_1} and the $\oplus$ denotes the modulo $2$ addition. It can be easily verified that $|q\rangle=|-\rangle$ recovers the oracle operation on the register's state, so that $U_f(|x\rangle \otimes  |-\rangle) = (O|x\rangle) \otimes |-\rangle$. 
The physical implementation of $U_f$ generally requires the use of multiple `work qubits' \cite{N&C,harmut,sinha2010quantum} in the oracle workspace (see Fig.~\ref{fig:Noisy_grover}). Suppose the workspace has $\bar{n}$ work qubits, each initialized in $|0\rangle$ state and then evolved with the auxiliary and the $n$ register qubits under some combination of several $1$-qubit and $2$-qubit gates, depending on the particular $U_f$ \cite{barenco1995}. The 2-qubit gates, due to technical constraints \cite{craik2017,Preskill2018,erhard2019,lienhard2019,BAKER_murali,molavi2022} of physical implementation, require the concerned qubits to be in close proximity \cite{chhangte,burgholzer2022}. This increases the possibility of spatially correlated errors \cite{Kwiatkowski,LViola} on those qubits. Unwanted cross-talks \cite{gambetta_2012,proctor2019,sarovar2020,murali2020,pengzhao} could creep in while the qubits are \textit{idle}, i.e, in between any two Grover steps -- when no gates are applied on the register. For these types of errors and noises, some of the register qubits face noisy evolutions.

Let us consider that between two Grover steps, a cross-talk error occurs due to a stochastic interaction Hamiltonian \cite{onorati2017} of the form
\begin{equation}
    \Theta^{(r,\nu)} =  \xi_t (h^{(r)} \otimes \sigma_z^{(\nu)}),
    \label{eq:theta}
\end{equation}
with $h^{(r)}$ acting on the $r$-th qubit of the register and Pauli $\sigma_z^{(\nu)}$ acting on the $\nu$-th work qubit. The dimensionless coupling strength $\xi_t$ undergoes fluctuations at each time step. The Hamiltonians $h^{(r)}$ and $\Theta^{(r,\nu)}$ are taken to be 
dimensionless. The Hamiltonian $h^{(r)}$ leads to the local `noisy unitary' $U=\text{exp}(-ih^{(r)})$ on the register qubit at each time step, as described in Sec. \ref{noise}. For the composite setup of the register and the auxiliary qubit, we have $\text{exp}(-i\, \Theta^{(r,\nu)})=U^{\xi_t}\otimes|0\rangle\langle 0|\,+\, (U^\dagger)^{\xi_t}\otimes|1\rangle\langle 1|$, which is a unitary on the $r$-th register qubit controlled by the $\nu$-th work qubit \cite{wallraff,bradley} at each time step.

Now suppose that any $m$ of the total $n$-qubit register suffer the cross-talk error, given in Eq.~\eqref{eq:theta}, in the interval between two Grover iterations at time $t$. Hence, the total noisy Hamiltonian becomes $\Theta = \sum_{(r,\nu)} \Theta^{(r,\nu)}$, where the sum is over all the $(r,\nu)$ pairs corresponding to the $m$ noisy register qubits. In the physical implementation of Grover's algorithm, all the work qubits are unitarily brought back to $|0\rangle$ state after each oracle operation using uncomputation \cite{Uncomp_bennett,unco_aaron,amy_ross,paradis}. Thus, the state of the workspace before and after the complete oracle operation is $|0\rangle^{\otimes n}$. So, the joint state of the register $|\psi(t)\rangle$ and the oracle workspace evolve under the total noise unitary, due to the interaction Hamiltonian in Eq.~\eqref{eq:theta}, as $e^{-i\Theta}(|\psi(t)\rangle\otimes |0\rangle^{\otimes n})$. 
This situation is analogous to the case where the noise unitary $\bigchi_m$, introduced in the previous section, is being acted on the register qubit. Therefore, we can write 
\begin{equation}
    e^{-i\Theta}(|\psi(t)\rangle\otimes |0\rangle^{\otimes n}) = (\bigchi_m^{\ \xi_t} \ |\psi(t)\rangle)\otimes |0\rangle^{\otimes n}
\end{equation}
Here $\xi_t$ indicates that the noise can be time-correlated. The occurrence of this type of noise is demonstrated in Fig.~\ref{fig:Noisy_grover}.

In this paper, we consider the coupling strength $\xi_t$ to be a \textit{time-homogeneous} discrete-time Markov process \cite{doob1953stochastic,kemeny1960finite,grimmett2020probability}. Particularly, we choose the dichotomous Markov chain considered in~\cite{Macchia, Macchia_2, Daems,bena}. This kind of time-dependent coupling strength may arise due to a noisy coupling field, that couples the register and work qubits \cite{yu2006sudden,szankowski2015,aolita2015open}, and also due to a qubit in the environment \cite{Jurcevic_2022} or a spurious control field \cite{rudinger}. For our noise model, $\xi_t$ takes the two values 0 and 1 according to the following conditional probabilities: 
\begin{align}
& \text{Pr}(\xi_{t+1}=l|\xi_{t}=k) \nonumber\\
&= (1-\mu)\,\text{Pr}(\xi_{t}=k) + \mu\ \delta_{lk}\nonumber\\
&=\  (1-\mu)\,p_{k} + \mu\ \delta_{lk}  = p_{k|l}
\label{eq:xi_t_}
\end{align}
Here $p_k = \text{Pr}(\xi_{t}=k)$ denotes the probability of the event $k$ of the Markov process. $p_{k|l}$ denotes the conditional probability of event $k$, given that event $l$ happened in the previous time step. The parameter $\mu$ will be referred to as the `memory parameter' and it can take any real value from 0 (memory-less) to 1 (perfect memory).

This kind of correlated noise with partial memory can potentially be found in real quantum devices, and it has been shown to provide an enhancement in the transmission of classical information as compared to transmission through noisy channels without memory~\cite{Macchia}. In the next subsection, we describe the register's time evolution under this time-correlated noise. It will become evident that such scenario could arise if the noisy qubits in the quantum register get coupled to an external degree of freedom acting as a physical memory state \cite{Bowen_2004,Specht,kumar,bradley}.

\subsection{Time evolution under Markovian-correlated noise}
\label{Example}
In our noise model, 
a total unitary evolution by $\bigchi_m$ of locally evolving $m$ noisy qubits is a \textit{probabilistic} process, happening after each noiseless Grover evolution $G$. This noisy evolution is `probabilistic' in the sense that after a given Grover evolution $G$, the state of the register is a convex mixture of two possible states: one corresponding to no noise after evolution by $G$ and another corresponding to a noisy evolution by $\bigchi_m$ after $G$. Now, as we discussed in the previous section, it can happen that the probability of noise at a given time depends on the history of the register's noisy evolution \cite{Kre,Man_rev}. In this paper we consider the simplest of such situations -- where this noise is Markovian-correlated in time (see Appendix \ref{app:Markovian}). In this case, the evolution at each given instant is affected only by the immediately previous time step. This potentially important variety of noise with memory has not yet been studied before in case of the Grover algorithm. It is to be noted here that the results shown in the paper are not exclusive to only this kind of noise, and validity in this case will serve as an indication to the generality of the results.

\begin{figure}
    \centering
    \includegraphics[width=\linewidth]{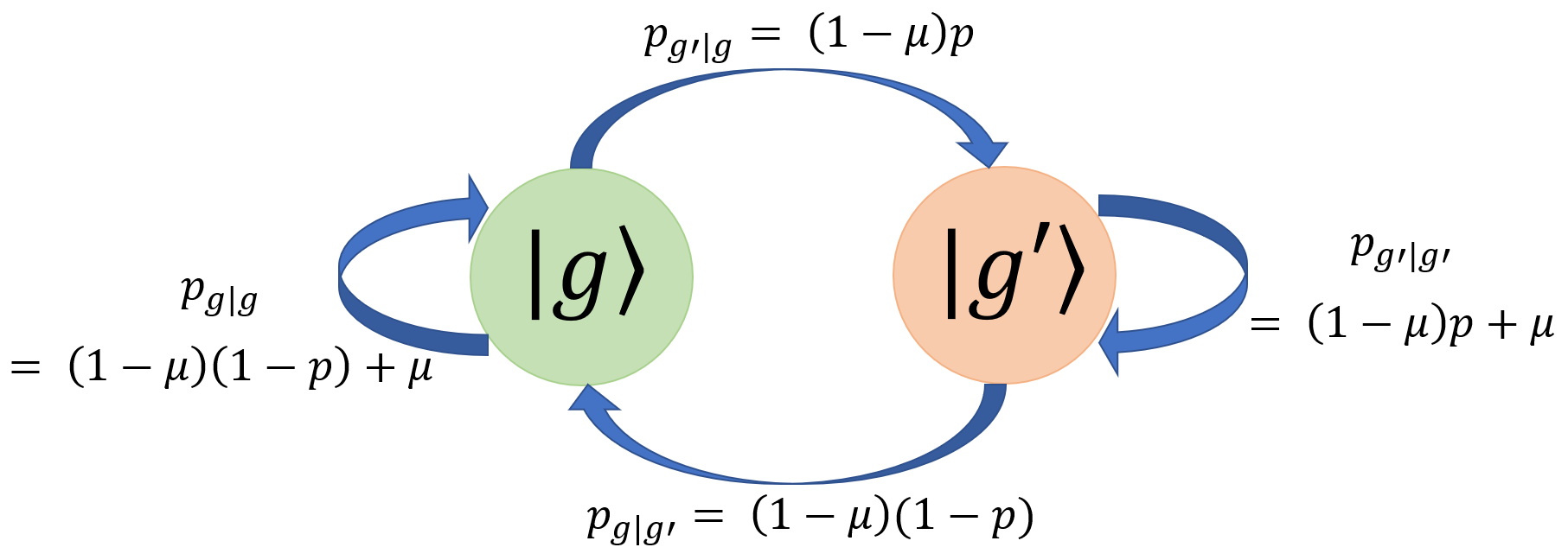}
    \caption{Schematic diagram of transitions of the walker's state at $t \ge 2$ with the conditional probabilities defined in Eq.~\eqref{eq:xi_t_}. Please see the text for details.}
    \label{fig:Markovian probabilities}
\end{figure}

Time evolution under the time-correlated noise is easier to describe if we incorporate an extra (not necessarily physical) degree of freedom, the \textit{walker}, which we can trace out after the evolution of the composite register-walker state. The walker helps to keep track of the fluctuating coupling strength $\xi_t$ introduced in Sec.~\ref{sec:physical_scenario}. The walker has two orthogonal states: $|g\rangle$ and $|g'\rangle$ and at each time step, it performs a transition between these two states with some well-defined probability. Particularly, it is in the state $|g'\rangle$ when $\xi_t=1$, and in $|g\rangle$ when $\xi_t=0$. A schematic diagram is shown in Fig.~\ref{fig:Markovian probabilities}. When it transitions to $|g'\rangle$, all the $m$ qubits connected to it are rotated by a unitary $U$ and the other $(n-m)$ are left as they were. When it transitions to $|g\rangle$, all the $n$ qubits connected to it are left as they were. Thus, at each time step, application of an ideal unitary Grover operator $G$ is followed by each of the following with some corresponding probabilities:
\begin{enumerate}
    \item[($I$)] any $m$ out of $n$ qubits are rotated by a unitary $U$, i.e., walker is in state $|g'\rangle$, or,
    \item[($II$)] all the $n$ qubits are left untouched, i.e., walker is in state $|g\rangle$.
\end{enumerate}
 To make the situation clearer, let us say that after the $(t-1)^{\text{th}}$ noisy Grover iteration, the register is in a state given by the density matrix $\rho_{t-1}$. After the noisy \(t^{\text{th}}\) iteration, the register will be a convex mixture of the following two possible states:
\begin{eqnarray*}
&(I)& \ \rho_t = G'\,\rho_{t-1}\,G'^{\dagger},\;\; \text{with}\;\; G'= \bigchi_{m}\,G,\\
&(II)& \ \rho_t = G\,\rho_{t-1}\,G^{\dagger}.
\end{eqnarray*}
These processes are dictated by the state of the walker, which in turn performs transitions according to the Markov process as described in Eq.~\eqref{eq:xi_t_}. When $\mu=0$, noise at each time step is independent of what happened in the previous step, since $p_{k|l}=p_k$. On the other hand, $\mu=1$ leads to $p_{k|k}=1$, meaning that in the case of \textit{perfect} memory, the walker state \textit{remains fixed} throughout the evolution. At $t=1$, i.e., on
the first Grover iteration, the probabilities of $(I)$ and $(II)$ are determined by the initial probabilities of the walker to be in states $|g'\rangle$ and $|g\rangle$ respectively. These probabilities are called \textit{stationary probabilities} and are taken to be $p_{g'} = p$ and $p_g = (1-p)$ respectively. Here $p$ can be referred to as the \textit{noise probability}.  
Note that, here $p$ and $(1-p)$ are equal to $p_1$ and $p_0$, respectively in Eq.~\eqref{eq:xi_t_}.

Before the application of the first Grover iteration, the $n$-qubit register is in the uniform superposition state corresponding to $\rho_0$ $\coloneqq$ $|s\rangle  \langle s| $. Thus the density matrix of the composite system containing the walker and the register before applying the first Grover iteration is $R_0 = \left(\frac{|g\rangle+| g'\rangle}{\sqrt{2}}\right)\left(\frac{\langle g|+\langle g'|}{\sqrt{2}}\right) \otimes |s\rangle\langle s|$. So, the state of the register after the first and subsequent Grover iterations will be obtained by tracing out the walker from $R_t$, i.e., $\rho_t = \text{Tr}_{walker} \{R_t\}$. In the following, the superoperators $\Phi^{g}[\bullet]$ and $\Phi^{g'}[\bullet]$ acting on an operator $\rho$ will represent unitary evolutions $G \rho \, G^\dagger$ and $G' \rho \, G'^\dagger$, respectively. The time evolution of $\rho_t$ can then be expressed using \textit{transition operators} $S_0$ and $S$ as
\begin{eqnarray}
&&R_1 = S_0 R_0 =\biggl( p_g \Bigl(|g\rangle\langle g| + |g\rangle\langle g'|\Bigr) \otimes \Phi^{g}[\bullet]\nonumber\\
&& \phantom{ami kon pothe}+ p_{g'}\Bigl( |g'\rangle\langle g| +  |g'\rangle\langle g'|\Bigl) \otimes \Phi^{g'}[\bullet]\biggr)R_0\nonumber\\
&&=\biggl( p_g \Bigl(|g\rangle\langle g| +|g\rangle\langle g'|\Bigr)\otimes \Phi^{g}[\rho_0]\nonumber\\
&&\phantom{ar kon kotha} + p_{g'}\Bigl(|g'\rangle\langle g| +  |g'\rangle\langle g'|\Bigr) \otimes \Phi^{g'}[\rho_0]\biggr).
\label{eq:R_1_}
\end{eqnarray}
Therefore, $\rho_1 = \text{Tr}_{walker} \{R_1\} = p_{g}\Phi^{g}[\rho_0]+p_{g'}  \Phi^{g'}[\rho_0]$.
\begin{eqnarray}
&& R_2 = S R_1=\biggl( p_{g|g}|g\rangle\langle g| \otimes \Phi^{g}[\bullet]+p_{g|g'}|g\rangle\langle g'| \otimes \Phi^{g}[\bullet]\nonumber\\
&&+p_{g'|g} |g'\rangle\langle g| \otimes \Phi^{g'}[\bullet] +p_{g'|g'}|g'\rangle\langle g'| \otimes \Phi^{g'}[\bullet]\biggr)R_1.\;\;\;\;\;\;\
\label{eq:R_2_}
\end{eqnarray}
Hence, $\rho_2 = \text{Tr}_{walker} \{R_2\}=\sum_{i,j}p_{i|j}p_j\Phi^{i}[\Phi^{j}[\rho_0]]$, where $i$ and $j$ can be $g$ or $g^{\prime}$.
Similarly, for $t\geqslant2$, we have
\begin{align}
    R_t &= S^{t-1} R_1,\nonumber\\
    \rho_t &= \text{Tr}_{walker}\{R_t\}.
\end{align}
The success probability, i.e., the probability to find the marked state at time $t$, is given as
\begin{equation}
    P(t) = |\langle w|\rho_t |w\rangle|.
    \label{29}
\end{equation}
 We show in the next subsection that in the absence of any time-correlations, our noise model reduces to some well-known decoherence processes. 

\subsection{\texorpdfstring{$\mu$} 
 \texorpdfstring{$ \ = 0$} and unital decoherence processes} \label{sec:unital}
In the case when the noise in consecutive steps do not have any time-correlations, i.e., $\mu=0$, Eq.~\eqref{eq:xi_t_} has the form $p_{g|g}=p_g=p_{g|g'}$ and $p_{g'|g}=p_{g'}=p_{g'|g'}$. Putting these in the expressions of $S_0$ and $S$, we get 
$S=S_0$. Taking the initial register state $\rho_0$ and the composite walker and register state $R_0$ as given in previous section, we get the register's state $\rho_1$ after the first noisy Grover iteration as
\begin{align}
    \rho_1 &= \text{Tr}_{walker} \{R_1\}= \text{Tr}_{walker} \{S_0\, R_0\}\nonumber\\
    & = (1-p)\Phi^{g}[\rho_0] + p\Phi^{g'}[\rho_0]\nonumber\\
    & = (1-p)\ (G\rho_0 G^{\dagger}) + p\ \bigchi_m (G \rho_0 G^{\dagger}) \bigchi_m^{\dagger}. \label{eq:14}
\end{align}
Thus, the noisy evolution after the first noiseless Grover operation is a quantum dynamical map \cite{sudarshan1961,choi1975,Lind,kraus1983} $\mathcal{E}$ given by the Kraus operators $K_1 = \sqrt{1-p}\ \mathds{1}_N$ and $K_2 = \sqrt{p}\ \bigchi_m$ so that $\mathcal{E}[\rho]=K_1\rho K_1^{\dagger}+K_2\rho K_2^{\dagger}$ and $\rho_1=\mathcal{E}[G\,\rho_0\, G^{\dagger}]$. Since $S_0=S$ in case of $\mu=0$, we have for $t\geq 1$,
\begin{equation}
    \rho_t = \mathcal{E}[G\,\rho_{t-1}\, G^{\dagger}]. \label{eq:E_G_r}
\end{equation}
Note that for $\mu>0$, an expression like Eq.~\eqref{eq:E_G_r} is not possible because of noise being conditioned on the application at the previous time step.

For $U=\sigma_x$, the noisy operation $\mathcal{E}$ thus becomes an $m$-qubit \textit{bit-flip} channel. Similarly, $U = \sigma_z$ leads to \textit{phase-damping} and $U = \sigma_y$ to \textit{bit-phase flip} channels. A comparison of the effects of these channels on the Grover algorithm was done extensively in \cite{Gawron}. In fact, all these are examples of unital channels (that is, $\sum_{i} K_i^{\dagger}K_i=\mathds{1}$) and \textit{any} unital channel can be expressed, like in Eq.~\eqref{eq:14}, as an affine combination of unitary channels \cite{Mendl}.

We have shown how the memory-less special case of the Markovian-correlated noise gives rise to some of the most relevant sources of decoherence in quantum registers \cite{proctor2019}. It is a general feature that the success probability of an algorithm reduces with increase in strength of noise, as seen e.g.
in~\cite{Biham}. Whereas, there is a possibility of identifying such noise for which the decrease in the success probability does not depend on the number of noisy qubits, $m$. For an example, see Appendix \ref{App_A}. If it is possible to choose between different noise generating unitaries in an experimental setup, it will be helpful to have those noise unitaries which do not decrease the success probability with increase in the noise strength. We can christen such noise unitaries as the `\textit{good noises}'.
In the succeeding section, we try to identify the form of 
such good noises.

\section{The set of ``good noises"}
\label{Proof}
To find what the good noises are, we will start with the most general single-qubit
unitary matrix (in the $\{|0\rangle,\ |1\rangle\}$ basis), viz.
\begin{equation}
    U = \begin{pmatrix} a & b\\ -\olsi{b} e^{i\theta} & \olsi{a} e^{i\theta}
    \end{pmatrix},
    \label{unitary}
\end{equation}
with $a,b\in \mathbb{C}, |a|^2+|b|^2=1 $, $\theta \in [0,2\pi)$, and $\olsi{z}$ denoting the complex conjugate of $z$. The good noise corresponds to the values of $a,b$ and $ \theta$, for which the success probability $P(t)$ (Eq.~\eqref{29}) remains unchanged on changing the value of $m$. If the total noise unitary $\bigchi_m$ acts on $m$ sites, we will denote $P(t)$ in that case as $P_m(t)$. The good noises will be found through elimination of $U$'s for which $P_m(t)$ changes with $m$.
Alongside, we will also show the independence of the positions of the $m$ noisy qubits as long as $U$ is one of the Pauli matrices.

To start with, we will check under what conditions the success probability at time $t=1$ remains constant under varying $m$. After that, we will extend our investigations for the times $t>1$. A detailed calculation of the search for good noises is given in Appendix \ref{app:proof} for $t=1$. The derivation begins with the aim to keep $P_m(t=1) = (1-p)|\langle w|G|s\rangle|^2+p|\langle w|G'|s\rangle|^2$ constant with the alteration of $m$, and we find that both $a$ and $b$ of Eq.~\eqref{unitary} can not be non-zero (see the derivation of Eqs.~\eqref{Cond-1-1} and \eqref{Cond-1-2}). The first condition for constructing a good noise comes to be
\begin{equation*}
    \text{\textit{Condition 1:} $|a|=1$ or $|b|=1$}.
\end{equation*}
This condition implies that $\bigchi_m$ must be a \textit{generalized permutation unitary matrix}. We can then introduce a state $|w'\rangle$ as follows:
\begin{smal}[\small]
\bigchi_m|w\rangle\langle w|\bigchi_m^{\dagger}=
\begin{cases}
 |w\rangle\langle w|, & \text{for} \quad |a|=1,\;\;\;\;\;\;\;\\
 |w'\rangle\langle w'|, & \text{for} \quad |b|=1.\;\;\;\;\;\;\;
 \label{eq:chi_m_w}
\end{cases}
\end{smal}
It turns out that the state's evolution can be written in the basis $\mathbb{B}$, where
\begin{smal}[\small]
\mathbb{B}=
\begin{cases}
 \{|s'_1\rangle,\ldots,|s'_{M}\rangle,|w\rangle\}, & \text{for} \ \ |a|=1, \;\;\;\;\;\;\;\\
 \{|s'_1\rangle,\ldots,|s'_{M}\rangle,|w\rangle, |w'\rangle \}, & \text{for} \ \ |b|=1.\;\;\;\;\;\;\;
 \label{eq:basis_b}
\end{cases}
\end{smal}
See the arguments around Eqs.~\eqref{decomp_1} and~\eqref{decomp_2} in Appendix~\ref{app:proof}. 
The basis set $\mathbb{B}$ is different from the $N$-dimensional computational basis set $\{|x\rangle\}$ used in Eq.~\eqref{eq:1s}. The basis elements, $|s'_i\rangle$, are constructed using the computation basis states $\{|d\rangle_i\}_{d\in \{x\}}$, as
\begin{equation*}
\quad |s'_i\rangle = \frac{1}{\sqrt{\Delta_i}}\ \underset{|w\rangle,|w'\rangle \notin \{|d\rangle_i\}}{\mathlarger{\sum}}|d\rangle_i,
\end{equation*}
with $\Delta_i = \text{size}(\{|d\rangle_i\})$, $\bigsqcup_{i} \{|d\rangle_i\}=\{|x\rangle\}$, and
$\langle s'_i|s'_j\rangle = \delta_{ij}$. 
The dimension of $\mathbb{B}$ is $M+1$ and $M+2$ for $|a|=1$ and $|b|=1$, respectively. For a matrix $U$ satisfying \textit{Condition 1}, there are two possibilities: its two non-zero elements are either equal, or unequal. As elaborated in Appendix \ref{app:proof}, this implies 
\begin{smal}[\small]
\text{dim}(\mathbb{B})=
\begin{cases}
 2 \text{ or } 3, & \text{for} \ \ |a|=1, \;\;\;\;\;\;\;\\
 3 \text{ or } 4, & \text{for} \ \ |b|=1,
 \label{eq:basis_M}
\end{cases}
\end{smal}
which then leads to another necessary condition:
\begin{equation*}
    \text{\textit{Condition 2:} $M = 1$ or $2$.} 
\end{equation*}
This requirement ensures that the dimension of the basis $\mathbb{B}$ remains constant for any given number of noise sites $m$. See Appendix~\ref{app:proof} for more delails.
The Conditions $1$ and $2$ narrow down the possible set of good noises to a restricted set of unitaries -- the Pauli matrices $e^{i\phi}\,\mathds{1}_2$, $e^{i\phi}\,\sigma_x$, $e^{i\phi}\,\sigma_y$ and $e^{i\phi}\,\sigma_z$, for any $\phi \in [0,2\pi)$. Basically, we have derived that the above three (excluding the trivial $\mathds{1}_2$) noises lead to $P_m(t=1)=P_{m+1}(t=1), \forall m$ and thus satisfying the criteria for being good noises.

We now check if these noise unitaries belong to the set of `good noises' \textit{for all times}, i.e., for $t>1$. From Eqs.~\eqref{eq:R_1_}, \eqref{eq:R_2_} and~\eqref{29}, we can see that the success probability at time $t$, for $m$ noisy qubits, can be written as $P_m(t)$ $= \sum_{\{\gamma_m(t)\}} p_{\gamma_{m}(t)} |\langle w|\gamma_{m}(t)|s\rangle|^2$. 
$\{\gamma_m(t)\}$ is the set of results from the multiplication of all possible length-$t$ configurations composed of the two unitaries $G$ and $G'$. For example, at $t=2$, $\{\gamma_m(2)\}=\{GG,GG^{\prime},G^{\prime}G,G^{\prime}G^{\prime}\}$. $\{p_{\gamma_m(t)}\}$ are the respective probabilities of each such `trajectory' $\gamma_m(t)$ in $\{\gamma_m(t)\}$. To satisfy the requirement of $P_m(t) = P_{m+1}(t), \forall m$ and all $t$, we need to have $|\langle w|\gamma_m(t)|s\rangle|=|\langle w|\gamma_{m+1}(t)|s\rangle|, \forall m$ for any time $t$. We can check that $\langle w|\gamma_m(t)|s\rangle$ have to be polynomials of order $t$ of the four variables: $\langle s|\bigchi_m|s\rangle$, $\langle s|\bigchi_m|w\rangle$, $\langle w|\bigchi_m|s\rangle$, and $\langle w|\bigchi_m|w\rangle$. Now, for $U\in \{\sigma_x,\sigma_y,\sigma_z\}$, we have $\bigchi_m^2=\mathds{1}_{N}$.
So, the constituent non-zero terms in $\langle w|\gamma_m(\tau)|s\rangle$ for any $t=\tau$ will have degrees with the same parity as $\tau$, i.e., the degrees of each term will belong to the set $\{\tau,\tau-2,\tau-4,\ldots\}$. For example, a trajectory $\gamma_m(2)=G^{\prime}G^{\prime}$ will correspond to the polynomial $\langle w|\gamma_m(2)|s\rangle$ of order 2. It contains terms of degree 2, such as $\langle w|\bigchi_m|s\rangle\langle s|\bigchi_m|s\rangle$, $\langle w|\bigchi_m|w\rangle^2$, etc., and terms of degree 0, such as $\langle w|w\rangle=1$ and $\langle w|s\rangle = \frac{1}{\sqrt{N}}$.

It can be shown that $\langle s|\bigchi_m|s\rangle =\frac{1}{N}\sum_{k=1}^{N}\psi_q = [\frac{a+b}{2}+e^{i\theta}\frac{\olsi{a}-\olsi{b}}{2}]^m$, where $\psi_q$ is introduced in Eq.~\eqref{deq:e10}. Also, $|\langle s|\bigchi_m|w\rangle|=|\langle w|\bigchi_m|s\rangle|=\frac{1}{\sqrt{N}}$, and $|\langle w|\bigchi_m|w\rangle|=|a|^{m}$. These results will be used in the following arguments for verifying the constancy of $|\langle w|\gamma_m(t)|s\rangle|$ with respect to $m$ for any given $t$, in case of the Pauli matrices.

For $U=\sigma_x$, we have $a=0$, $b=1$, and $\theta=\pi$. So, $\langle s|\bigchi_m|s\rangle = 1$, $\langle w|\bigchi_m|s\rangle =\frac{1}{\sqrt{N}}=\langle s|\bigchi_m|w\rangle$ and $\langle w|\bigchi_m|w\rangle = 0$, $\forall m$. Thus, $|\langle w|\gamma_m(t)|s\rangle|$ for any $t$ does not depend on $m$. This in turn implies that that $P_m(t)$ is independent of $m$ in case of $U = \sigma_x$.

For $U=\sigma_z$, we have $b=0$, $a=1$ and $\theta=\pi$ and so $\langle s|\bigchi_m|s\rangle=0$. It also comes from Eq. \eqref{eq:chi_m_w} that the magnitudes $|\langle w|\bigchi_m|s\rangle|$, $|\langle w|\bigchi_m |w\rangle|$, and $|\langle s|\bigchi_m|w\rangle|$ remain constant with respect to $m$. Moreover, $\text{sgn}(\langle w|\bigchi_m|s\rangle)$ $=$ $\text{sgn}(\langle w|\bigchi_m |w\rangle)$ $=$ $\text{sgn}(\langle s|\bigchi_m|w\rangle),\ \forall m,$ where $\text{sgn}(z) = \frac{z}{|z|}$ with $z\in \mathds{R}$. Thus, in case of $\sigma_z$, $|\langle w|\gamma_m(t)|s\rangle|$ for any $t$ depends on the three variables $\langle w|\bigchi_m|s\rangle$, $\langle w|\bigchi_m |w\rangle$, and $\langle s|\bigchi_m|w\rangle$. The magnitudes of these variables remain constant with $m$, but their signs, which do vary with $m$, are nevertheless equal among themselves. 
We have shown above that the constituent terms of the polynomials are of same parity (all even or all odd), whereby we can infer that the value of $|\langle w|\gamma_m(t)|s\rangle|$ is not affected by $m$. Hence, our claim for $\sigma_z$ to be a good noise thus also holds for any time $t$.

In case of $U=\sigma_y$, we have $a=0$, $b=-i$ and $\theta=\pi$. So, $\langle s|\bigchi_m|s\rangle=0$ and $\langle w|\bigchi_m|w\rangle=0$. We also have $\langle w|\bigchi_m|s\rangle = (-1)^m \langle s|\bigchi_m|w\rangle$. Thus, $|\langle w|\gamma_m(t)|s\rangle|$ for any $t$ depends only on $\langle s|\bigchi_m|w\rangle$ and $(-1)^{m}$. For example, one of the elements of $\{\gamma_m(2)\}$ in case of $P_m(2)$ is $(GG')$ -- for which $\lvert\langle w|GG'|s\rangle\rvert ^2=\frac{1}{N}|(1-\frac{4}{N})^2+\frac{4}{N}(-1)^m |^2$. Because of the presence of the $(-1)^m$ factor in some of the terms of any polynomial $|\langle w|\gamma_m|s\rangle|$ for $U=\sigma_y$, we can infer that the success probabilities $P_m(t)=P_{m+2}(t),\ \forall m$. That is, the success probability at any given time is not constant for all $m$'s like in the case of $\sigma_x$ or $\sigma_z$ -- instead, $m$'s of equal parity do have the same success probability among themselves at any given time.

So for example, if we have total of $n=50$ qubits in the register performing the search algorithm, it turns out that the evaluation of the success probability in case of $m = 22$ noise sites and that in case of $m=41$ noise sites will be indistinguishable if the qubits in those sites evolve under the good noises, i.e., $U \in \{\sigma_x,\sigma_z\}$. The success probabilities in the cases where $m=10$, $m=40$ or $m=50$, will be exactly the same in case of $U=\sigma_y$. Similarly, the cases of $m=9$, $m=33$ or $m=45$ will be indistinguishable among themselves when $U=\sigma_y$. It is to be noted that there may be some unitary $U$, other than these Pauli matrices, which makes the success probability independent of $m$ for some particular time $t$ and not at other times. The Pauli matrices $\sigma_x$ and $\sigma_z$ are special in the sense that when $U$ is one of these, the success probability becomes independent of $m$, \textit{for all \(t\)}.

Another important observation is that none of the conditions used above put restrictions on what the positions of the $m$ unitaries are, out of the total $n$ positions. The coefficients $c_i$ (in Eqs.~\eqref{decomp_1} and \eqref{decomp_2}) remain the same for any arrangement of the $m$ noisy qubits. So, the success probability does not depend on the positions of the qubits which evolve under the noise unitary $U\in \{\sigma_x, \sigma_y, \sigma_z\}$. This result is also supported by Fig.~\ref{fig:Comparison_sites}. We will now investigate the effects of Markovian-correlated noise on the Grover's search algorithm numerically, and the results are gathered in the following section.
\begin{figure}[b]
    \includegraphics[width=\linewidth]{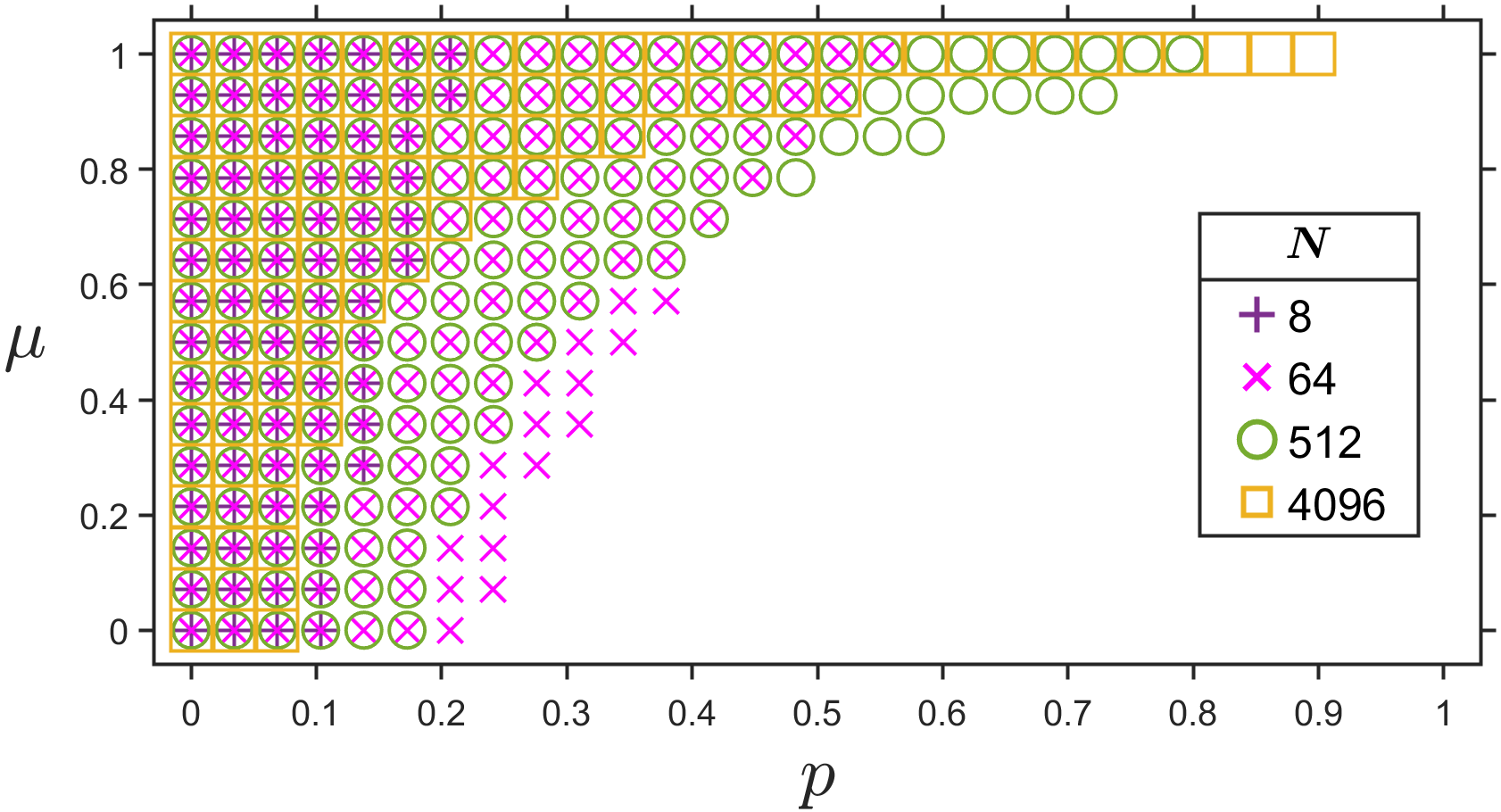}
    \caption{The regime of memory parameter and noise probability for which the noisy Grover algorithm is at least as good (measured by the quantity $\widetilde{P}$) as the classical search algorithm. The different colored boxes represent different sizes ($N$) of the search space. We find that the quantum search can withstand more noise if $\mu$ is higher. The advantage of the quantum algorithm becomes more evident with increasing $N$. All the quantities plotted are dimensionless.}
    \label{fig:performance}
\end{figure}

\section{Examples}
Before numerically showing the invariance of success probabilities proved in the previous section in presence of the good noises, we will first identify the parameter regime in our noise model for which the algorithm performs better than classical search.

\subsection{Performance of the noisy algorithm}
\label{performance}
The preservation of success probability upon increasing the number of noise sites is a potentially important feature. Nevertheless, increasing the noise probability still has a detrimental effect on the performance of the algorithm, as will be evident in the analysis in the next subsection. Since probability of finding the marked element in the Grover's algorithm is given by a success probability which never reaches unity in the noisy scenario, the algorithm needs to be re-run multiple times to find the element with some confidence \cite{Altaba,Gawron}.

Suppose that a noisy register, searching for a marked state out of total $N$ states, reaches its success probability maximum $P$ at time $T$. A classical search would find the element in $\frac{N}{2}$ time steps on average. Thus, assuming $T<\frac{N}{2}$, the quantum algorithm reaches its global maximum approximately $q =  \frac{N}{2T}$ times faster than the classical one. But it being likely that $P$ is much less than 1 for a noisy algorithm, we can claim that the quantum algorithm is at least as good as the classical one, only if, after running the noisy algorithm $q$ times, the probability $\widetilde{P}=(1-(1-P)^{q})$ of finding the marked element at least once, is close to unity. Here we take a probability of 0.95 to be the lower bound of such confidence.

In Fig.~\ref{fig:performance}, we have shown the values of $\mu$ and $p$ for which the register, under $U=\sigma_x$ noise, searching from a collection of $N$ elements is at least as good as the classical algorithm.
We can see that a higher memory $\mu$ helps the algorithm to perform better than its classical counterpart up to much higher noise probabilities. Another observation from the figure is that the quantum advantage becomes more prominent in case of larger database sizes $N$.

\subsection{Patterns of success probability}
\label{numerics}
In this subsection, we will first show that the invariance of the success probabilities in case of Pauli noise unitaries, persists irrespective of any time-correlation in the noise. Then, the independence from positions of the noise sites in case of the good noises and the effect of memory on the algorithm is shown numerically.

\begin{figure}[h]
    \includegraphics[width=\linewidth]{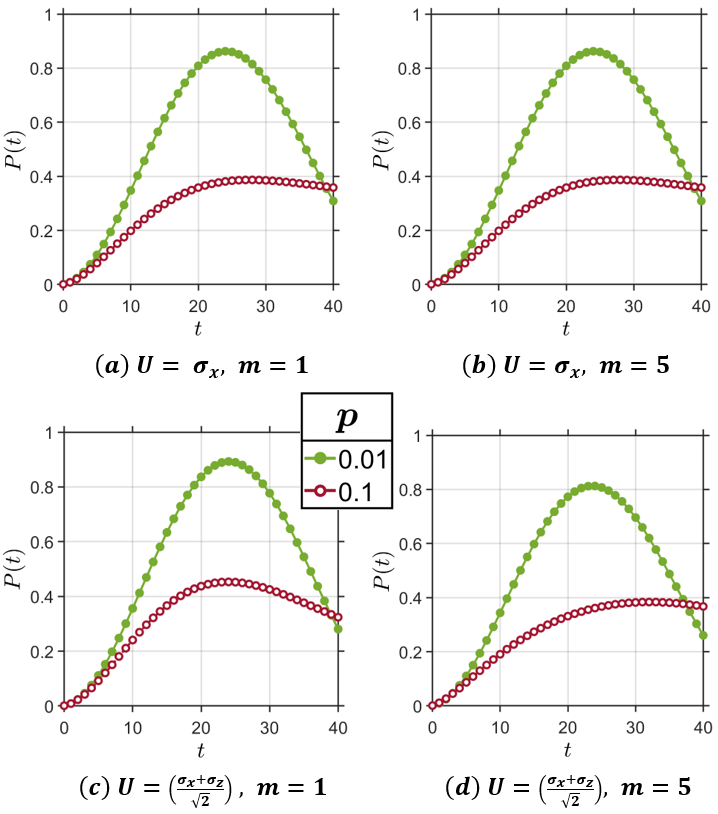}
    \caption{Success probabilities of Grover search algorithm in presence of noise without any time-correlation (i.e., $\mu=0$). We have plotted $P(t)$ on the vertical axis and the number, $t$, of Grover steps along the horizontal axis. The noise occurs with probability $p$ at each Grover step. The plots are for a 10-qubit register with $m$ noisy qubits (details in Sec.~\ref{sub:noise_w_mem}). 
    The inset shows symbols used in the plots for the noise probability $p$: 0.01 (filled, green circles), 0.1 (empty, brown circles). The plots (a) - (d) are for different \(U\)'s and \(m\)'s as displayed below each plot.
    All quantities used are dimensionless.}
    \label{fig:Memoryless}
\end{figure}

\subsubsection{Noise without memory}
\label{sub:noise_w_mem}
\begin{figure}[h]
    \includegraphics[width=\linewidth]{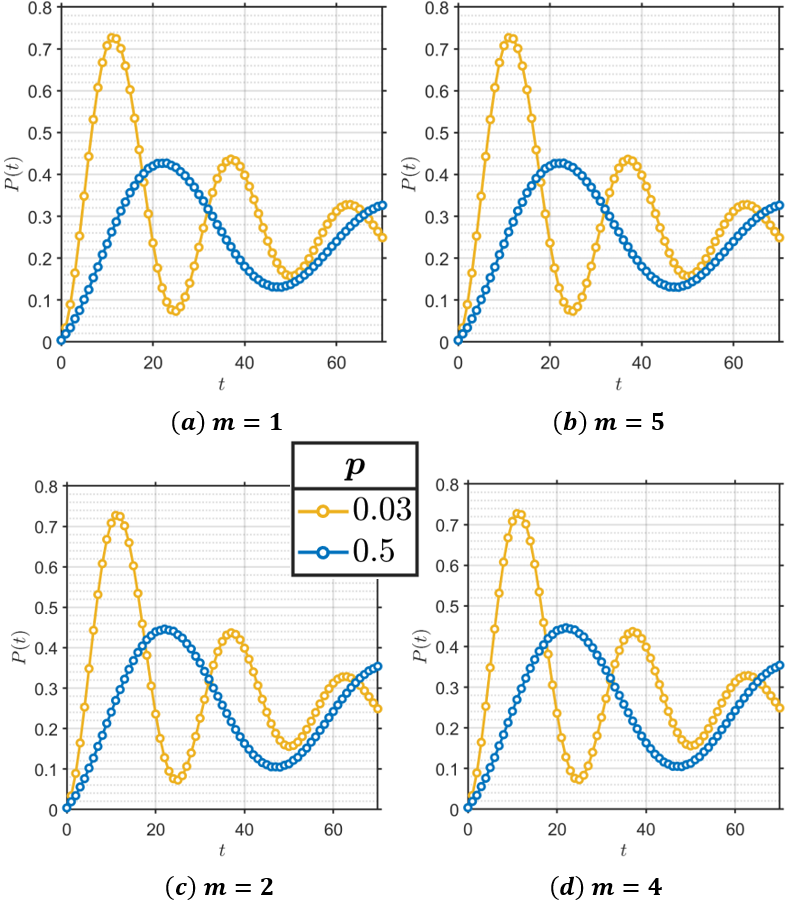}
    \caption{Success probabilities of the Grover algorithm
    in presence of noise without any time-correlation, i.e., $\mu=0$. We have plotted $P(t)$ on the vertical axis and the number, $t$, of Grover steps along the horizontal axis. The plots are for \(n=8\) and $U=\sigma_y$ (details in Sec. \ref{sub:noise_w_mem}). The inset shows symbols used for the noise probability $p$: 0.03 (light, yellow curve), 0.1 (dark, blue curve).
    The different noise strengths \(m\) are displayed below each plot (a) - (d). As expected, the first maximum of $P(t)$ is higher in case of the lower noise probability $p = 0.03$. All quantities used are dimensionless.}
    \label{fig:s_Y}
\end{figure}
The case of $\mu=0$, discussed in Sec.~\ref{sec:unital}, is a noise without any memory or time-correlation. So at each time step, the probability for the Grover operation to become noisy is $p_{g'}=p$. In Fig.~\ref{fig:Memoryless}, we compare the behavior in case of two noise unitaries $U$. The noise sites are the first $m$ qubits in the register, i.e., $\bigchi_m= U^{\otimes m}\otimes \mathds{1}_2 ^{\otimes (10-m)}$.

The case of $U=\sigma_x$ here corresponds to an \textit{$m$-qubit bit-flip channel}, as was shown in Sec.~\ref{sec:unital}. We see that the success probability's evolution, $P(t)$, for a given noise probability $p$, is unchanged when the number of noisy qubits is increased from $m=1$ to $m=5$ for $U=\sigma_x$. We contrast this with the evolution of $P(t)$ in case of $U=(\sigma_x+\sigma_z)/\sqrt{2}$, i.e., the Hadamard operator. This $U$ is a linear combination of two Pauli matrices and thus is not a good noise. $P(t)$ in presence of this noise changes when the number of noise sites is increased from $m=1$ to $m=5$, as expected.

We have also plotted in Fig.~\ref{fig:s_Y} the success probability's evolution for $m=1,\ 2,\ 4,\ 5$ in presence of noise unitary $U=\sigma_y$ and $\mu=0$, on an 8-qubit register. In other words, the register is under an $m$-qubit bit-phase flip channel occurring with probability $p$ after each noiseless Grover operation. As discussed in Sec.~\ref{Proof}, the behavior of $P(t)$, for any given $p$ and $n$, is exactly the same for odd number of total noise sites, i.e., for $m=1$ and $m=5$ in the figure. The same is true among noise strengths of even parity, $m = 2$ and $m=4$. In Fig.~\ref{fig:Comparison_sites}, we will also see that the locations of the noisy qubits are not important in case of the good noises $U=\sigma_x$, $\sigma_y$ or $\sigma_y$. In the next section, we study how the success probability evolution is affected by the presence of time-correlations in the noise.

\subsubsection{Noise with finite time-correlation}

\begin{figure}
    \includegraphics[width=\linewidth]{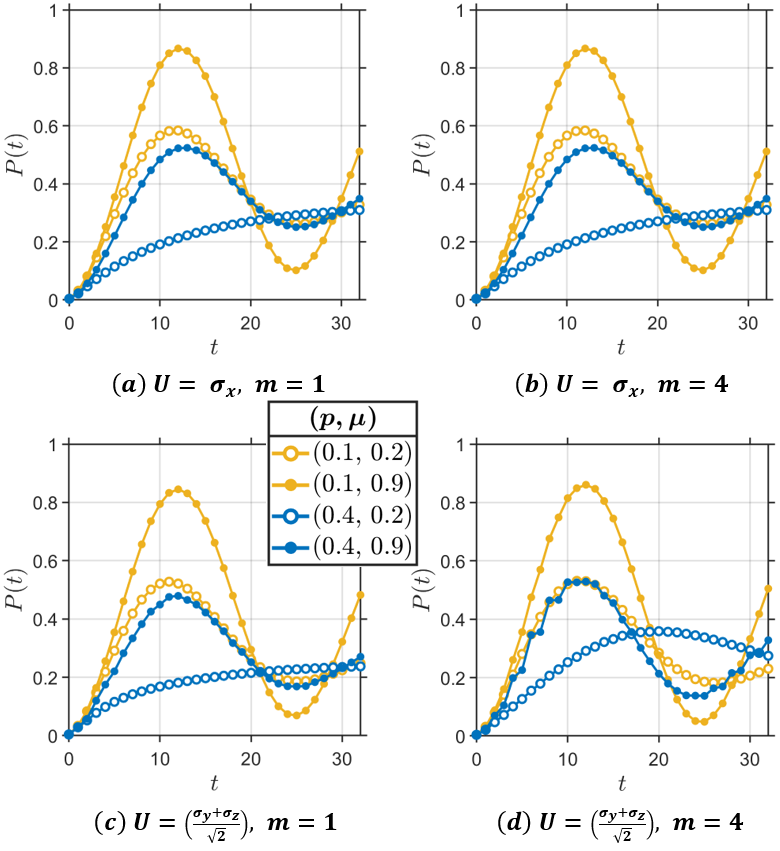}
    \caption{Success probabilities of Grover's search algorithm
    in presence of time-correlated noise. We have plotted $P(t)$ on the vertical axis and the number, $t$, of Grover steps along the horizontal axis. The plots are for a register with \(n=8\) qubits, out of which $m$ are noisy. 
    The inset table exhibits symbols used in the plots for different pairs of values of the noise probability $p$ and memory parameter $\mu$: $p = 0.1$ (light, yellow curve), $p = 0.4$ (dark, blue curve), $\mu = 0.2$ (empty circle), $\mu = 0.9$ (filled circle).
    The sub-figures (a) - (d) are for different \(U\)'s and \(m\)'s, as displayed below each plot.
    All quantities used are dimensionless.}
    \label{fig:Success probabilities}
\end{figure}

In Fig.~\ref{fig:Success probabilities}, the success probability $P(t)$ of Grover's search algorithm for non-zero (positive) memory $\mu$ and for $n=8$ qubits (256 elements in the search database) for two different noise unitaries are depicted. Here we have used the form of noise as $\bigchi_m= \mathds{1}_2 ^{\otimes (8-m)}\otimes U^{\otimes m}$ with $m=1$ and $4$. We can observe that the success probability $P(t)$ depends on the noise probability $p$ and the memory parameter $\mu$. It is obvious that the success probability reduces with increasing noise probability, and we can see from all the four panels that for a very high noise probability, the oscillatory behaviour of $P(t)$ tends to vanish.
\begin{figure}
    \centering
    \includegraphics[width=1.01\linewidth]{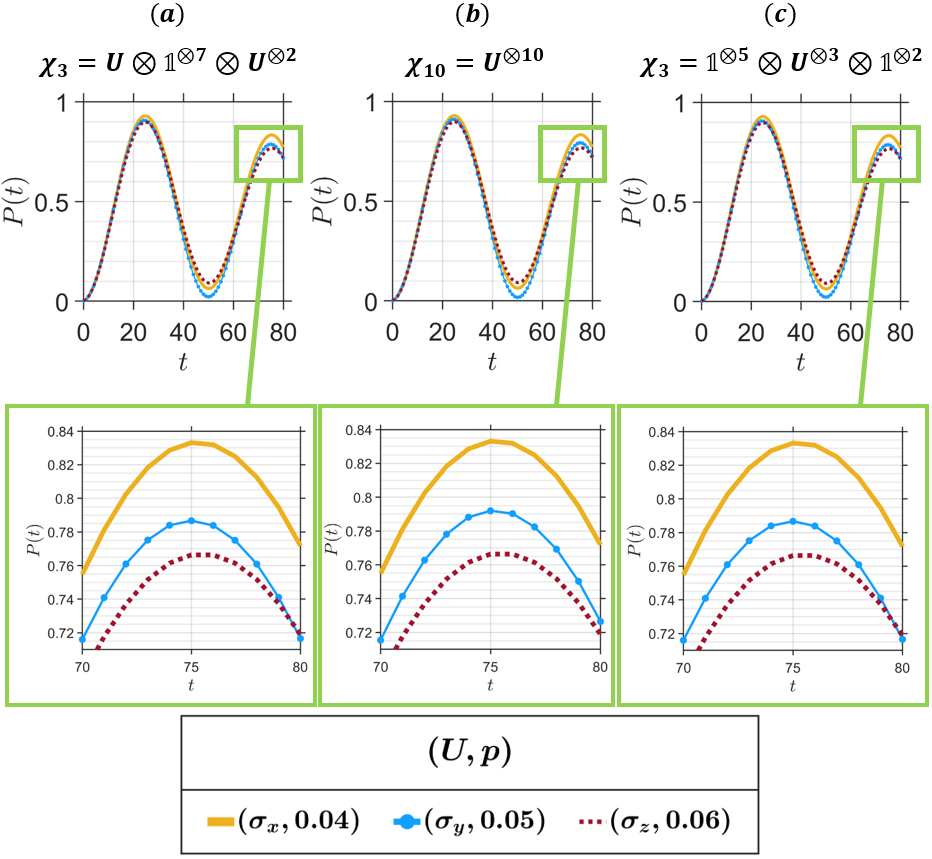}
    \caption{Success probabilities of Grover algorithm for good noises. Here \(n=10\) and $\mu=0.9$. Each of the colored curves are for different $p$ values and noise unitaries $U$, as shown at the bottom. The noise matrices $\bigchi_m$ are displayed above each plot. The plots inside the green boxes show magnified views of the indicated smaller portions of the plots above. The total number of noisy qubits in subplots (a) and (c) is $m=3$ (number of odd parity). In subplot (b), $m=10$ (number of even parity). For $U=\sigma_y$, its corresponding curve (blue, continuous line with dots) in subplot (b) is slightly different from those in case of $U=\sigma_y$ in (a) and (c). This supports our claim that $P(t)$ in case of $U=\sigma_y$ depends on the parity of $m$, unlike the other two Pauli matrices. All quantities used are dimensionless.}
    \label{fig:Comparison_sites}
\end{figure}
\begin{figure}
    \centering
    \includegraphics[width=.96\linewidth]{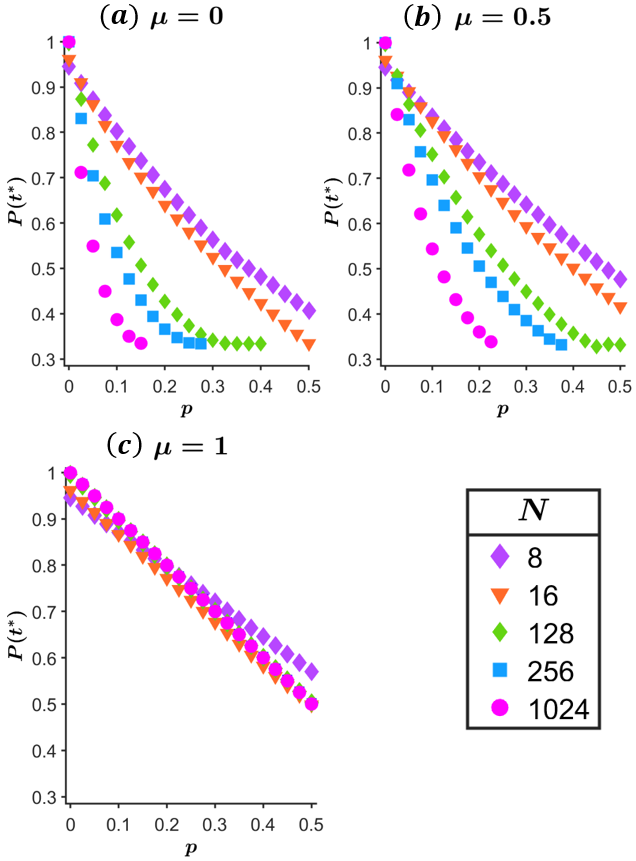}
    \caption{Effects of memory, database size and noise probability on the algorithm's success probability. We have presented here the values of $P(t^*)$, with respect to the noise probability $p$, for different $\mu$'s. The algorithm is performed on $n=\log_2 N$ qubits. Here the noise unitary considered is $U=\sigma_x$. All quantities used are dimensionless.}
    \label{fig:P_first max}
\end{figure}
It can be seen from Fig.~\ref{fig:Success probabilities}(a) and \ref{fig:Success probabilities}(b) that for a good noise $U=\sigma_x$, $P(t)$ for a given $p$ and $\mu$, remains unaffected when we change the number of noise sites $m$ on which $U$ is applied. Whereas for an unitary $U=(\sigma_y+\sigma_z)/\sqrt{2}$, which was shown not to be a good noise before, the success probability $P(t)$ changes with the noise strength $m$. Compare panels \ref{fig:Success probabilities}(c) and \ref{fig:Success probabilities}(d). We can see that in case of a noise with low memory $\mu=0.2$ and a high noise probability $p=0.4$, the success probability evolution of the algorithm almost disappears. The noisy Grover's search algorithm achieves greater efficiency for lower values of $p$ and higher values of $\mu$. Moreover, for higher values of $p$ for which the oscillation of $P(t)$ completely vanishes, the correlated noise helps in achieving higher success probabilities. For example, compare the lines corresponding to $(p,\mu)=(0.4,0.2) \ \text{and} \ (0.4,0.9)$ in the figure. The time evolution for $U=\sigma_x$ in case of perfect memory ($\mu=1$) is analyzed in Appendix \ref{perfect memory}.

The success probabilities for the good noises are plotted with respect to time in Fig.~\ref{fig:Comparison_sites} for different locations and number of noise sites. As we have noted previously in Sec.~\ref{Proof}, the positions of the noisy qubits do not matter if $U$ is a good noise. But it was also shown that the parity of the total number of noise sites is important in case of $U=\sigma_y$. In Figs.~\ref{fig:Comparison_sites}(a) and \ref{fig:Comparison_sites}(c) both, the parity of the total number $m$ of noise sites is the same. Only the positions of the noise sites are different. As expected, the respective profiles of $P(t)$ in case of all three Pauli matrices are exactly the same in Figs.~\ref{fig:Comparison_sites}(a) and \ref{fig:Comparison_sites}(c). In Fig.~\ref{fig:Comparison_sites}(b) all the qubits in the register are noisy and $m=10$ being an even number, the behavior of $P(t)$ in case of $\sigma_y$ is not exactly the same as in the other two sub-figures where $m$ is odd for both. Whereas $P(t)$ in case of $U=\sigma_x$ and $\sigma_z$ remains unaltered in all three sub-figures of Fig.~\ref{fig:Comparison_sites}.

Fig.~\ref{fig:P_first max} gives an overview of the effects of memory, database size and noise probability on the algorithm's success. Here we plot the success probabilities at their first maxima $P(t=t^*)$ with respect to the noise probability $p$ for $U=\sigma_x$. The effect of memory is contrasted in the three subplots. As observed in Fig.~\ref{fig:Success probabilities}, here also we can see that for a given amount of noise probability $p$, a higher memory $\mu $ of the noise helps the noisy algorithm to reach a higher success probability.

\section{Conclusion}
\label{Conclusion}
The Grover's algorithm can be employed to achieve a quadratic speed-up over classical methods in an unstructured search.
While this gives an advantage, a practical quantum circuit will undoubtedly be affected by different types of noise and several studies have already been pursued on the effects of such noises on the algorithm's performance. In our study, we consider the quantum register performing the algorithm to be under a local unitary noise which can also be correlated in time. In the interval between any two Grover operations, there is some probability for the noise to act on the register. In this setting, we find that the success probability of the algorithm at all times remains unchanged with respect to the number of noisy qubits, if and only if the local noisy evolutions are given by some special unitaries. We call these unitaries the `good noises.' These noises are shown to reduce to multi-qubit bit-flip or phase-damping errors and in some cases bit-phase flip errors in the absence of time-correlations. Locations of the noisy qubits are also shown to be not relevant in case of the good noises. This can be a potentially useful information in an actual implementation of the search algorithm on a register. The result that two of the Pauli noises behave in a different way than the third, can be explained by the symmetry-breaking in Grover algorithm due to the choice of the initial state of the algorithm's register (which is a product of eigenvectors of the Pauli \(\sigma_z\) operator) and the ensuing Hadamard rotation (which connects the \(\sigma_x\) and  \(\sigma_z\) eigenbases). Numerically, we have been also able to show that a time-correlated noise could lead to a better performance of the noisy algorithm.

\acknowledgements
We acknowledge partial support from the Department of Science and Technology, Government of India through the QuEST grant
(grant number DST/ICPS/QUST/Theme-3/2019/120).

\appendix
\section{Invariance of success probability with respect to noise strength: a special case}
\label{App_A}
In case of $U = \sigma_x$, we get  $\bigchi_m |s\rangle = |s\rangle$. So, we can express all the states in terms of the orthogonal basis vector set $\{|s_1'\rangle, |w\rangle, |w'\rangle\}$, with $|s_1'\rangle = \frac{1}{\sqrt{N-2}}\sum_{x\neq w,w'}|x\rangle.$

In this basis, $\langle s|=\begin{psmallmatrix}
  \sqrt{\frac{N-2}{N}} & \frac{1}{\sqrt{N}} & \frac{1}{\sqrt{N}}
\end{psmallmatrix} 
$. From Eqs.~\eqref{eq:3} and \eqref{eq:6},
\begin{equation} 
G=\begin{psmallmatrix} 
2\frac{N-2}{N}-1 & -2\frac{\sqrt{N-2}}{N} & 2\frac{\sqrt{N-2}}{N}\\
2\frac{\sqrt{N-2}}{N} & -\frac{2}{N}+1 & \frac{2}{N}\\
2\frac{\sqrt{N-2}}{N} & -\frac{2}{N} & \frac{2}{N}-1
\end{psmallmatrix},
\end{equation}
\begin{equation}
    G' = \begin{psmallmatrix} 
2\frac{N-2}{N}-1 & -2\frac{\sqrt{N-2}}{N} & 2\frac{\sqrt{N-2}}{N}\\
2\frac{\sqrt{N-2}}{N} & -\frac{2}{N} & \frac{2}{N}-1\\
2\frac{\sqrt{N-2}}{N} & -\frac{2}{N}+1 & \frac{2}{N}
\end{psmallmatrix}.
\label{eq-4.15}
\end{equation}
It is evident from the expressions above that, at least for the case $U = \sigma_x$, although changing $m$ does change the forms of the basis vectors $|w'\rangle$ and $|s_1'\rangle$ in the computational basis, elements of all the states or operators like $|s\rangle$ or $G'$ remain the same in 
$\{|s_1'\rangle, |w\rangle, |w'\rangle\}$ basis. Thus, increasing or decreasing the number ($m$) of noise sites does not affect the success probability (Eq.~\eqref{29}) of the algorithm in case of $U=\sigma_x$ and $m\geq 1$.
\section{Example of a Markovian correlated channel}
\label{app:Markovian}
An important example of noise with memory is the Markovian correlated Pauli channel investigated in \cite{Macchia, Macchia_2, Daems}. In that paper they studied the classical capacity of channels with partial memory. More specifically, they considered a channel that applies $\pi$-rotations along random sets of axes $l_1,l_2,\ldots,l_n$ on a sequence of $n$ qubits, with joint probability $p_{l_1l_2\ldots l_n}$, where $\sum_{l_1,l_2,\ldots,l_n}^{}p_{l_1l_2 \ldots l_n} = 1$. They also assumed that the rotation about axes $l_1,\ldots,l_n$ form a Markov chain so that \begin{equation}p_{l_1 \ldots l_n} = p_{l_1}p_{l_2|l_1}\ldots p_{l_n|l_{n-1}},\label{eq:1}\end{equation} 
where $p_{i|j}$ denotes the conditional probability of rotation about $i$-axis given that the previous one was about $j$-axis. The conditional probabilities are given as 
\begin{equation}
p_{i|j} = (1-\mu)\,p_i + \mu\, \delta_{i,j}.
\label{eq:2} 
\end{equation} 
Here $\mu$ corresponds to the relaxation time or ``\textit{memory}''. For example, if $\mu = 1,$ the same rotation axis $l_1$ is used at all subsequent rotations, i.e., $l_1 l_1\ldots l_1$ on the qubits.


\section{Details of calculation for \texorpdfstring{$P_m(t=1)$} ~~of Sec. \ref{Proof}}
\label{app:proof}
In connection to search for good noises, we detail here the conditions on $U$ for keeping $P_m(t=1) =$ $(1-p)|\langle w|G|s\rangle|^2$ $+p|\langle w|G'|s\rangle|^2$ constant with changing $m$.
We have \begin{smal}[\small]
\lvert\langle w|G'|s\rangle\rvert ^2 
&= \left\lvert\left(1-\frac{4}{N}\right)\langle w|\bigchi_{m}|s\rangle+\frac{2}{\sqrt{N}}\langle w|\bigchi_{m}|w\rangle\right\rvert^2\nonumber\\ 
&= \frac{1}{N}\Biggl\lvert \left(1-\frac{4}{N}\right)\sum_{j=1}^{N}(\bigchi_{m})_{w,j}+2(\bigchi_{m})_{w,w}\Biggr\rvert^2 \label{deq:ch}
\end{smal}
where, $\bigchi_m |s\rangle = \frac{1}{\sqrt{N}}\begin{psmallmatrix}\sum_{j=1}^{N}(\bigchi_{m})_{1,j}\\.\\.\\.\\\sum_{j=1}^{N}(\bigchi_{m})_{N,j}\end{psmallmatrix}$. It can be shown that
\begin{equation}
    \sum_{j=1}^{N}(\bigchi_m)_{k,j} = e^{iq\theta}(a+b)^{m-q}(\olsi{a}-\olsi{b})^q \ :=\psi_q
    \label{deq:e10}
\end{equation}
and $(\bigchi_m)_{k,k} = e^{iq\theta}a^{m-q}\olsi{a}^q$, 
where $q \in [0,m]$ and $q$ depends on $k$. Here each $\psi_q$ appears $\left(\frac{N}{2^m}\right)\begin{psmallmatrix}
m\\q
\end{psmallmatrix}$ times in $\bigchi_m|s\rangle$.
Since $|\langle w|G|s\rangle|^2$ is independent of $m$, we can conclude from the expression of $|\langle w|G'|s\rangle|^2$ in Eq.~\eqref{deq:ch} that for getting $P_{m+1}(1) = P_m(1)$, we need either $|a|=0$ or $|b|=0$. Therefore we get our first condition for constructing a good noise which gives the constraints, Eqs.~\eqref{Cond-1-1} and \eqref{Cond-1-2}. So, a good noise needs to obey
\begin{center}
    \textit{Condition 1:} $|a|=1$ or $|b|=1$
\end{center}
{\small
 \begin{numcases}{\text{\normalsize Thus,}\quad U=}
  \begin{pmatrix} a & 0\\ 0 & \olsi{a} e^{i\theta}\end{pmatrix}, \quad \text{for} \quad |a|=1,\label{Cond-1-1}\\
  \begin{pmatrix} 0 & b\\-\olsi{b} e^{i\theta} & 0\end{pmatrix}, \quad \text{for} \quad |b|=1.
  \label{Cond-1-2}
\end{numcases}}\\
Hence $\bigchi_m$ has to be a \textit{generalized permutation unitary matrix}. We can now define a new state $|w'\rangle$ so that 
\begin{smal}[\small]
\bigchi_m|w\rangle\langle w|\bigchi_m^{\dagger}=
\begin{cases}
 |w\rangle\langle w|, & \text{for} \quad |a|=1,\;\;\;\;\;\;\;\\
 |w'\rangle\langle w'|, & \text{for} \quad |b|=1.\;\;\;\;\;\;\;
\end{cases}
\end{smal}
Using this we can write \ $\bigchi_m|s\rangle= \dfrac{1}{\sqrt{N}} \times$
{\small\begin{numcases}{}
    \left(\mathlarger{\sum}_{i=1}^{M}c_i \sqrt{\Delta_i}|s'_i\rangle+\alpha|w\rangle\right),&for $|a|=1$,\label{decomp_1}\;\;\;\;\\
    \left(\mathlarger{\sum}_{i=1}^{M}c_i
   \sqrt{\Delta_i}|s'_i\rangle+\alpha|w\rangle +\beta|w'\rangle\right),&for $|b|=1$,\;\;\;\;\label{decomp_2}
\end{numcases}}\vspace{-.7cm}
\begin{equation*}
\text{where}    \quad |s'_i\rangle = \frac{1}{\sqrt{\Delta_i}}\underset{\{|d_i\rangle\}\neq|w\rangle,|w'\rangle}{\mathlarger{\sum}}|d_i\rangle,
\end{equation*}  $\Delta_i = \text{size}(\{|d\rangle_i\})$, $\bigsqcup_{i} \{|d\rangle_i\}=\{|x\rangle\}$,
$\langle s'_i|s'_j\rangle = \delta_{ij}$, $|c_i|=1=|\alpha|=|\beta|$, i.e., we get the basis $\mathbb{B}=\{|s'_1\rangle,|s'_2\rangle,\ldots,|s'_{M}\rangle,|w\rangle\}$ of dimension $(M+1)$ from Eq.~\eqref{decomp_1} and $\mathbb{B}=\{|s'_1\rangle,|s'_2\rangle,\ldots,|s'_{M}\rangle,|w\rangle, |w'\rangle \}$ of dimension $(M+2)$ from Eq.~\eqref{decomp_2}.

Now, there are two possibilities for a unitary $U$ of the form in Eqs.~\eqref{Cond-1-1} and~\eqref{Cond-1-2}: its two non-zero elements are either  (Case $(i)$) equal, or (Case $(ii)$) unequal. Case $(i)$ suggests that $M=1$, and 
directly leads to the constraints, given in Eqs.~\eqref{cond-2-1} and~\eqref{cond-2-2}, which have to be satisfied by $U$ to be a good noise. In Case $(ii)$, we need to put further restrictions on $U$ for the success probability to stay conserved with $m$. We should not have $\dim(\mathbb{B})$ changing with $m$. Thus, the number of distinct $c_i$'s in Eqs.~\eqref{decomp_1} and~\eqref{decomp_2} must remain constant with $m$. There are total $M$ of these coefficients for both $|a|=1$ and $|b|=1$. For $m=1$, e.g., for $\bigchi_1=U\otimes \mathds{1}_{N/2}$ in Case $(ii)$, there are only two distinct non-zero elements in $\bigchi_1$ -- because $U$ has two distinct non-zero elements. This implies
$M=2$. Since $M$ should remain constant with $m$, Case $(ii)$  leads to the conditions given in Eqs.~\eqref{cond-2-3}, \eqref{cond-2-4}, \eqref{cond-2-5}. To summarise, we have the following necessary (but not sufficient) condition for $U$ to be a good noise:
\begin{center}
    \textit{Condition 2:} $M$ = 1 or 2  
 \end{center}
 \begin{widetext}
 {\small
 \begin{numcases}{\text{\normalsize Thus,}\quad \bigchi_m|s\rangle =}
 c\left(\,\sqrt{\frac{N-1}{N}}|s_1'\rangle +\frac{1}{\sqrt{N}}|w\rangle\right), & for $  |a| = 1, M=1$\label{cond-2-1}\\
 c\,\left(\sqrt{\frac{N-2}{N}}|s_1'\rangle +\frac{1}{\sqrt{N}}|w\rangle+\frac{1}{\sqrt{N}}|w'\rangle\right), & for $  |b| = 1, M=1$\label{cond-2-2}\end{numcases}\\}
 {\small
 \begin{numcases}{\text{\normalsize and,}\quad\bigchi_m|s\rangle =}
 c_1\,\left( \sqrt{\frac{N-2}{2N}}|s_1'\rangle+\frac{1}{\sqrt{N}}|w\rangle\right)+\frac{c_2}{\sqrt{2}}|s_2'\rangle, & for $ |a|=1, M=2$ \label{cond-2-3} \\
c_1\, \left(\sqrt{\frac{N-4}{2N}}|s_1'\rangle+\frac{1}{\sqrt{N}}|w\rangle+\frac{1}{\sqrt{N}}|w'\rangle\right)+\frac{c_2}{\sqrt{2}}|s_2'\rangle, & for $ |b|=1, M=2,\alpha=\beta$ \label{cond-2-4}\\
c_1\,\left( \sqrt{\frac{N-2}{2N}}|s_1'\rangle+\frac{1}{\sqrt{N}}|w\rangle\right)+c_2\,\left(\sqrt{\frac{N-2}{2N}}|s_2'\rangle+\frac{1}{\sqrt{N}}|w'\rangle\right),& for $ |b|=1, M=2,\alpha\neq\beta$ \label{cond-2-5}\end{numcases}}
\end{widetext}
So, only the $U$'s that satisfy one of the Eqs.~\eqref{cond-2-1}-\eqref{cond-2-5}, are the unitaries corresponding to the good noise for which $P(t)$ does not depend on the number of noise sites $m$.
 It can be shown that 
 $\psi_q$ appears $(\frac{N}{2^m})\begin{psmallmatrix}
 m\\q
 \end{psmallmatrix}$ times in the column vector $\bigchi_m|s\rangle$. We have the following observations.
 \begin{enumerate}
\item[(1)] If $U$ satisfies Eq.~\eqref{cond-2-1}, then $b=0$ and $\psi_q = c, \forall q $. Solving for $a$ and $\theta$ gives $a = e^{i\phi}= \sqrt[m]{c},\ \theta = 2\phi$, i.e., $U = \sqrt[m]{c}\begin{psmallmatrix}1 & 0\\0 & 1\end{psmallmatrix} = \sqrt[m]{c}\; \mathds{1}_2$. 

\item[(2)] If $U$ satisfies Eq.~\eqref{cond-2-3}, then it turns out that we need $(a)\;\psi_q = \psi_{q+2} = c_1, \forall q$ even and $(b)\;\psi_q = \psi_{q+2} = c_2, \forall q$ odd. That is because, $\begin{psmallmatrix}
 m\\0
 \end{psmallmatrix}+\begin{psmallmatrix}
 m\\2
 \end{psmallmatrix}+\begin{psmallmatrix}
 m\\4
 \end{psmallmatrix}+\ldots = \begin{psmallmatrix}
 m\\1
 \end{psmallmatrix}+\begin{psmallmatrix}
 m\\3
 \end{psmallmatrix}+\begin{psmallmatrix}
 m\\5
 \end{psmallmatrix}+\ldots = 2^{m-1}$, i.e., the sum of multiplicities of elements in $\bigchi_m|s\rangle$ from the set $\{\psi_q|q\ \text{even}\}$ is equal to that in case of elements from the set $\{\psi_q|q\ \text{odd}\}$. Since $c_1 \neq c_2$, solving $(a)$ and $(b)$ 
for $a$ and $\theta$ give the solution, $a = \sqrt[m]{c_1},\ \theta = 2\phi-\pi$. The solution corresponds to $c_1=-c_2$, i.e., $U = \sqrt[m]{c}\begin{psmallmatrix}1 & 0\\0 & -1\end{psmallmatrix} = \sqrt[m]{c}\;\sigma_z$.
\item[(3)] If $U$ satisfies Eq.~\eqref{cond-2-2}, then $a=0$ and $\psi_q = c, \forall q $. Solving for $b$ and $\theta$ gives $b = e^{i\phi}= \sqrt[m]{c},\ \theta = 2\phi - \pi$, i.e., $U = \sqrt[m]{c}\begin{psmallmatrix}0 & 1\\1 & 0\end{psmallmatrix} = \sqrt[m]{c} \;\sigma_x$.
\item[(4)] If $U$ satisfies Eqs.~\eqref{cond-2-4} or \eqref{cond-2-5}, a similar analysis as above can be performed and the solution is $U = \sqrt[m]{c}\begin{psmallmatrix}0 & 1\\-1 & 0\end{psmallmatrix} = \sqrt[m]{c}\;i\;\sigma_y$.

 \end{enumerate}
Here $\sqrt[m]{c}$ is only a constant phase factor. We can see from the above discussion for $P_m(t=1)$, the candidates for good noise are the unitaries $e^{i\phi}\,\mathds{1}_2$, $e^{i\phi}\,\sigma_x$, $e^{i\phi}\,\sigma_y$ and $e^{i\phi}\,\sigma_z$, for any $\phi \in [0,2\pi)$.

\section{Evolution of success probability for perfect memory for $U=\sigma_x$}
\label{perfect memory}
Here, we consider the case when $\mu = 1$, i.e., \textit{perfect memory}. On the first noisy iteration (i.e., $t=1$), $G$ occurs with probability $(1-p)$ and $G'$ with $p$. Let us assume at $t = 1$, $G$ is applied. Due to perfect memory, for all $t\geqslant 2$, the same operator $G$ will be applied. This scenario corresponds to an ideal \textit{noiseless} Grover algorithm. The success probability in this case will be denoted as $P(t)$ and the marked state is reached at $t \approx \frac{\pi}{4}\sqrt{N}$ \cite{N&C}.

If $G'$ is applied at $t=1$, for $t\geqslant2$ the state of the whole $n$-qubit register would be $|\psi(t)\rangle = G'\, ^t\ |s\rangle$. 
Using the form of $G'$ in Eq.~\eqref{eq-4.15} for $U = \sigma_x$,
\begin{align}
    \langle w|G'\,^t|s\rangle 
    =\frac{(-1)^{t+1}}{\sqrt{N}}\, \operatorname{\mathbb{I}m}\left[\left(\tan\left(\frac{\theta}{2}\right)-i\ \right)e^{i t \theta}\right]\nonumber
\end{align}
where $\theta = \cos^{-1}(\frac{2}{N})$ and $\operatorname{\mathbb{I}m}[\bullet]$ denotes the imaginary part of a complex number. Then the success probability at time $t$ in this case is 
\begin{align}
    P'(t) = |\langle w|G'\,^t|s\rangle|^2  = \frac{\cos^2(\theta t)}{N} \left(\tan\left(\frac{\theta}{2}\right)\tan(\theta t)-1\right)^2.
    \label{C6}
\end{align}
Combining the above two cases, the success probability of a noisy algorithm at time $t$, with noise probability $p$ and perfect memory $\mu=1$, then becomes $(1-p)\, P(t)+p\, P'(t)$.


\bibliography{References}

\begin{thebibliography}{127}%
\makeatletter
\providecommand \@ifxundefined [1]{%
 \@ifx{#1\undefined}
}%
\providecommand \@ifnum [1]{%
 \ifnum #1\expandafter \@firstoftwo
 \else \expandafter \@secondoftwo
 \fi
}%
\providecommand \@ifx [1]{%
 \ifx #1\expandafter \@firstoftwo
 \else \expandafter \@secondoftwo
 \fi
}%
\providecommand \natexlab [1]{#1}%
\providecommand \enquote  [1]{``#1''}%
\providecommand \bibnamefont  [1]{#1}%
\providecommand \bibfnamefont [1]{#1}%
\providecommand \citenamefont [1]{#1}%
\providecommand \href@noop [0]{\@secondoftwo}%
\providecommand \href [0]{\begingroup \@sanitize@url \@href}%
\providecommand \@href[1]{\@@startlink{#1}\@@href}%
\providecommand \@@href[1]{\endgroup#1\@@endlink}%
\providecommand \@sanitize@url [0]{\catcode `\\12\catcode `\$12\catcode
  `\&12\catcode `\#12\catcode `\^12\catcode `\_12\catcode `\%12\relax}%
\providecommand \@@startlink[1]{}%
\providecommand \@@endlink[0]{}%
\providecommand \url  [0]{\begingroup\@sanitize@url \@url }%
\providecommand \@url [1]{\endgroup\@href {#1}{\urlprefix }}%
\providecommand \urlprefix  [0]{URL }%
\providecommand \Eprint [0]{\href }%
\providecommand \doibase [0]{https://doi.org/}%
\providecommand \selectlanguage [0]{\@gobble}%
\providecommand \bibinfo  [0]{\@secondoftwo}%
\providecommand \bibfield  [0]{\@secondoftwo}%
\providecommand \translation [1]{[#1]}%
\providecommand \BibitemOpen [0]{}%
\providecommand \bibitemStop [0]{}%
\providecommand \bibitemNoStop [0]{.\EOS\space}%
\providecommand \EOS [0]{\spacefactor3000\relax}%
\providecommand \BibitemShut  [1]{\csname bibitem#1\endcsname}%
\let\auto@bib@innerbib\@empty
\bibitem [{\citenamefont {Deutsch}\ and\ \citenamefont
  {Jozsa}(1992)}]{Deutsch}%
  \BibitemOpen
  \bibfield  {author} {\bibinfo {author} {\bibfnamefont {D.}~\bibnamefont
  {Deutsch}}\ and\ \bibinfo {author} {\bibfnamefont {R.}~\bibnamefont
  {Jozsa}},\ }\bibfield  {title} {\bibinfo {title} {Rapid solution of problems
  by quantum computation},\ }\href {https://doi.org/10.1098/rspa.1992.0167}
  {\bibfield  {journal} {\bibinfo  {journal} {Proceedings of the Royal Society
  of London. Series A: Mathematical and Physical Sciences}\ }\textbf {\bibinfo
  {volume} {439}},\ \bibinfo {pages} {553} (\bibinfo {year}
  {1992})}\BibitemShut {NoStop}%
\bibitem [{\citenamefont {Deutsch}\ and\ \citenamefont
  {Penrose}(1985)}]{Deutsch_1}%
  \BibitemOpen
  \bibfield  {author} {\bibinfo {author} {\bibfnamefont {D.}~\bibnamefont
  {Deutsch}}\ and\ \bibinfo {author} {\bibfnamefont {R.}~\bibnamefont
  {Penrose}},\ }\bibfield  {title} {\bibinfo {title} {Quantum theory, the
  {Church-Turing} principle and the universal quantum computer},\ }\href
  {https://doi.org/10.1098/rspa.1985.0070} {\bibfield  {journal} {\bibinfo
  {journal} {Proceedings of the Royal Society of London. A. Mathematical and
  Physical Sciences}\ }\textbf {\bibinfo {volume} {400}},\ \bibinfo {pages}
  {97} (\bibinfo {year} {1985})}\BibitemShut {NoStop}%
\bibitem [{\citenamefont {Shor}(1994)}]{Shor}%
  \BibitemOpen
  \bibfield  {author} {\bibinfo {author} {\bibfnamefont {P.}~\bibnamefont
  {Shor}},\ }\bibfield  {title} {\bibinfo {title} {Algorithms for quantum
  computation: discrete logarithms and factoring},\ }in\ \href
  {https://doi.org/10.1109/SFCS.1994.365700} {\emph {\bibinfo {booktitle}
  {Proceedings 35th Annual Symposium on Foundations of Computer Science}}}\
  (\bibinfo {year} {1994})\ pp.\ \bibinfo {pages} {124--134}\BibitemShut
  {NoStop}%
\bibitem [{\citenamefont {Ekert}\ and\ \citenamefont {Jozsa}(1996)}]{Shor1}%
  \BibitemOpen
  \bibfield  {author} {\bibinfo {author} {\bibfnamefont {A.}~\bibnamefont
  {Ekert}}\ and\ \bibinfo {author} {\bibfnamefont {R.}~\bibnamefont {Jozsa}},\
  }\bibfield  {title} {\bibinfo {title} {Quantum computation and shor's
  factoring algorithm},\ }\href {https://doi.org/10.1103/RevModPhys.68.733}
  {\bibfield  {journal} {\bibinfo  {journal} {Rev. Mod. Phys.}\ }\textbf
  {\bibinfo {volume} {68}},\ \bibinfo {pages} {733} (\bibinfo {year}
  {1996})}\BibitemShut {NoStop}%
\bibitem [{\citenamefont {Grover}(1996)}]{Grover}%
  \BibitemOpen
  \bibfield  {author} {\bibinfo {author} {\bibfnamefont {L.~K.}\ \bibnamefont
  {Grover}},\ }\bibfield  {title} {\bibinfo {title} {A fast quantum mechanical
  algorithm for database search},\ }in\ \href
  {https://doi.org/10.1145/237814.237866} {\emph {\bibinfo {booktitle}
  {Proceedings of the Twenty-Eighth Annual ACM Symposium on Theory of
  Computing}}},\ \bibinfo {series and number} {STOC '96}\ (\bibinfo
  {publisher} {Association for Computing Machinery},\ \bibinfo {address} {New
  York, NY, USA},\ \bibinfo {year} {1996})\ p.\ \bibinfo {pages}
  {212–219}\BibitemShut {NoStop}%
\bibitem [{\citenamefont {Grover}(1997)}]{Grover_2}%
  \BibitemOpen
  \bibfield  {author} {\bibinfo {author} {\bibfnamefont {L.~K.}\ \bibnamefont
  {Grover}},\ }\bibfield  {title} {\bibinfo {title} {Quantum mechanics helps in
  searching for a needle in a haystack},\ }\href
  {https://doi.org/10.1103/PhysRevLett.79.325} {\bibfield  {journal} {\bibinfo
  {journal} {Phys. Rev. Lett.}\ }\textbf {\bibinfo {volume} {79}},\ \bibinfo
  {pages} {325} (\bibinfo {year} {1997})}\BibitemShut {NoStop}%
\bibitem [{\citenamefont {Boyer}\ \emph {et~al.}(1998)\citenamefont {Boyer},
  \citenamefont {Brassard}, \citenamefont {Høyer},\ and\ \citenamefont
  {Tapp}}]{Boyer}%
  \BibitemOpen
  \bibfield  {author} {\bibinfo {author} {\bibfnamefont {M.}~\bibnamefont
  {Boyer}}, \bibinfo {author} {\bibfnamefont {G.}~\bibnamefont {Brassard}},
  \bibinfo {author} {\bibfnamefont {P.}~\bibnamefont {Høyer}},\ and\ \bibinfo
  {author} {\bibfnamefont {A.}~\bibnamefont {Tapp}},\ }\bibfield  {title}
  {\bibinfo {title} {Tight bounds on quantum searching},\ }\href@noop {}
  {\bibfield  {journal} {\bibinfo  {journal} {Fortschritte der Physik}\
  }\textbf {\bibinfo {volume} {46}},\ \bibinfo {pages} {493} (\bibinfo {year}
  {1998})}\BibitemShut {NoStop}%
\bibitem [{\citenamefont {Biham}\ \emph
  {et~al.}(1999{\natexlab{a}})\citenamefont {Biham}, \citenamefont {Biham},
  \citenamefont {Biron}, \citenamefont {Grassl},\ and\ \citenamefont
  {Lidar}}]{Lidar}%
  \BibitemOpen
  \bibfield  {author} {\bibinfo {author} {\bibfnamefont {E.}~\bibnamefont
  {Biham}}, \bibinfo {author} {\bibfnamefont {O.}~\bibnamefont {Biham}},
  \bibinfo {author} {\bibfnamefont {D.}~\bibnamefont {Biron}}, \bibinfo
  {author} {\bibfnamefont {M.}~\bibnamefont {Grassl}},\ and\ \bibinfo {author}
  {\bibfnamefont {D.~A.}\ \bibnamefont {Lidar}},\ }\bibfield  {title} {\bibinfo
  {title} {{G}rover's quantum search algorithm for an arbitrary initial
  amplitude distribution},\ }\href {https://doi.org/10.1103/PhysRevA.60.2742}
  {\bibfield  {journal} {\bibinfo  {journal} {Phys. Rev. A}\ }\textbf {\bibinfo
  {volume} {60}},\ \bibinfo {pages} {2742} (\bibinfo {year}
  {1999}{\natexlab{a}})}\BibitemShut {NoStop}%
\bibitem [{\citenamefont {Shenvi}\ \emph {et~al.}(2003)\citenamefont {Shenvi},
  \citenamefont {Kempe},\ and\ \citenamefont {Whaley}}]{Shenvi}%
  \BibitemOpen
  \bibfield  {author} {\bibinfo {author} {\bibfnamefont {N.}~\bibnamefont
  {Shenvi}}, \bibinfo {author} {\bibfnamefont {J.}~\bibnamefont {Kempe}},\ and\
  \bibinfo {author} {\bibfnamefont {K.~B.}\ \bibnamefont {Whaley}},\ }\bibfield
   {title} {\bibinfo {title} {Quantum random-walk search algorithm},\ }\href
  {https://doi.org/10.1103/PhysRevA.67.052307} {\bibfield  {journal} {\bibinfo
  {journal} {Phys. Rev. A}\ }\textbf {\bibinfo {volume} {67}},\ \bibinfo
  {pages} {052307} (\bibinfo {year} {2003})}\BibitemShut {NoStop}%
\bibitem [{\citenamefont {Ambainis}\ \emph {et~al.}(2005)\citenamefont
  {Ambainis}, \citenamefont {Kempe},\ and\ \citenamefont {Rivosh}}]{Kempe}%
  \BibitemOpen
  \bibfield  {author} {\bibinfo {author} {\bibfnamefont {A.}~\bibnamefont
  {Ambainis}}, \bibinfo {author} {\bibfnamefont {J.}~\bibnamefont {Kempe}},\
  and\ \bibinfo {author} {\bibfnamefont {A.}~\bibnamefont {Rivosh}},\
  }\bibfield  {title} {\bibinfo {title} {Coins make quantum walks faster},\
  }in\ \href@noop {} {\emph {\bibinfo {booktitle} {Proceedings of the Sixteenth
  Annual ACM-SIAM Symposium on Discrete Algorithms}}},\ \bibinfo {series and
  number} {SODA '05}\ (\bibinfo  {publisher} {Society for Industrial and
  Applied Mathematics},\ \bibinfo {address} {USA},\ \bibinfo {year} {2005})\
  p.\ \bibinfo {pages} {1099–1108}\BibitemShut {NoStop}%
\bibitem [{\citenamefont {Manin}(1980)}]{manin}%
  \BibitemOpen
  \bibfield  {author} {\bibinfo {author} {\bibfnamefont {Y.}~\bibnamefont
  {Manin}},\ }\href@noop {} {\emph {\bibinfo {title} {Computable and
  Uncomputable (in Russian)}}}\ (\bibinfo  {publisher} {Sovetskoye Radio,
  Moscow},\ \bibinfo {year} {1980})\ p.\ \bibinfo {pages} {128}\BibitemShut
  {NoStop}%
\bibitem [{\citenamefont {Feynman}(1982)}]{Feynman}%
  \BibitemOpen
  \bibfield  {author} {\bibinfo {author} {\bibfnamefont {R.}~\bibnamefont
  {Feynman}},\ }\bibfield  {title} {\bibinfo {title} {Simulating physics with
  computers},\ }\href {https://doi.org/10.1007/BF02650179} {\bibfield
  {journal} {\bibinfo  {journal} {Int J Theor Phys}\ }\textbf {\bibinfo
  {volume} {21}},\ \bibinfo {pages} {467–488} (\bibinfo {year}
  {1982})}\BibitemShut {NoStop}%
\bibitem [{\citenamefont {Lloyd}(1996)}]{lloyd}%
  \BibitemOpen
  \bibfield  {author} {\bibinfo {author} {\bibfnamefont {S.}~\bibnamefont
  {Lloyd}},\ }\bibfield  {title} {\bibinfo {title} {Universal quantum
  simulators},\ }\href@noop {} {\bibfield  {journal} {\bibinfo  {journal}
  {Science}\ ,\ \bibinfo {pages} {1073}} (\bibinfo {year} {1996})}\BibitemShut
  {NoStop}%
\bibitem [{\citenamefont {Bernien}\ \emph {et~al.}(2017)\citenamefont
  {Bernien}, \citenamefont {Schwartz}, \citenamefont {Keesling}, \citenamefont
  {Levine}, \citenamefont {Omran}, \citenamefont {Pichler}, \citenamefont
  {Choi}, \citenamefont {Zibrov}, \citenamefont {Endres}, \citenamefont
  {Greiner} \emph {et~al.}}]{bernien}%
  \BibitemOpen
  \bibfield  {author} {\bibinfo {author} {\bibfnamefont {H.}~\bibnamefont
  {Bernien}}, \bibinfo {author} {\bibfnamefont {S.}~\bibnamefont {Schwartz}},
  \bibinfo {author} {\bibfnamefont {A.}~\bibnamefont {Keesling}}, \bibinfo
  {author} {\bibfnamefont {H.}~\bibnamefont {Levine}}, \bibinfo {author}
  {\bibfnamefont {A.}~\bibnamefont {Omran}}, \bibinfo {author} {\bibfnamefont
  {H.}~\bibnamefont {Pichler}}, \bibinfo {author} {\bibfnamefont
  {S.}~\bibnamefont {Choi}}, \bibinfo {author} {\bibfnamefont {A.~S.}\
  \bibnamefont {Zibrov}}, \bibinfo {author} {\bibfnamefont {M.}~\bibnamefont
  {Endres}}, \bibinfo {author} {\bibfnamefont {M.}~\bibnamefont {Greiner}},
  \emph {et~al.},\ }\bibfield  {title} {\bibinfo {title} {Probing many-body
  dynamics on a 51-atom quantum simulator},\ }\href@noop {} {\bibfield
  {journal} {\bibinfo  {journal} {Nature}\ }\textbf {\bibinfo {volume} {551}},\
  \bibinfo {pages} {579} (\bibinfo {year} {2017})}\BibitemShut {NoStop}%
\bibitem [{\citenamefont {Zhang}\ \emph {et~al.}(2017)\citenamefont {Zhang},
  \citenamefont {Pagano}, \citenamefont {Hess}, \citenamefont {Kyprianidis},
  \citenamefont {Becker}, \citenamefont {Kaplan}, \citenamefont {Gorshkov},
  \citenamefont {Gong},\ and\ \citenamefont {Monroe}}]{zhan}%
  \BibitemOpen
  \bibfield  {author} {\bibinfo {author} {\bibfnamefont {J.}~\bibnamefont
  {Zhang}}, \bibinfo {author} {\bibfnamefont {G.}~\bibnamefont {Pagano}},
  \bibinfo {author} {\bibfnamefont {P.~W.}\ \bibnamefont {Hess}}, \bibinfo
  {author} {\bibfnamefont {A.}~\bibnamefont {Kyprianidis}}, \bibinfo {author}
  {\bibfnamefont {P.}~\bibnamefont {Becker}}, \bibinfo {author} {\bibfnamefont
  {H.}~\bibnamefont {Kaplan}}, \bibinfo {author} {\bibfnamefont {A.~V.}\
  \bibnamefont {Gorshkov}}, \bibinfo {author} {\bibfnamefont {Z.-X.}\
  \bibnamefont {Gong}},\ and\ \bibinfo {author} {\bibfnamefont
  {C.}~\bibnamefont {Monroe}},\ }\bibfield  {title} {\bibinfo {title}
  {Observation of a many-body dynamical phase transition with a 53-qubit
  quantum simulator},\ }\href@noop {} {\bibfield  {journal} {\bibinfo
  {journal} {Nature}\ }\textbf {\bibinfo {volume} {551}},\ \bibinfo {pages}
  {601} (\bibinfo {year} {2017})}\BibitemShut {NoStop}%
\bibitem [{\citenamefont {Neill}\ \emph {et~al.}(2018)\citenamefont {Neill},
  \citenamefont {Roushan}, \citenamefont {Kechedzhi}, \citenamefont {Boixo},
  \citenamefont {Isakov}, \citenamefont {Smelyanskiy}, \citenamefont {Megrant},
  \citenamefont {Chiaro}, \citenamefont {Dunsworth}, \citenamefont {Arya} \emph
  {et~al.}}]{neill2018}%
  \BibitemOpen
  \bibfield  {author} {\bibinfo {author} {\bibfnamefont {C.}~\bibnamefont
  {Neill}}, \bibinfo {author} {\bibfnamefont {P.}~\bibnamefont {Roushan}},
  \bibinfo {author} {\bibfnamefont {K.}~\bibnamefont {Kechedzhi}}, \bibinfo
  {author} {\bibfnamefont {S.}~\bibnamefont {Boixo}}, \bibinfo {author}
  {\bibfnamefont {S.~V.}\ \bibnamefont {Isakov}}, \bibinfo {author}
  {\bibfnamefont {V.}~\bibnamefont {Smelyanskiy}}, \bibinfo {author}
  {\bibfnamefont {A.}~\bibnamefont {Megrant}}, \bibinfo {author} {\bibfnamefont
  {B.}~\bibnamefont {Chiaro}}, \bibinfo {author} {\bibfnamefont
  {A.}~\bibnamefont {Dunsworth}}, \bibinfo {author} {\bibfnamefont
  {K.}~\bibnamefont {Arya}}, \emph {et~al.},\ }\bibfield  {title} {\bibinfo
  {title} {A blueprint for demonstrating quantum supremacy with superconducting
  qubits},\ }\href@noop {} {\bibfield  {journal} {\bibinfo  {journal}
  {Science}\ }\textbf {\bibinfo {volume} {360}},\ \bibinfo {pages} {195}
  (\bibinfo {year} {2018})}\BibitemShut {NoStop}%
\bibitem [{\citenamefont {Brassard}\ \emph {et~al.}(2002)\citenamefont
  {Brassard}, \citenamefont {Høyer},\ and\ \citenamefont {Mosca}}]{Brassard}%
  \BibitemOpen
  \bibfield  {author} {\bibinfo {author} {\bibfnamefont {G.}~\bibnamefont
  {Brassard}}, \bibinfo {author} {\bibfnamefont {P.}~\bibnamefont {Høyer}},\
  and\ \bibinfo {author} {\bibfnamefont {M.}~\bibnamefont {Mosca}},\ }\bibfield
   {title} {\bibinfo {title} {Quantum amplitude amplification and estimation}\
  }(\bibinfo {year} {2002})\ pp.\ \bibinfo {pages} {53--74}\BibitemShut
  {NoStop}%
\bibitem [{\citenamefont {Chuang}\ \emph {et~al.}(1998)\citenamefont {Chuang},
  \citenamefont {Gershenfeld},\ and\ \citenamefont {Kubinec}}]{Chuang_NMR}%
  \BibitemOpen
  \bibfield  {author} {\bibinfo {author} {\bibfnamefont {I.~L.}\ \bibnamefont
  {Chuang}}, \bibinfo {author} {\bibfnamefont {N.}~\bibnamefont
  {Gershenfeld}},\ and\ \bibinfo {author} {\bibfnamefont {M.}~\bibnamefont
  {Kubinec}},\ }\bibfield  {title} {\bibinfo {title} {Experimental
  implementation of fast quantum searching},\ }\href
  {https://doi.org/10.1103/PhysRevLett.80.3408} {\bibfield  {journal} {\bibinfo
   {journal} {Phys. Rev. Lett.}\ }\textbf {\bibinfo {volume} {80}},\ \bibinfo
  {pages} {3408} (\bibinfo {year} {1998})}\BibitemShut {NoStop}%
\bibitem [{\citenamefont {Zalka}(1999)}]{Zalka}%
  \BibitemOpen
  \bibfield  {author} {\bibinfo {author} {\bibfnamefont {C.}~\bibnamefont
  {Zalka}},\ }\bibfield  {title} {\bibinfo {title} {{G}rover's quantum
  searching algorithm is optimal},\ }\href
  {https://doi.org/10.1103/PhysRevA.60.2746} {\bibfield  {journal} {\bibinfo
  {journal} {Phys. Rev. A}\ }\textbf {\bibinfo {volume} {60}},\ \bibinfo
  {pages} {2746} (\bibinfo {year} {1999})}\BibitemShut {NoStop}%
\bibitem [{\citenamefont {Biham}\ \emph
  {et~al.}(1999{\natexlab{b}})\citenamefont {Biham}, \citenamefont {Biham},
  \citenamefont {Biron}, \citenamefont {Grassl},\ and\ \citenamefont
  {Lidar}}]{Biham2}%
  \BibitemOpen
  \bibfield  {author} {\bibinfo {author} {\bibfnamefont {E.}~\bibnamefont
  {Biham}}, \bibinfo {author} {\bibfnamefont {O.}~\bibnamefont {Biham}},
  \bibinfo {author} {\bibfnamefont {D.}~\bibnamefont {Biron}}, \bibinfo
  {author} {\bibfnamefont {M.}~\bibnamefont {Grassl}},\ and\ \bibinfo {author}
  {\bibfnamefont {D.~A.}\ \bibnamefont {Lidar}},\ }\bibfield  {title} {\bibinfo
  {title} {{G}rover's quantum search algorithm for an arbitrary initial
  amplitude distribution},\ }\href {https://doi.org/10.1103/PhysRevA.60.2742}
  {\bibfield  {journal} {\bibinfo  {journal} {Phys. Rev. A}\ }\textbf {\bibinfo
  {volume} {60}},\ \bibinfo {pages} {2742} (\bibinfo {year}
  {1999}{\natexlab{b}})}\BibitemShut {NoStop}%
\bibitem [{\citenamefont {Abrams}\ and\ \citenamefont
  {Williams}(1999)}]{abram}%
  \BibitemOpen
  \bibfield  {author} {\bibinfo {author} {\bibfnamefont {D.~S.}\ \bibnamefont
  {Abrams}}\ and\ \bibinfo {author} {\bibfnamefont {C.~P.}\ \bibnamefont
  {Williams}},\ }\bibfield  {title} {\bibinfo {title} {Fast quantum algorithms
  for numerical integrals and stochastic processes},\ }\href@noop {} {\bibfield
   {journal} {\bibinfo  {journal} {arXiv preprint quant-ph/9908083}\ }
  (\bibinfo {year} {1999})}\BibitemShut {NoStop}%
\bibitem [{\citenamefont {GuiLu}\ \emph {et~al.}(1999)\citenamefont {GuiLu},
  \citenamefont {WeiLin}, \citenamefont {YanSong},\ and\ \citenamefont
  {Li}}]{GuiLu}%
  \BibitemOpen
  \bibfield  {author} {\bibinfo {author} {\bibfnamefont {L.}~\bibnamefont
  {GuiLu}}, \bibinfo {author} {\bibfnamefont {Z.}~\bibnamefont {WeiLin}},
  \bibinfo {author} {\bibfnamefont {L.}~\bibnamefont {YanSong}},\ and\ \bibinfo
  {author} {\bibfnamefont {N.}~\bibnamefont {Li}},\ }\bibfield  {title}
  {\bibinfo {title} {Arbitrary phase rotation of the marked state cannot be
  used for grover's quantum search algorithm},\ }\href@noop {} {\bibfield
  {journal} {\bibinfo  {journal} {Communications in Theoretical Physics}\
  }\textbf {\bibinfo {volume} {32}},\ \bibinfo {pages} {335} (\bibinfo {year}
  {1999})}\BibitemShut {NoStop}%
\bibitem [{\citenamefont {Kwiat}\ \emph {et~al.}(2000)\citenamefont {Kwiat},
  \citenamefont {Mitchell}, \citenamefont {Schwindt},\ and\ \citenamefont
  {White}}]{Kwiat}%
  \BibitemOpen
  \bibfield  {author} {\bibinfo {author} {\bibfnamefont {P.~G.}\ \bibnamefont
  {Kwiat}}, \bibinfo {author} {\bibfnamefont {J.~R.}\ \bibnamefont {Mitchell}},
  \bibinfo {author} {\bibfnamefont {P.~D.~D.}\ \bibnamefont {Schwindt}},\ and\
  \bibinfo {author} {\bibfnamefont {A.~G.}\ \bibnamefont {White}},\ }\bibfield
  {title} {\bibinfo {title} {{G}rover's search algorithm: An optical
  approach},\ }\href {https://doi.org/10.1080/09500340008244040} {\bibfield
  {journal} {\bibinfo  {journal} {Journal of Modern Optics}\ }\textbf {\bibinfo
  {volume} {47}},\ \bibinfo {pages} {257} (\bibinfo {year} {2000})}\BibitemShut
  {NoStop}%
\bibitem [{\citenamefont {Hao-Sheng}\ and\ \citenamefont
  {Le-Man}(2000)}]{Sheng}%
  \BibitemOpen
  \bibfield  {author} {\bibinfo {author} {\bibfnamefont {Z.}~\bibnamefont
  {Hao-Sheng}}\ and\ \bibinfo {author} {\bibfnamefont {K.}~\bibnamefont
  {Le-Man}},\ }\bibfield  {title} {\bibinfo {title} {Preparation of {GHZ}
  states via {G}rover's quantum searching algorithm},\ }\href@noop {}
  {\bibfield  {journal} {\bibinfo  {journal} {Chinese Physics Letters}\
  }\textbf {\bibinfo {volume} {17}},\ \bibinfo {pages} {410} (\bibinfo {year}
  {2000})}\BibitemShut {NoStop}%
\bibitem [{\citenamefont {Long}(2001)}]{Long}%
  \BibitemOpen
  \bibfield  {author} {\bibinfo {author} {\bibfnamefont {G.~L.}\ \bibnamefont
  {Long}},\ }\bibfield  {title} {\bibinfo {title} {{G}rover algorithm with zero
  theoretical failure rate},\ }\href
  {https://doi.org/10.1103/PhysRevA.64.022307} {\bibfield  {journal} {\bibinfo
  {journal} {Phys. Rev. A}\ }\textbf {\bibinfo {volume} {64}},\ \bibinfo
  {pages} {022307} (\bibinfo {year} {2001})}\BibitemShut {NoStop}%
\bibitem [{\citenamefont {Biham}\ and\ \citenamefont
  {Kenigsberg}(2002)}]{Biham3}%
  \BibitemOpen
  \bibfield  {author} {\bibinfo {author} {\bibfnamefont {E.}~\bibnamefont
  {Biham}}\ and\ \bibinfo {author} {\bibfnamefont {D.}~\bibnamefont
  {Kenigsberg}},\ }\bibfield  {title} {\bibinfo {title} {{G}rover's quantum
  search algorithm for an arbitrary initial mixed state},\ }\href
  {https://doi.org/10.1103/PhysRevA.66.062301} {\bibfield  {journal} {\bibinfo
  {journal} {Phys. Rev. A}\ }\textbf {\bibinfo {volume} {66}},\ \bibinfo
  {pages} {062301} (\bibinfo {year} {2002})}\BibitemShut {NoStop}%
\bibitem [{\citenamefont {Heinrich}(2002)}]{Heinrich}%
  \BibitemOpen
  \bibfield  {author} {\bibinfo {author} {\bibfnamefont {S.}~\bibnamefont
  {Heinrich}},\ }\bibfield  {title} {\bibinfo {title} {Quantum summation with
  an application to integration},\ }\href
  {https://doi.org/10.1006/jcom.2001.0629} {\bibfield  {journal} {\bibinfo
  {journal} {J. Complex.}\ }\textbf {\bibinfo {volume} {18}},\ \bibinfo {pages}
  {1–50} (\bibinfo {year} {2002})}\BibitemShut {NoStop}%
\bibitem [{\citenamefont {Roland}\ and\ \citenamefont {Cerf}(2003)}]{Roland}%
  \BibitemOpen
  \bibfield  {author} {\bibinfo {author} {\bibfnamefont {J.}~\bibnamefont
  {Roland}}\ and\ \bibinfo {author} {\bibfnamefont {N.~J.}\ \bibnamefont
  {Cerf}},\ }\bibfield  {title} {\bibinfo {title} {Quantum-circuit model of
  {H}amiltonian search algorithms},\ }\href
  {https://doi.org/10.1103/PhysRevA.68.062311} {\bibfield  {journal} {\bibinfo
  {journal} {Phys. Rev. A}\ }\textbf {\bibinfo {volume} {68}},\ \bibinfo
  {pages} {062311} (\bibinfo {year} {2003})}\BibitemShut {NoStop}%
\bibitem [{\citenamefont {Xiao}\ and\ \citenamefont {Jones}(2005)}]{Xiao}%
  \BibitemOpen
  \bibfield  {author} {\bibinfo {author} {\bibfnamefont {L.}~\bibnamefont
  {Xiao}}\ and\ \bibinfo {author} {\bibfnamefont {J.~A.}\ \bibnamefont
  {Jones}},\ }\bibfield  {title} {\bibinfo {title} {Error tolerance in an {NMR}
  implementation of {G}rover's fixed-point quantum search algorithm},\ }\href
  {https://doi.org/10.1103/PhysRevA.72.032326} {\bibfield  {journal} {\bibinfo
  {journal} {Phys. Rev. A}\ }\textbf {\bibinfo {volume} {72}},\ \bibinfo
  {pages} {032326} (\bibinfo {year} {2005})}\BibitemShut {NoStop}%
\bibitem [{\citenamefont {Jones}\ \emph {et~al.}(1998)\citenamefont {Jones},
  \citenamefont {Mosca},\ and\ \citenamefont {Hansen}}]{Jones}%
  \BibitemOpen
  \bibfield  {author} {\bibinfo {author} {\bibfnamefont {J.}~\bibnamefont
  {Jones}}, \bibinfo {author} {\bibfnamefont {M.}~\bibnamefont {Mosca}},\ and\
  \bibinfo {author} {\bibfnamefont {R.}~\bibnamefont {Hansen}},\ }\bibfield
  {title} {\bibinfo {title} {Implementation of a quantum search algorithm on a
  quantum computer},\ }\href {https://doi.org/https://doi.org/10.1038/30687}
  {\bibfield  {journal} {\bibinfo  {journal} {Nature}\ }\textbf {\bibinfo
  {volume} {393}},\ \bibinfo {pages} {344–346} (\bibinfo {year}
  {1998})}\BibitemShut {NoStop}%
\bibitem [{\citenamefont {Vandersypen}\ \emph {et~al.}(2000)\citenamefont
  {Vandersypen}, \citenamefont {Steffen}, \citenamefont {Sherwood},
  \citenamefont {Yannoni}, \citenamefont {Breyta},\ and\ \citenamefont
  {Chuang}}]{Vander}%
  \BibitemOpen
  \bibfield  {author} {\bibinfo {author} {\bibfnamefont {L.~M.~K.}\
  \bibnamefont {Vandersypen}}, \bibinfo {author} {\bibfnamefont
  {M.}~\bibnamefont {Steffen}}, \bibinfo {author} {\bibfnamefont {M.~H.}\
  \bibnamefont {Sherwood}}, \bibinfo {author} {\bibfnamefont {C.~S.}\
  \bibnamefont {Yannoni}}, \bibinfo {author} {\bibfnamefont {G.}~\bibnamefont
  {Breyta}},\ and\ \bibinfo {author} {\bibfnamefont {I.~L.}\ \bibnamefont
  {Chuang}},\ }\bibfield  {title} {\bibinfo {title} {Implementation of a
  three-quantum-bit search algorithm},\ }\href
  {https://doi.org/10.1063/1.125846} {\bibfield  {journal} {\bibinfo  {journal}
  {Applied Physics Letters}\ }\textbf {\bibinfo {volume} {76}},\ \bibinfo
  {pages} {646} (\bibinfo {year} {2000})}\BibitemShut {NoStop}%
\bibitem [{\citenamefont {Ermakov}\ and\ \citenamefont {Fung}(2002)}]{Ermakov}%
  \BibitemOpen
  \bibfield  {author} {\bibinfo {author} {\bibfnamefont {V.~L.}\ \bibnamefont
  {Ermakov}}\ and\ \bibinfo {author} {\bibfnamefont {B.~M.}\ \bibnamefont
  {Fung}},\ }\bibfield  {title} {\bibinfo {title} {Experimental realization of
  a continuous version of the {G}rover algorithm},\ }\href
  {https://doi.org/10.1103/PhysRevA.66.042310} {\bibfield  {journal} {\bibinfo
  {journal} {Phys. Rev. A}\ }\textbf {\bibinfo {volume} {66}},\ \bibinfo
  {pages} {042310} (\bibinfo {year} {2002})}\BibitemShut {NoStop}%
\bibitem [{\citenamefont {Bhattacharya}\ \emph {et~al.}(2002)\citenamefont
  {Bhattacharya}, \citenamefont {van Linden van~den Heuvell},\ and\
  \citenamefont {Spreeuw}}]{Bhattacharya}%
  \BibitemOpen
  \bibfield  {author} {\bibinfo {author} {\bibfnamefont {N.}~\bibnamefont
  {Bhattacharya}}, \bibinfo {author} {\bibfnamefont {H.~B.}\ \bibnamefont {van
  Linden van~den Heuvell}},\ and\ \bibinfo {author} {\bibfnamefont {R.~J.~C.}\
  \bibnamefont {Spreeuw}},\ }\bibfield  {title} {\bibinfo {title}
  {Implementation of quantum search algorithm using classical fourier optics},\
  }\href {https://doi.org/10.1103/PhysRevLett.88.137901} {\bibfield  {journal}
  {\bibinfo  {journal} {Phys. Rev. Lett.}\ }\textbf {\bibinfo {volume} {88}},\
  \bibinfo {pages} {137901} (\bibinfo {year} {2002})}\BibitemShut {NoStop}%
\bibitem [{\citenamefont {Jing-Fu}\ \emph {et~al.}(2003)\citenamefont
  {Jing-Fu}, \citenamefont {Zhi-Heng}, \citenamefont {Zhi-Wei},\ and\
  \citenamefont {Lu}}]{Zhang}%
  \BibitemOpen
  \bibfield  {author} {\bibinfo {author} {\bibfnamefont {Z.}~\bibnamefont
  {Jing-Fu}}, \bibinfo {author} {\bibfnamefont {L.}~\bibnamefont {Zhi-Heng}},
  \bibinfo {author} {\bibfnamefont {D.}~\bibnamefont {Zhi-Wei}},\ and\ \bibinfo
  {author} {\bibfnamefont {S.}~\bibnamefont {Lu}},\ }\bibfield  {title}
  {\bibinfo {title} {{NMR} analogue of the generalized {G}rover's algorithm of
  multiple marked states and its application},\ }\href@noop {} {\bibfield
  {journal} {\bibinfo  {journal} {Chinese Physics}\ }\textbf {\bibinfo {volume}
  {12}},\ \bibinfo {pages} {700} (\bibinfo {year} {2003})}\BibitemShut
  {NoStop}%
\bibitem [{\citenamefont {Walther}\ \emph {et~al.}(2005)\citenamefont
  {Walther}, \citenamefont {Resch}, \citenamefont {Rudolph}, \citenamefont
  {Schenck}, \citenamefont {Weinfurter}, \citenamefont {Vedral}, \citenamefont
  {Aspelmeyer},\ and\ \citenamefont {Zeilinger}}]{WaltherP}%
  \BibitemOpen
  \bibfield  {author} {\bibinfo {author} {\bibfnamefont {P.}~\bibnamefont
  {Walther}}, \bibinfo {author} {\bibfnamefont {K.}~\bibnamefont {Resch}},
  \bibinfo {author} {\bibfnamefont {T.}~\bibnamefont {Rudolph}}, \bibinfo
  {author} {\bibfnamefont {E.}~\bibnamefont {Schenck}}, \bibinfo {author}
  {\bibfnamefont {H.}~\bibnamefont {Weinfurter}}, \bibinfo {author}
  {\bibfnamefont {V.}~\bibnamefont {Vedral}}, \bibinfo {author} {\bibfnamefont
  {M.}~\bibnamefont {Aspelmeyer}},\ and\ \bibinfo {author} {\bibfnamefont
  {A.}~\bibnamefont {Zeilinger}},\ }\bibfield  {title} {\bibinfo {title}
  {Experimental one-way quantum computing},\ }\href
  {https://doi.org/10.1038/nature03347} {\bibfield  {journal} {\bibinfo
  {journal} {Nature}\ }\textbf {\bibinfo {volume} {434}},\ \bibinfo {pages}
  {169} (\bibinfo {year} {2005})}\BibitemShut {NoStop}%
\bibitem [{\citenamefont {Brickman}\ \emph {et~al.}(2005)\citenamefont
  {Brickman}, \citenamefont {Haljan}, \citenamefont {Lee}, \citenamefont
  {Acton}, \citenamefont {Deslauriers},\ and\ \citenamefont
  {Monroe}}]{Brickman}%
  \BibitemOpen
  \bibfield  {author} {\bibinfo {author} {\bibfnamefont {K.-A.}\ \bibnamefont
  {Brickman}}, \bibinfo {author} {\bibfnamefont {P.~C.}\ \bibnamefont
  {Haljan}}, \bibinfo {author} {\bibfnamefont {P.~J.}\ \bibnamefont {Lee}},
  \bibinfo {author} {\bibfnamefont {M.}~\bibnamefont {Acton}}, \bibinfo
  {author} {\bibfnamefont {L.}~\bibnamefont {Deslauriers}},\ and\ \bibinfo
  {author} {\bibfnamefont {C.}~\bibnamefont {Monroe}},\ }\bibfield  {title}
  {\bibinfo {title} {Implementation of {G}rover's quantum search algorithm in a
  scalable system},\ }\href {https://doi.org/10.1103/PhysRevA.72.050306}
  {\bibfield  {journal} {\bibinfo  {journal} {Phys. Rev. A}\ }\textbf {\bibinfo
  {volume} {72}},\ \bibinfo {pages} {050306} (\bibinfo {year}
  {2005})}\BibitemShut {NoStop}%
\bibitem [{\citenamefont {DiCarlo}\ \emph {et~al.}(2009)\citenamefont
  {DiCarlo}, \citenamefont {Chow}, \citenamefont {Gambetta}, \citenamefont
  {Bishop}, \citenamefont {Johnson}, \citenamefont {Schuster}, \citenamefont
  {Majer}, \citenamefont {Blais}, \citenamefont {Frunzio}, \citenamefont
  {Girvin},\ and\ \citenamefont {Schoelkopf}}]{DiCarlo}%
  \BibitemOpen
  \bibfield  {author} {\bibinfo {author} {\bibfnamefont {L.}~\bibnamefont
  {DiCarlo}}, \bibinfo {author} {\bibfnamefont {J.}~\bibnamefont {Chow}},
  \bibinfo {author} {\bibfnamefont {J.}~\bibnamefont {Gambetta}}, \bibinfo
  {author} {\bibfnamefont {L.}~\bibnamefont {Bishop}}, \bibinfo {author}
  {\bibfnamefont {B.}~\bibnamefont {Johnson}}, \bibinfo {author} {\bibfnamefont
  {D.}~\bibnamefont {Schuster}}, \bibinfo {author} {\bibfnamefont
  {J.}~\bibnamefont {Majer}}, \bibinfo {author} {\bibfnamefont
  {A.}~\bibnamefont {Blais}}, \bibinfo {author} {\bibfnamefont
  {L.}~\bibnamefont {Frunzio}}, \bibinfo {author} {\bibfnamefont
  {S.}~\bibnamefont {Girvin}},\ and\ \bibinfo {author} {\bibfnamefont
  {R.}~\bibnamefont {Schoelkopf}},\ }\bibfield  {title} {\bibinfo {title}
  {Demonstration of two-qubit algorithms with a superconducting quantum
  processor},\ }\href {https://doi.org/10.1038/nature08121} {\bibfield
  {journal} {\bibinfo  {journal} {Nature}\ }\textbf {\bibinfo {volume} {460}},\
  \bibinfo {pages} {240} (\bibinfo {year} {2009})}\BibitemShut {NoStop}%
\bibitem [{\citenamefont {Figgatt}\ \emph {et~al.}(2017)\citenamefont
  {Figgatt}, \citenamefont {Maslov}, \citenamefont {Landsman} \emph
  {et~al.}}]{Figg}%
  \BibitemOpen
  \bibfield  {author} {\bibinfo {author} {\bibfnamefont {C.}~\bibnamefont
  {Figgatt}}, \bibinfo {author} {\bibfnamefont {D.}~\bibnamefont {Maslov}},
  \bibinfo {author} {\bibfnamefont {K.}~\bibnamefont {Landsman}}, \emph
  {et~al.},\ }\bibfield  {title} {\bibinfo {title} {Complete 3-qubit {G}rover
  search on a programmable quantum computer},\ }\href
  {https://doi.org/https://doi.org/10.1038/s41467-017-01904-7} {\bibfield
  {journal} {\bibinfo  {journal} {Nat Commun}\ ,\ \bibinfo {pages} {1918}}
  (\bibinfo {year} {2017})}\BibitemShut {NoStop}%
\bibitem [{\citenamefont {Bernstein}\ and\ \citenamefont
  {Vazirani}(1997)}]{Bernstein}%
  \BibitemOpen
  \bibfield  {author} {\bibinfo {author} {\bibfnamefont {E.}~\bibnamefont
  {Bernstein}}\ and\ \bibinfo {author} {\bibfnamefont {U.}~\bibnamefont
  {Vazirani}},\ }\bibfield  {title} {\bibinfo {title} {Quantum complexity
  theory},\ }\href@noop {} {\bibfield  {journal} {\bibinfo  {journal} {SIAM J.
  Comput.}\ }\textbf {\bibinfo {volume} {26}},\ \bibinfo {pages} {1411}
  (\bibinfo {year} {1997})}\BibitemShut {NoStop}%
\bibitem [{\citenamefont {Bassi}\ and\ \citenamefont {Deckert}(2008)}]{bassi}%
  \BibitemOpen
  \bibfield  {author} {\bibinfo {author} {\bibfnamefont {A.}~\bibnamefont
  {Bassi}}\ and\ \bibinfo {author} {\bibfnamefont {D.-A.}\ \bibnamefont
  {Deckert}},\ }\bibfield  {title} {\bibinfo {title} {Noise gates for
  decoherent quantum circuits},\ }\href
  {https://doi.org/10.1103/PhysRevA.77.032323} {\bibfield  {journal} {\bibinfo
  {journal} {Phys. Rev. A}\ }\textbf {\bibinfo {volume} {77}},\ \bibinfo
  {pages} {032323} (\bibinfo {year} {2008})}\BibitemShut {NoStop}%
\bibitem [{\citenamefont {Preskill}(2000)}]{Preskill}%
  \BibitemOpen
  \bibfield  {author} {\bibinfo {author} {\bibfnamefont {J.}~\bibnamefont
  {Preskill}},\ }\bibfield  {title} {\bibinfo {title} {Course information for
  physics 219/computer science 219 quantum computation},\ }\href@noop {}
  {\bibfield  {journal} {\bibinfo  {journal}
  {http://theory.caltech.edu/~preskill/ph229/}\ } (\bibinfo {year}
  {2000})}\BibitemShut {NoStop}%
\bibitem [{\citenamefont {Nielsen}\ and\ \citenamefont {Chuang}(2011)}]{N&C}%
  \BibitemOpen
  \bibfield  {author} {\bibinfo {author} {\bibfnamefont {M.~A.}\ \bibnamefont
  {Nielsen}}\ and\ \bibinfo {author} {\bibfnamefont {I.~L.}\ \bibnamefont
  {Chuang}},\ }\href@noop {} {\emph {\bibinfo {title} {Quantum Computation and
  Quantum Information: 10th Anniversary Edition}}}\ (\bibinfo  {publisher}
  {Cambridge University Press},\ \bibinfo {year} {2011})\BibitemShut {NoStop}%
\bibitem [{\citenamefont {Barnes}\ and\ \citenamefont {Warren}(1999)}]{Barnes}%
  \BibitemOpen
  \bibfield  {author} {\bibinfo {author} {\bibfnamefont {J.~P.}\ \bibnamefont
  {Barnes}}\ and\ \bibinfo {author} {\bibfnamefont {W.~S.}\ \bibnamefont
  {Warren}},\ }\bibfield  {title} {\bibinfo {title} {Decoherence and
  programmable quantum computation},\ }\href
  {https://doi.org/10.1103/PhysRevA.60.4363} {\bibfield  {journal} {\bibinfo
  {journal} {Phys. Rev. A}\ }\textbf {\bibinfo {volume} {60}},\ \bibinfo
  {pages} {4363} (\bibinfo {year} {1999})}\BibitemShut {NoStop}%
\bibitem [{\citenamefont {Pablo-Norman}\ and\ \citenamefont
  {Ruiz-Altaba}(1999)}]{Altaba}%
  \BibitemOpen
  \bibfield  {author} {\bibinfo {author} {\bibfnamefont {B.}~\bibnamefont
  {Pablo-Norman}}\ and\ \bibinfo {author} {\bibfnamefont {M.}~\bibnamefont
  {Ruiz-Altaba}},\ }\bibfield  {title} {\bibinfo {title} {Noise in {G}rover's
  quantum search algorithm},\ }\href
  {https://doi.org/10.1103/PhysRevA.61.012301} {\bibfield  {journal} {\bibinfo
  {journal} {Phys. Rev. A}\ }\textbf {\bibinfo {volume} {61}},\ \bibinfo
  {pages} {012301} (\bibinfo {year} {1999})}\BibitemShut {NoStop}%
\bibitem [{\citenamefont {Azuma}(2002)}]{Azuma}%
  \BibitemOpen
  \bibfield  {author} {\bibinfo {author} {\bibfnamefont {H.}~\bibnamefont
  {Azuma}},\ }\bibfield  {title} {\bibinfo {title} {Decoherence in {G}rover's
  quantum algorithm: Perturbative approach},\ }\href
  {https://doi.org/10.1103/PhysRevA.65.042311} {\bibfield  {journal} {\bibinfo
  {journal} {Phys. Rev. A}\ }\textbf {\bibinfo {volume} {65}},\ \bibinfo
  {pages} {042311} (\bibinfo {year} {2002})}\BibitemShut {NoStop}%
\bibitem [{\citenamefont {Long}\ \emph {et~al.}(2000)\citenamefont {Long},
  \citenamefont {Li}, \citenamefont {Zhang},\ and\ \citenamefont {Tu}}]{Tu}%
  \BibitemOpen
  \bibfield  {author} {\bibinfo {author} {\bibfnamefont {G.~L.}\ \bibnamefont
  {Long}}, \bibinfo {author} {\bibfnamefont {Y.~S.}\ \bibnamefont {Li}},
  \bibinfo {author} {\bibfnamefont {W.~L.}\ \bibnamefont {Zhang}},\ and\
  \bibinfo {author} {\bibfnamefont {C.~C.}\ \bibnamefont {Tu}},\ }\bibfield
  {title} {\bibinfo {title} {Dominant gate imperfection in {G}rover's quantum
  search algorithm},\ }\href {https://doi.org/10.1103/PhysRevA.61.042305}
  {\bibfield  {journal} {\bibinfo  {journal} {Phys. Rev. A}\ }\textbf {\bibinfo
  {volume} {61}},\ \bibinfo {pages} {042305} (\bibinfo {year}
  {2000})}\BibitemShut {NoStop}%
\bibitem [{\citenamefont {Bae}\ and\ \citenamefont {Kwon}(2003)}]{Bae}%
  \BibitemOpen
  \bibfield  {author} {\bibinfo {author} {\bibfnamefont {J.}~\bibnamefont
  {Bae}}\ and\ \bibinfo {author} {\bibfnamefont {Y.}~\bibnamefont {Kwon}},\
  }\bibfield  {title} {\bibinfo {title} {Perturbations can enhance quantum
  search},\ }\href {https://doi.org/https://doi.org/10.1023/A:1027343321366}
  {\bibfield  {journal} {\bibinfo  {journal} {International Journal of
  Theoretical Physics}\ }\textbf {\bibinfo {volume} {42}},\ \bibinfo {pages}
  {2075–2080} (\bibinfo {year} {2003})}\BibitemShut {NoStop}%
\bibitem [{\citenamefont {Chen}\ \emph {et~al.}(2003)\citenamefont {Chen},
  \citenamefont {Kaszlikowski}, \citenamefont {Kwek},\ and\ \citenamefont
  {Oh}}]{kwek}%
  \BibitemOpen
  \bibfield  {author} {\bibinfo {author} {\bibfnamefont {J.}~\bibnamefont
  {Chen}}, \bibinfo {author} {\bibfnamefont {D.}~\bibnamefont {Kaszlikowski}},
  \bibinfo {author} {\bibfnamefont {L.}~\bibnamefont {Kwek}},\ and\ \bibinfo
  {author} {\bibfnamefont {C.}~\bibnamefont {Oh}},\ }\bibfield  {title}
  {\bibinfo {title} {Searching a database under decoherence},\ }\href
  {https://doi.org/https://doi.org/10.1016/S0375-9601(02)01637-7} {\bibfield
  {journal} {\bibinfo  {journal} {Physics Letters A}\ }\textbf {\bibinfo
  {volume} {306}},\ \bibinfo {pages} {296} (\bibinfo {year}
  {2003})}\BibitemShut {NoStop}%
\bibitem [{\citenamefont {Shapira}\ \emph {et~al.}(2003)\citenamefont
  {Shapira}, \citenamefont {Mozes},\ and\ \citenamefont {Biham}}]{Biham}%
  \BibitemOpen
  \bibfield  {author} {\bibinfo {author} {\bibfnamefont {D.}~\bibnamefont
  {Shapira}}, \bibinfo {author} {\bibfnamefont {S.}~\bibnamefont {Mozes}},\
  and\ \bibinfo {author} {\bibfnamefont {O.}~\bibnamefont {Biham}},\ }\bibfield
   {title} {\bibinfo {title} {Effect of unitary noise on {G}rover's quantum
  search algorithm},\ }\href {https://doi.org/10.1103/PhysRevA.67.042301}
  {\bibfield  {journal} {\bibinfo  {journal} {Phys. Rev. A}\ }\textbf {\bibinfo
  {volume} {67}},\ \bibinfo {pages} {042301} (\bibinfo {year}
  {2003})}\BibitemShut {NoStop}%
\bibitem [{\citenamefont {Gawron}\ \emph {et~al.}(2012)\citenamefont {Gawron},
  \citenamefont {Klamka},\ and\ \citenamefont {Winiarczyk}}]{Gawron}%
  \BibitemOpen
  \bibfield  {author} {\bibinfo {author} {\bibfnamefont {P.}~\bibnamefont
  {Gawron}}, \bibinfo {author} {\bibfnamefont {J.}~\bibnamefont {Klamka}},\
  and\ \bibinfo {author} {\bibfnamefont {R.}~\bibnamefont {Winiarczyk}},\
  }\bibfield  {title} {\bibinfo {title} {Noise effects in the quantum search
  algorithm from the viewpoint of computational complexity},\ }\href
  {https://doi.org/doi:10.2478/v10006-012-0037-2} {\bibfield  {journal}
  {\bibinfo  {journal} {International Journal of Applied Mathematics and
  Computer Science}\ }\textbf {\bibinfo {volume} {22}},\ \bibinfo {pages} {493}
  (\bibinfo {year} {2012})}\BibitemShut {NoStop}%
\bibitem [{\citenamefont {Reitzner}\ and\ \citenamefont
  {Hillery}(2019)}]{Reitzner}%
  \BibitemOpen
  \bibfield  {author} {\bibinfo {author} {\bibfnamefont {D.}~\bibnamefont
  {Reitzner}}\ and\ \bibinfo {author} {\bibfnamefont {M.}~\bibnamefont
  {Hillery}},\ }\bibfield  {title} {\bibinfo {title} {{G}rover search under
  localized dephasing},\ }\href {https://doi.org/10.1103/PhysRevA.99.012339}
  {\bibfield  {journal} {\bibinfo  {journal} {Phys. Rev. A}\ }\textbf {\bibinfo
  {volume} {99}},\ \bibinfo {pages} {012339} (\bibinfo {year}
  {2019})}\BibitemShut {NoStop}%
\bibitem [{\citenamefont {Salas}(2008)}]{Salas}%
  \BibitemOpen
  \bibfield  {author} {\bibinfo {author} {\bibfnamefont {P.}~\bibnamefont
  {Salas}},\ }\bibfield  {title} {\bibinfo {title} {Noise effect on {G}rover
  algorithm},\ }\href {https://doi.org/doi:10.2478/v10006-012-0037-2}
  {\bibfield  {journal} {\bibinfo  {journal} {Eur. Phys. J. D}\ }\textbf
  {\bibinfo {volume} {46}},\ \bibinfo {pages} {365–373} (\bibinfo {year}
  {(2008)})}\BibitemShut {NoStop}%
\bibitem [{\citenamefont {Hasegawa}(2009)}]{Hasegawa}%
  \BibitemOpen
  \bibfield  {author} {\bibinfo {author} {\bibfnamefont {J.}~\bibnamefont
  {Hasegawa}},\ }\bibfield  {title} {\bibinfo {title} {Variety of effects of
  decoherence in quantum algorithms},\ }\href
  {https://doi.org/10.1587/transfun.E92.A.1284} {\bibfield  {journal} {\bibinfo
   {journal} {IEICE Transactions on Fundamentals of Electronics, Communications
  and Computer Sciences}\ }\textbf {\bibinfo {volume} {E92.A}},\ \bibinfo
  {pages} {1284} (\bibinfo {year} {2009})}\BibitemShut {NoStop}%
\bibitem [{\citenamefont {Cohn}\ \emph {et~al.}(2016)\citenamefont {Cohn},
  \citenamefont {De~Oliveira}, \citenamefont {Buksman},\ and\ \citenamefont
  {De~Lacalle}}]{Cohn}%
  \BibitemOpen
  \bibfield  {author} {\bibinfo {author} {\bibfnamefont {I.}~\bibnamefont
  {Cohn}}, \bibinfo {author} {\bibfnamefont {A.~L.~F.}\ \bibnamefont
  {De~Oliveira}}, \bibinfo {author} {\bibfnamefont {E.}~\bibnamefont
  {Buksman}},\ and\ \bibinfo {author} {\bibfnamefont {J.~G.~L.}\ \bibnamefont
  {De~Lacalle}},\ }\bibfield  {title} {\bibinfo {title} {{G}rover’s search
  with local and total depolarizing channel errors: Complexity analysis},\
  }\href {https://doi.org/10.1142/S021974991650009X} {\bibfield  {journal}
  {\bibinfo  {journal} {International Journal of Quantum Information}\ }\textbf
  {\bibinfo {volume} {14}},\ \bibinfo {pages} {1650009} (\bibinfo {year}
  {2016})}\BibitemShut {NoStop}%
\bibitem [{\citenamefont {Vrana}\ \emph {et~al.}(2014)\citenamefont {Vrana},
  \citenamefont {Reeb}, \citenamefont {Reitzner},\ and\ \citenamefont
  {Wolf}}]{Wolf}%
  \BibitemOpen
  \bibfield  {author} {\bibinfo {author} {\bibfnamefont {P.}~\bibnamefont
  {Vrana}}, \bibinfo {author} {\bibfnamefont {D.}~\bibnamefont {Reeb}},
  \bibinfo {author} {\bibfnamefont {D.}~\bibnamefont {Reitzner}},\ and\
  \bibinfo {author} {\bibfnamefont {M.}~\bibnamefont {Wolf}},\ }\bibfield
  {title} {\bibinfo {title} {Fault-ignorant quantum search},\ }\href
  {https://doi.org/https://doi.org/10.1088/1367-2630/16/7/073033} {\bibfield
  {journal} {\bibinfo  {journal} {New J. Phys.}\ }\textbf {\bibinfo {volume}
  {16}},\ \bibinfo {pages} {073033} (\bibinfo {year} {2014})}\BibitemShut
  {NoStop}%
\bibitem [{\citenamefont {Steane}(1996)}]{Steane2}%
  \BibitemOpen
  \bibfield  {author} {\bibinfo {author} {\bibfnamefont {A.}~\bibnamefont
  {Steane}},\ }\bibfield  {title} {\bibinfo {title} {Multiple-particle
  interference and quantum error correction},\ }\href
  {https://doi.org/10.1098/rspa.1996.0136} {\bibfield  {journal} {\bibinfo
  {journal} {Proceedings of the Royal Society of London. Series A:
  Mathematical, Physical and Engineering Sciences}\ }\textbf {\bibinfo {volume}
  {452}},\ \bibinfo {pages} {2551} (\bibinfo {year} {1996})}\BibitemShut
  {NoStop}%
\bibitem [{\citenamefont {Botsinis}\ \emph {et~al.}(2016)\citenamefont
  {Botsinis}, \citenamefont {Babar},\ and\ \citenamefont {Alanis}}]{Botsinis}%
  \BibitemOpen
  \bibfield  {author} {\bibinfo {author} {\bibfnamefont {P.}~\bibnamefont
  {Botsinis}}, \bibinfo {author} {\bibfnamefont {Z.}~\bibnamefont {Babar}},\
  and\ \bibinfo {author} {\bibfnamefont {D.~e.~a.}\ \bibnamefont {Alanis}},\
  }\bibfield  {title} {\bibinfo {title} {Quantum error correction protects
  quantum search algorithms against decoherence},\ }\href
  {https://doi.org/https://doi.org/10.1038/srep38095} {\bibfield  {journal}
  {\bibinfo  {journal} {Sci Rep}\ ,\ \bibinfo {pages} {38095}} (\bibinfo {year}
  {2016})}\BibitemShut {NoStop}%
\bibitem [{\citenamefont {Alicki}\ \emph {et~al.}(2002)\citenamefont {Alicki},
  \citenamefont {Horodecki}, \citenamefont {Horodecki},\ and\ \citenamefont
  {Horodecki}}]{alicki_}%
  \BibitemOpen
  \bibfield  {author} {\bibinfo {author} {\bibfnamefont {R.}~\bibnamefont
  {Alicki}}, \bibinfo {author} {\bibfnamefont {M.}~\bibnamefont {Horodecki}},
  \bibinfo {author} {\bibfnamefont {P.}~\bibnamefont {Horodecki}},\ and\
  \bibinfo {author} {\bibfnamefont {R.}~\bibnamefont {Horodecki}},\ }\bibfield
  {title} {\bibinfo {title} {Dynamical description of quantum computing:
  generic nonlocality of quantum noise},\ }\href@noop {} {\bibfield  {journal}
  {\bibinfo  {journal} {Physical Review A}\ }\textbf {\bibinfo {volume} {65}},\
  \bibinfo {pages} {062101} (\bibinfo {year} {2002})}\BibitemShut {NoStop}%
\bibitem [{\citenamefont {Bialczak}\ \emph {et~al.}(2007)\citenamefont
  {Bialczak}, \citenamefont {McDermott}, \citenamefont {Ansmann}, \citenamefont
  {Hofheinz}, \citenamefont {Katz}, \citenamefont {Lucero}, \citenamefont
  {Neeley}, \citenamefont {O'Connell}, \citenamefont {Wang}, \citenamefont
  {Cleland},\ and\ \citenamefont {Martinis}}]{bialczak}%
  \BibitemOpen
  \bibfield  {author} {\bibinfo {author} {\bibfnamefont {R.~C.}\ \bibnamefont
  {Bialczak}}, \bibinfo {author} {\bibfnamefont {R.}~\bibnamefont {McDermott}},
  \bibinfo {author} {\bibfnamefont {M.}~\bibnamefont {Ansmann}}, \bibinfo
  {author} {\bibfnamefont {M.}~\bibnamefont {Hofheinz}}, \bibinfo {author}
  {\bibfnamefont {N.}~\bibnamefont {Katz}}, \bibinfo {author} {\bibfnamefont
  {E.}~\bibnamefont {Lucero}}, \bibinfo {author} {\bibfnamefont
  {M.}~\bibnamefont {Neeley}}, \bibinfo {author} {\bibfnamefont {A.~D.}\
  \bibnamefont {O'Connell}}, \bibinfo {author} {\bibfnamefont {H.}~\bibnamefont
  {Wang}}, \bibinfo {author} {\bibfnamefont {A.~N.}\ \bibnamefont {Cleland}},\
  and\ \bibinfo {author} {\bibfnamefont {J.~M.}\ \bibnamefont {Martinis}},\
  }\bibfield  {title} {\bibinfo {title} {$1/f$ flux noise in {J}osephson phase
  qubits},\ }\href {https://doi.org/10.1103/PhysRevLett.99.187006} {\bibfield
  {journal} {\bibinfo  {journal} {Phys. Rev. Lett.}\ }\textbf {\bibinfo
  {volume} {99}},\ \bibinfo {pages} {187006} (\bibinfo {year}
  {2007})}\BibitemShut {NoStop}%
\bibitem [{\citenamefont {Bylander}\ \emph {et~al.}(2011)\citenamefont
  {Bylander}, \citenamefont {Gustavsson}, \citenamefont {Yan}, \citenamefont
  {Yoshihara}, \citenamefont {Harrabi}, \citenamefont {Fitch}, \citenamefont
  {Cory}, \citenamefont {Nakamura}, \citenamefont {Tsai},\ and\ \citenamefont
  {Oliver}}]{bylander}%
  \BibitemOpen
  \bibfield  {author} {\bibinfo {author} {\bibfnamefont {J.}~\bibnamefont
  {Bylander}}, \bibinfo {author} {\bibfnamefont {S.}~\bibnamefont
  {Gustavsson}}, \bibinfo {author} {\bibfnamefont {F.}~\bibnamefont {Yan}},
  \bibinfo {author} {\bibfnamefont {F.}~\bibnamefont {Yoshihara}}, \bibinfo
  {author} {\bibfnamefont {K.}~\bibnamefont {Harrabi}}, \bibinfo {author}
  {\bibfnamefont {G.}~\bibnamefont {Fitch}}, \bibinfo {author} {\bibfnamefont
  {D.~G.}\ \bibnamefont {Cory}}, \bibinfo {author} {\bibfnamefont
  {Y.}~\bibnamefont {Nakamura}}, \bibinfo {author} {\bibfnamefont {J.-S.}\
  \bibnamefont {Tsai}},\ and\ \bibinfo {author} {\bibfnamefont {W.~D.}\
  \bibnamefont {Oliver}},\ }\bibfield  {title} {\bibinfo {title} {Noise
  spectroscopy through dynamical decoupling with a superconducting flux
  qubit},\ }\href@noop {} {\bibfield  {journal} {\bibinfo  {journal} {Nature
  Physics}\ }\textbf {\bibinfo {volume} {7}},\ \bibinfo {pages} {565} (\bibinfo
  {year} {2011})}\BibitemShut {NoStop}%
\bibitem [{\citenamefont {Paladino}\ \emph {et~al.}(2014)\citenamefont
  {Paladino}, \citenamefont {Galperin}, \citenamefont {Falci},\ and\
  \citenamefont {Altshuler}}]{Paladino}%
  \BibitemOpen
  \bibfield  {author} {\bibinfo {author} {\bibfnamefont {E.}~\bibnamefont
  {Paladino}}, \bibinfo {author} {\bibfnamefont {Y.~M.}\ \bibnamefont
  {Galperin}}, \bibinfo {author} {\bibfnamefont {G.}~\bibnamefont {Falci}},\
  and\ \bibinfo {author} {\bibfnamefont {B.~L.}\ \bibnamefont {Altshuler}},\
  }\bibfield  {title} {\bibinfo {title} {$1/f$ noise: Implications for
  solid-state quantum information},\ }\href
  {https://doi.org/10.1103/RevModPhys.86.361} {\bibfield  {journal} {\bibinfo
  {journal} {Rev. Mod. Phys.}\ }\textbf {\bibinfo {volume} {86}},\ \bibinfo
  {pages} {361} (\bibinfo {year} {2014})}\BibitemShut {NoStop}%
\bibitem [{\citenamefont {Bose}(2003)}]{Bose}%
  \BibitemOpen
  \bibfield  {author} {\bibinfo {author} {\bibfnamefont {S.}~\bibnamefont
  {Bose}},\ }\bibfield  {title} {\bibinfo {title} {Quantum communication
  through an unmodulated spin chain},\ }\href
  {https://doi.org/10.1103/PhysRevLett.91.207901} {\bibfield  {journal}
  {\bibinfo  {journal} {Phys. Rev. Lett.}\ }\textbf {\bibinfo {volume} {91}},\
  \bibinfo {pages} {207901} (\bibinfo {year} {2003})}\BibitemShut {NoStop}%
\bibitem [{\citenamefont {Haake}(1973)}]{haake}%
  \BibitemOpen
  \bibfield  {author} {\bibinfo {author} {\bibfnamefont {F.}~\bibnamefont
  {Haake}},\ }\bibfield  {title} {\bibinfo {title} {Statistical treatment of
  open systems by generalized master equations},\ }\href@noop {} {\bibfield
  {journal} {\bibinfo  {journal} {Springer Tracts in Modern Physics}\ }\textbf
  {\bibinfo {volume} {66}},\ \bibinfo {pages} {98} (\bibinfo {year}
  {1973})}\BibitemShut {NoStop}%
\bibitem [{\citenamefont {Daffer}\ \emph {et~al.}(2004)\citenamefont {Daffer},
  \citenamefont {W{\'{o} }dkiewicz}, \citenamefont {Cresser},\ and\
  \citenamefont {McIver}}]{Daffer}%
  \BibitemOpen
  \bibfield  {author} {\bibinfo {author} {\bibfnamefont {S.}~\bibnamefont
  {Daffer}}, \bibinfo {author} {\bibfnamefont {K.}~\bibnamefont {W{\'{o}
  }dkiewicz}}, \bibinfo {author} {\bibfnamefont {J.~D.}\ \bibnamefont
  {Cresser}},\ and\ \bibinfo {author} {\bibfnamefont {J.~K.}\ \bibnamefont
  {McIver}},\ }\bibfield  {title} {\bibinfo {title} {Depolarizing channel as a
  completely positive map with memory},\ }\bibfield  {journal} {\bibinfo
  {journal} {Physical Review A}\ }\textbf {\bibinfo {volume} {70}},\ \href
  {https://doi.org/10.1103/physreva.70.010304} {10.1103/physreva.70.010304}
  (\bibinfo {year} {2004})\BibitemShut {NoStop}%
\bibitem [{\citenamefont {Maniscalco}\ and\ \citenamefont
  {Petruccione}(2006)}]{Maniscalco_2006}%
  \BibitemOpen
  \bibfield  {author} {\bibinfo {author} {\bibfnamefont {S.}~\bibnamefont
  {Maniscalco}}\ and\ \bibinfo {author} {\bibfnamefont {F.}~\bibnamefont
  {Petruccione}},\ }\bibfield  {title} {\bibinfo {title} {Non-markovian
  dynamics of a qubit},\ }\bibfield  {journal} {\bibinfo  {journal} {Physical
  Review A}\ }\textbf {\bibinfo {volume} {73}},\ \href
  {https://doi.org/10.1103/physreva.73.012111} {10.1103/physreva.73.012111}
  (\bibinfo {year} {2006})\BibitemShut {NoStop}%
\bibitem [{\citenamefont {Buscemi}\ and\ \citenamefont
  {Bordone}(2013)}]{buscemi}%
  \BibitemOpen
  \bibfield  {author} {\bibinfo {author} {\bibfnamefont {F.}~\bibnamefont
  {Buscemi}}\ and\ \bibinfo {author} {\bibfnamefont {P.}~\bibnamefont
  {Bordone}},\ }\bibfield  {title} {\bibinfo {title} {Time evolution of
  tripartite quantum discord and entanglement under local and nonlocal random
  telegraph noise},\ }\href@noop {} {\bibfield  {journal} {\bibinfo  {journal}
  {Physical Review A}\ }\textbf {\bibinfo {volume} {87}},\ \bibinfo {pages}
  {042310} (\bibinfo {year} {2013})}\BibitemShut {NoStop}%
\bibitem [{\citenamefont {Ali}\ \emph {et~al.}(2014)\citenamefont {Ali},
  \citenamefont {Lo},\ and\ \citenamefont {Zhang}}]{Ali_}%
  \BibitemOpen
  \bibfield  {author} {\bibinfo {author} {\bibfnamefont {M.~M.}\ \bibnamefont
  {Ali}}, \bibinfo {author} {\bibfnamefont {P.-Y.}\ \bibnamefont {Lo}},\ and\
  \bibinfo {author} {\bibfnamefont {W.-M.}\ \bibnamefont {Zhang}},\ }\bibfield
  {title} {\bibinfo {title} {Exact decoherence dynamics of $1/f$ noise},\
  }\href {https://doi.org/10.1088/1367-2630/16/10/103010} {\bibfield  {journal}
  {\bibinfo  {journal} {New Journal of Physics}\ }\textbf {\bibinfo {volume}
  {16}},\ \bibinfo {pages} {103010} (\bibinfo {year} {2014})}\BibitemShut
  {NoStop}%
\bibitem [{\citenamefont {Benedetti}\ \emph {et~al.}(2014)\citenamefont
  {Benedetti}, \citenamefont {Paris},\ and\ \citenamefont
  {Maniscalco}}]{benedetti2014non}%
  \BibitemOpen
  \bibfield  {author} {\bibinfo {author} {\bibfnamefont {C.}~\bibnamefont
  {Benedetti}}, \bibinfo {author} {\bibfnamefont {M.~G.}\ \bibnamefont
  {Paris}},\ and\ \bibinfo {author} {\bibfnamefont {S.}~\bibnamefont
  {Maniscalco}},\ }\bibfield  {title} {\bibinfo {title} {Non-markovianity of
  colored noisy channels},\ }\href@noop {} {\bibfield  {journal} {\bibinfo
  {journal} {Physical Review A}\ }\textbf {\bibinfo {volume} {89}},\ \bibinfo
  {pages} {012114} (\bibinfo {year} {2014})}\BibitemShut {NoStop}%
\bibitem [{\citenamefont {Addis}\ \emph {et~al.}(2016)\citenamefont {Addis},
  \citenamefont {Karpat}, \citenamefont {Macchiavello},\ and\ \citenamefont
  {Maniscalco}}]{Addis_2016}%
  \BibitemOpen
  \bibfield  {author} {\bibinfo {author} {\bibfnamefont {C.}~\bibnamefont
  {Addis}}, \bibinfo {author} {\bibfnamefont {G.}~\bibnamefont {Karpat}},
  \bibinfo {author} {\bibfnamefont {C.}~\bibnamefont {Macchiavello}},\ and\
  \bibinfo {author} {\bibfnamefont {S.}~\bibnamefont {Maniscalco}},\ }\bibfield
   {title} {\bibinfo {title} {Dynamical memory effects in correlated quantum
  channels},\ }\bibfield  {journal} {\bibinfo  {journal} {Physical Review A}\
  }\textbf {\bibinfo {volume} {94}},\ \href
  {https://doi.org/10.1103/physreva.94.032121} {10.1103/physreva.94.032121}
  (\bibinfo {year} {2016})\BibitemShut {NoStop}%
\bibitem [{\citenamefont {Schultz}\ \emph {et~al.}(2021)\citenamefont
  {Schultz}, \citenamefont {Quiroz}, \citenamefont {Titum},\ and\ \citenamefont
  {Clader}}]{schwarma}%
  \BibitemOpen
  \bibfield  {author} {\bibinfo {author} {\bibfnamefont {K.}~\bibnamefont
  {Schultz}}, \bibinfo {author} {\bibfnamefont {G.}~\bibnamefont {Quiroz}},
  \bibinfo {author} {\bibfnamefont {P.}~\bibnamefont {Titum}},\ and\ \bibinfo
  {author} {\bibfnamefont {B.~D.}\ \bibnamefont {Clader}},\ }\bibfield  {title}
  {\bibinfo {title} {Sch{WARMA}: A model-based approach for time-correlated
  noise in quantum circuits},\ }\href
  {https://doi.org/10.1103/PhysRevResearch.3.033229} {\bibfield  {journal}
  {\bibinfo  {journal} {Phys. Rev. Research}\ }\textbf {\bibinfo {volume}
  {3}},\ \bibinfo {pages} {033229} (\bibinfo {year} {2021})}\BibitemShut
  {NoStop}%
\bibitem [{\citenamefont {Harper}\ \emph {et~al.}(2020)\citenamefont {Harper},
  \citenamefont {Flammia},\ and\ \citenamefont {Wallman}}]{flammia}%
  \BibitemOpen
  \bibfield  {author} {\bibinfo {author} {\bibfnamefont {R.}~\bibnamefont
  {Harper}}, \bibinfo {author} {\bibfnamefont {S.~T.}\ \bibnamefont
  {Flammia}},\ and\ \bibinfo {author} {\bibfnamefont {J.~J.}\ \bibnamefont
  {Wallman}},\ }\bibfield  {title} {\bibinfo {title} {Efficient learning of
  quantum noise},\ }\href@noop {} {\bibfield  {journal} {\bibinfo  {journal}
  {Nature Physics}\ }\textbf {\bibinfo {volume} {16}},\ \bibinfo {pages} {1184}
  (\bibinfo {year} {2020})}\BibitemShut {NoStop}%
\bibitem [{\citenamefont {Aliferis}\ \emph {et~al.}(2005)\citenamefont
  {Aliferis}, \citenamefont {Gottesman},\ and\ \citenamefont
  {Preskill}}]{aliferis}%
  \BibitemOpen
  \bibfield  {author} {\bibinfo {author} {\bibfnamefont {P.}~\bibnamefont
  {Aliferis}}, \bibinfo {author} {\bibfnamefont {D.}~\bibnamefont
  {Gottesman}},\ and\ \bibinfo {author} {\bibfnamefont {J.}~\bibnamefont
  {Preskill}},\ }\bibfield  {title} {\bibinfo {title} {Quantum accuracy
  threshold for concatenated distance-3 codes},\ }\href@noop {} {\bibfield
  {journal} {\bibinfo  {journal} {arXiv preprint quant-ph/0504218}\ } (\bibinfo
  {year} {2005})}\BibitemShut {NoStop}%
\bibitem [{\citenamefont {Aharonov}\ \emph {et~al.}(2006)\citenamefont
  {Aharonov}, \citenamefont {Kitaev},\ and\ \citenamefont
  {Preskill}}]{daharonov}%
  \BibitemOpen
  \bibfield  {author} {\bibinfo {author} {\bibfnamefont {D.}~\bibnamefont
  {Aharonov}}, \bibinfo {author} {\bibfnamefont {A.}~\bibnamefont {Kitaev}},\
  and\ \bibinfo {author} {\bibfnamefont {J.}~\bibnamefont {Preskill}},\
  }\bibfield  {title} {\bibinfo {title} {Fault-tolerant quantum computation
  with long-range correlated noise},\ }\href@noop {} {\bibfield  {journal}
  {\bibinfo  {journal} {Physical review letters}\ }\textbf {\bibinfo {volume}
  {96}},\ \bibinfo {pages} {050504} (\bibinfo {year} {2006})}\BibitemShut
  {NoStop}%
\bibitem [{\citenamefont {Clemens}\ \emph {et~al.}(2004)\citenamefont
  {Clemens}, \citenamefont {Siddiqui},\ and\ \citenamefont
  {Gea-Banacloche}}]{Clemens}%
  \BibitemOpen
  \bibfield  {author} {\bibinfo {author} {\bibfnamefont {J.~P.}\ \bibnamefont
  {Clemens}}, \bibinfo {author} {\bibfnamefont {S.}~\bibnamefont {Siddiqui}},\
  and\ \bibinfo {author} {\bibfnamefont {J.}~\bibnamefont {Gea-Banacloche}},\
  }\bibfield  {title} {\bibinfo {title} {Quantum error correction against
  correlated noise},\ }\href {https://doi.org/10.1103/PhysRevA.69.062313}
  {\bibfield  {journal} {\bibinfo  {journal} {Phys. Rev. A}\ }\textbf {\bibinfo
  {volume} {69}},\ \bibinfo {pages} {062313} (\bibinfo {year}
  {2004})}\BibitemShut {NoStop}%
\bibitem [{\citenamefont {Klesse}\ and\ \citenamefont {Frank}(2005)}]{Klesse}%
  \BibitemOpen
  \bibfield  {author} {\bibinfo {author} {\bibfnamefont {R.}~\bibnamefont
  {Klesse}}\ and\ \bibinfo {author} {\bibfnamefont {S.}~\bibnamefont {Frank}},\
  }\bibfield  {title} {\bibinfo {title} {Quantum error correction in spatially
  correlated quantum noise},\ }\href
  {https://doi.org/10.1103/PhysRevLett.95.230503} {\bibfield  {journal}
  {\bibinfo  {journal} {Phys. Rev. Lett.}\ }\textbf {\bibinfo {volume} {95}},\
  \bibinfo {pages} {230503} (\bibinfo {year} {2005})}\BibitemShut {NoStop}%
\bibitem [{\citenamefont {Novais}\ \emph {et~al.}(2008)\citenamefont {Novais},
  \citenamefont {Mucciolo},\ and\ \citenamefont {Baranger}}]{Novais}%
  \BibitemOpen
  \bibfield  {author} {\bibinfo {author} {\bibfnamefont {E.}~\bibnamefont
  {Novais}}, \bibinfo {author} {\bibfnamefont {E.~R.}\ \bibnamefont
  {Mucciolo}},\ and\ \bibinfo {author} {\bibfnamefont {H.~U.}\ \bibnamefont
  {Baranger}},\ }\bibfield  {title} {\bibinfo {title} {Hamiltonian formulation
  of quantum error correction and correlated noise: Effects of syndrome
  extraction in the long-time limit},\ }\href
  {https://doi.org/10.1103/PhysRevA.78.012314} {\bibfield  {journal} {\bibinfo
  {journal} {Phys. Rev. A}\ }\textbf {\bibinfo {volume} {78}},\ \bibinfo
  {pages} {012314} (\bibinfo {year} {2008})}\BibitemShut {NoStop}%
\bibitem [{\citenamefont {Cafaro}\ and\ \citenamefont
  {Mancini}(2010)}]{Cafaro}%
  \BibitemOpen
  \bibfield  {author} {\bibinfo {author} {\bibfnamefont {C.}~\bibnamefont
  {Cafaro}}\ and\ \bibinfo {author} {\bibfnamefont {S.}~\bibnamefont
  {Mancini}},\ }\bibfield  {title} {\bibinfo {title} {Quantum stabilizer codes
  for correlated and asymmetric depolarizing errors},\ }\href
  {https://doi.org/10.1103/PhysRevA.82.012306} {\bibfield  {journal} {\bibinfo
  {journal} {Phys. Rev. A}\ }\textbf {\bibinfo {volume} {82}},\ \bibinfo
  {pages} {012306} (\bibinfo {year} {2010})}\BibitemShut {NoStop}%
\bibitem [{\citenamefont {Clader}\ \emph {et~al.}(2021)\citenamefont {Clader},
  \citenamefont {Trout}, \citenamefont {Barnes}, \citenamefont {Schultz},
  \citenamefont {Quiroz},\ and\ \citenamefont {Titum}}]{Clader}%
  \BibitemOpen
  \bibfield  {author} {\bibinfo {author} {\bibfnamefont {B.~D.}\ \bibnamefont
  {Clader}}, \bibinfo {author} {\bibfnamefont {C.~J.}\ \bibnamefont {Trout}},
  \bibinfo {author} {\bibfnamefont {J.~P.}\ \bibnamefont {Barnes}}, \bibinfo
  {author} {\bibfnamefont {K.}~\bibnamefont {Schultz}}, \bibinfo {author}
  {\bibfnamefont {G.}~\bibnamefont {Quiroz}},\ and\ \bibinfo {author}
  {\bibfnamefont {P.}~\bibnamefont {Titum}},\ }\bibfield  {title} {\bibinfo
  {title} {Impact of correlations and heavy tails on quantum error
  correction},\ }\href {https://doi.org/10.1103/PhysRevA.103.052428} {\bibfield
   {journal} {\bibinfo  {journal} {Phys. Rev. A}\ }\textbf {\bibinfo {volume}
  {103}},\ \bibinfo {pages} {052428} (\bibinfo {year} {2021})}\BibitemShut
  {NoStop}%
\bibitem [{\citenamefont {Macchiavello}\ and\ \citenamefont
  {Palma}(2002)}]{Macchia}%
  \BibitemOpen
  \bibfield  {author} {\bibinfo {author} {\bibfnamefont {C.}~\bibnamefont
  {Macchiavello}}\ and\ \bibinfo {author} {\bibfnamefont {G.~M.}\ \bibnamefont
  {Palma}},\ }\bibfield  {title} {\bibinfo {title} {Entanglement-enhanced
  information transmission over a quantum channel with correlated noise},\
  }\href {https://doi.org/10.1103/PhysRevA.65.050301} {\bibfield  {journal}
  {\bibinfo  {journal} {Phys. Rev. A}\ }\textbf {\bibinfo {volume} {65}},\
  \bibinfo {pages} {050301} (\bibinfo {year} {2002})}\BibitemShut {NoStop}%
\bibitem [{\citenamefont {Macchiavello}\ \emph {et~al.}(2004)\citenamefont
  {Macchiavello}, \citenamefont {Palma},\ and\ \citenamefont
  {Virmani}}]{Macchia_2}%
  \BibitemOpen
  \bibfield  {author} {\bibinfo {author} {\bibfnamefont {C.}~\bibnamefont
  {Macchiavello}}, \bibinfo {author} {\bibfnamefont {G.~M.}\ \bibnamefont
  {Palma}},\ and\ \bibinfo {author} {\bibfnamefont {S.}~\bibnamefont
  {Virmani}},\ }\bibfield  {title} {\bibinfo {title} {Transition behavior in
  the channel capacity of two-quibit channels with memory},\ }\href
  {https://doi.org/10.1103/PhysRevA.69.010303} {\bibfield  {journal} {\bibinfo
  {journal} {Phys. Rev. A}\ }\textbf {\bibinfo {volume} {69}},\ \bibinfo
  {pages} {010303} (\bibinfo {year} {2004})}\BibitemShut {NoStop}%
\bibitem [{\citenamefont {Daems}(2007)}]{Daems}%
  \BibitemOpen
  \bibfield  {author} {\bibinfo {author} {\bibfnamefont {D.}~\bibnamefont
  {Daems}},\ }\bibfield  {title} {\bibinfo {title} {Entanglement-enhanced
  transmission of classical information in {P}auli channels with memory: Exact
  solution},\ }\href {https://doi.org/10.1103/PhysRevA.76.012310} {\bibfield
  {journal} {\bibinfo  {journal} {Phys. Rev. A}\ }\textbf {\bibinfo {volume}
  {76}},\ \bibinfo {pages} {012310} (\bibinfo {year} {2007})}\BibitemShut
  {NoStop}%
\bibitem [{\citenamefont {Kitaev}\ \emph {et~al.}(2002)\citenamefont {Kitaev},
  \citenamefont {Shen},\ and\ \citenamefont {Vyalyi}}]{Kitaev}%
  \BibitemOpen
  \bibfield  {author} {\bibinfo {author} {\bibfnamefont {A.~Y.}\ \bibnamefont
  {Kitaev}}, \bibinfo {author} {\bibfnamefont {A.~H.}\ \bibnamefont {Shen}},\
  and\ \bibinfo {author} {\bibfnamefont {M.~N.}\ \bibnamefont {Vyalyi}},\
  }\href@noop {} {\emph {\bibinfo {title} {Classical and Quantum
  Computation}}}\ (\bibinfo  {publisher} {American Mathematical Society
  Providence, Rhode Island},\ \bibinfo {year} {2002})\BibitemShut {NoStop}%
\bibitem [{\citenamefont {Preskill}(2018)}]{Preskill2018}%
  \BibitemOpen
  \bibfield  {author} {\bibinfo {author} {\bibfnamefont {J.}~\bibnamefont
  {Preskill}},\ }\bibfield  {title} {\bibinfo {title} {Quantum {C}omputing in
  the {NISQ} era and beyond},\ }\href
  {https://doi.org/10.22331/q-2018-08-06-79} {\bibfield  {journal} {\bibinfo
  {journal} {{Quantum}}\ }\textbf {\bibinfo {volume} {2}},\ \bibinfo {pages}
  {79} (\bibinfo {year} {2018})}\BibitemShut {NoStop}%
\bibitem [{\citenamefont {Murali}\ \emph
  {et~al.}(2019{\natexlab{a}})\citenamefont {Murali}, \citenamefont {Linke},
  \citenamefont {Martonosi}, \citenamefont {Abhari}, \citenamefont {Nguyen},\
  and\ \citenamefont {Alderete}}]{murali2019}%
  \BibitemOpen
  \bibfield  {author} {\bibinfo {author} {\bibfnamefont {P.}~\bibnamefont
  {Murali}}, \bibinfo {author} {\bibfnamefont {N.~M.}\ \bibnamefont {Linke}},
  \bibinfo {author} {\bibfnamefont {M.}~\bibnamefont {Martonosi}}, \bibinfo
  {author} {\bibfnamefont {A.~J.}\ \bibnamefont {Abhari}}, \bibinfo {author}
  {\bibfnamefont {N.~H.}\ \bibnamefont {Nguyen}},\ and\ \bibinfo {author}
  {\bibfnamefont {C.~H.}\ \bibnamefont {Alderete}},\ }\bibfield  {title}
  {\bibinfo {title} {Full-stack, real-system quantum computer studies:
  Architectural comparisons and design insights},\ }in\ \href@noop {} {\emph
  {\bibinfo {booktitle} {2019 ACM/IEEE 46th Annual International Symposium on
  Computer Architecture (ISCA)}}}\ (\bibinfo {organization} {IEEE},\ \bibinfo
  {year} {2019})\ pp.\ \bibinfo {pages} {527--540}\BibitemShut {NoStop}%
\bibitem [{\citenamefont {Kwiatkowski}\ and\ \citenamefont
  {Cywi\ifmmode~\acute{n}\else \'{n}\fi{}ski}(2018)}]{Kwiatkowski}%
  \BibitemOpen
  \bibfield  {author} {\bibinfo {author} {\bibfnamefont {D.}~\bibnamefont
  {Kwiatkowski}}\ and\ \bibinfo {author} {\bibfnamefont {L.}~\bibnamefont
  {Cywi\ifmmode~\acute{n}\else \'{n}\fi{}ski}},\ }\bibfield  {title} {\bibinfo
  {title} {Decoherence of two entangled spin qubits coupled to an interacting
  sparse nuclear spin bath: Application to nitrogen vacancy centers},\ }\href
  {https://doi.org/10.1103/PhysRevB.98.155202} {\bibfield  {journal} {\bibinfo
  {journal} {Phys. Rev. B}\ }\textbf {\bibinfo {volume} {98}},\ \bibinfo
  {pages} {155202} (\bibinfo {year} {2018})}\BibitemShut {NoStop}%
\bibitem [{\citenamefont {von L\"upke}\ \emph {et~al.}(2020)\citenamefont {von
  L\"upke}, \citenamefont {Beaudoin}, \citenamefont {Norris}, \citenamefont
  {Sung}, \citenamefont {Winik}, \citenamefont {Qiu}, \citenamefont
  {Kjaergaard}, \citenamefont {Kim}, \citenamefont {Yoder}, \citenamefont
  {Gustavsson}, \citenamefont {Viola},\ and\ \citenamefont {Oliver}}]{LViola}%
  \BibitemOpen
  \bibfield  {author} {\bibinfo {author} {\bibfnamefont {U.}~\bibnamefont {von
  L\"upke}}, \bibinfo {author} {\bibfnamefont {F.}~\bibnamefont {Beaudoin}},
  \bibinfo {author} {\bibfnamefont {L.~M.}\ \bibnamefont {Norris}}, \bibinfo
  {author} {\bibfnamefont {Y.}~\bibnamefont {Sung}}, \bibinfo {author}
  {\bibfnamefont {R.}~\bibnamefont {Winik}}, \bibinfo {author} {\bibfnamefont
  {J.~Y.}\ \bibnamefont {Qiu}}, \bibinfo {author} {\bibfnamefont
  {M.}~\bibnamefont {Kjaergaard}}, \bibinfo {author} {\bibfnamefont
  {D.}~\bibnamefont {Kim}}, \bibinfo {author} {\bibfnamefont {J.}~\bibnamefont
  {Yoder}}, \bibinfo {author} {\bibfnamefont {S.}~\bibnamefont {Gustavsson}},
  \bibinfo {author} {\bibfnamefont {L.}~\bibnamefont {Viola}},\ and\ \bibinfo
  {author} {\bibfnamefont {W.~D.}\ \bibnamefont {Oliver}},\ }\bibfield  {title}
  {\bibinfo {title} {Two-qubit spectroscopy of {S}patiotemporally correlated
  quantum noise in superconducting qubits},\ }\href
  {https://doi.org/10.1103/PRXQuantum.1.010305} {\bibfield  {journal} {\bibinfo
   {journal} {PRX Quantum}\ }\textbf {\bibinfo {volume} {1}},\ \bibinfo {pages}
  {010305} (\bibinfo {year} {2020})}\BibitemShut {NoStop}%
\bibitem [{\citenamefont {Gambetta}\ \emph {et~al.}(2012)\citenamefont
  {Gambetta}, \citenamefont {C\'orcoles}, \citenamefont {Merkel}, \citenamefont
  {Johnson}, \citenamefont {Smolin}, \citenamefont {Chow}, \citenamefont
  {Ryan}, \citenamefont {Rigetti}, \citenamefont {Poletto}, \citenamefont
  {Ohki}, \citenamefont {Ketchen},\ and\ \citenamefont
  {Steffen}}]{gambetta_2012}%
  \BibitemOpen
  \bibfield  {author} {\bibinfo {author} {\bibfnamefont {J.~M.}\ \bibnamefont
  {Gambetta}}, \bibinfo {author} {\bibfnamefont {A.~D.}\ \bibnamefont
  {C\'orcoles}}, \bibinfo {author} {\bibfnamefont {S.~T.}\ \bibnamefont
  {Merkel}}, \bibinfo {author} {\bibfnamefont {B.~R.}\ \bibnamefont {Johnson}},
  \bibinfo {author} {\bibfnamefont {J.~A.}\ \bibnamefont {Smolin}}, \bibinfo
  {author} {\bibfnamefont {J.~M.}\ \bibnamefont {Chow}}, \bibinfo {author}
  {\bibfnamefont {C.~A.}\ \bibnamefont {Ryan}}, \bibinfo {author}
  {\bibfnamefont {C.}~\bibnamefont {Rigetti}}, \bibinfo {author} {\bibfnamefont
  {S.}~\bibnamefont {Poletto}}, \bibinfo {author} {\bibfnamefont {T.~A.}\
  \bibnamefont {Ohki}}, \bibinfo {author} {\bibfnamefont {M.~B.}\ \bibnamefont
  {Ketchen}},\ and\ \bibinfo {author} {\bibfnamefont {M.}~\bibnamefont
  {Steffen}},\ }\bibfield  {title} {\bibinfo {title} {Characterization of
  addressability by simultaneous randomized benchmarking},\ }\href
  {https://doi.org/10.1103/PhysRevLett.109.240504} {\bibfield  {journal}
  {\bibinfo  {journal} {Phys. Rev. Lett.}\ }\textbf {\bibinfo {volume} {109}},\
  \bibinfo {pages} {240504} (\bibinfo {year} {2012})}\BibitemShut {NoStop}%
\bibitem [{\citenamefont {Proctor}\ \emph {et~al.}(2019)\citenamefont
  {Proctor}, \citenamefont {Carignan-Dugas}, \citenamefont {Rudinger},
  \citenamefont {Nielsen}, \citenamefont {Blume-Kohout},\ and\ \citenamefont
  {Young}}]{proctor2019}%
  \BibitemOpen
  \bibfield  {author} {\bibinfo {author} {\bibfnamefont {T.~J.}\ \bibnamefont
  {Proctor}}, \bibinfo {author} {\bibfnamefont {A.}~\bibnamefont
  {Carignan-Dugas}}, \bibinfo {author} {\bibfnamefont {K.}~\bibnamefont
  {Rudinger}}, \bibinfo {author} {\bibfnamefont {E.}~\bibnamefont {Nielsen}},
  \bibinfo {author} {\bibfnamefont {R.}~\bibnamefont {Blume-Kohout}},\ and\
  \bibinfo {author} {\bibfnamefont {K.}~\bibnamefont {Young}},\ }\bibfield
  {title} {\bibinfo {title} {Direct randomized benchmarking for multiqubit
  devices},\ }\href@noop {} {\bibfield  {journal} {\bibinfo  {journal}
  {Physical review letters}\ }\textbf {\bibinfo {volume} {123}},\ \bibinfo
  {pages} {030503} (\bibinfo {year} {2019})}\BibitemShut {NoStop}%
\bibitem [{\citenamefont {Sarovar}\ \emph {et~al.}(2020)\citenamefont
  {Sarovar}, \citenamefont {Proctor}, \citenamefont {Rudinger}, \citenamefont
  {Young}, \citenamefont {Nielsen},\ and\ \citenamefont
  {Blume-Kohout}}]{sarovar2020}%
  \BibitemOpen
  \bibfield  {author} {\bibinfo {author} {\bibfnamefont {M.}~\bibnamefont
  {Sarovar}}, \bibinfo {author} {\bibfnamefont {T.}~\bibnamefont {Proctor}},
  \bibinfo {author} {\bibfnamefont {K.}~\bibnamefont {Rudinger}}, \bibinfo
  {author} {\bibfnamefont {K.}~\bibnamefont {Young}}, \bibinfo {author}
  {\bibfnamefont {E.}~\bibnamefont {Nielsen}},\ and\ \bibinfo {author}
  {\bibfnamefont {R.}~\bibnamefont {Blume-Kohout}},\ }\bibfield  {title}
  {\bibinfo {title} {Detecting crosstalk errors in quantum information
  processors},\ }\href@noop {} {\bibfield  {journal} {\bibinfo  {journal}
  {Quantum}\ }\textbf {\bibinfo {volume} {4}},\ \bibinfo {pages} {321}
  (\bibinfo {year} {2020})}\BibitemShut {NoStop}%
\bibitem [{\citenamefont {Murali}\ \emph {et~al.}(2020)\citenamefont {Murali},
  \citenamefont {McKay}, \citenamefont {Martonosi},\ and\ \citenamefont
  {Javadi-Abhari}}]{murali2020}%
  \BibitemOpen
  \bibfield  {author} {\bibinfo {author} {\bibfnamefont {P.}~\bibnamefont
  {Murali}}, \bibinfo {author} {\bibfnamefont {D.~C.}\ \bibnamefont {McKay}},
  \bibinfo {author} {\bibfnamefont {M.}~\bibnamefont {Martonosi}},\ and\
  \bibinfo {author} {\bibfnamefont {A.}~\bibnamefont {Javadi-Abhari}},\
  }\bibfield  {title} {\bibinfo {title} {Software mitigation of crosstalk on
  noisy intermediate-scale quantum computers},\ }in\ \href@noop {} {\emph
  {\bibinfo {booktitle} {Proceedings of the Twenty-Fifth International
  Conference on Architectural Support for Programming Languages and Operating
  Systems}}}\ (\bibinfo {year} {2020})\ pp.\ \bibinfo {pages}
  {1001--1016}\BibitemShut {NoStop}%
\bibitem [{\citenamefont {Zhao}\ \emph {et~al.}(2022)\citenamefont {Zhao},
  \citenamefont {Linghu}, \citenamefont {Li}, \citenamefont {Xu}, \citenamefont
  {Wang}, \citenamefont {Xue}, \citenamefont {Jin},\ and\ \citenamefont
  {Yu}}]{pengzhao}%
  \BibitemOpen
  \bibfield  {author} {\bibinfo {author} {\bibfnamefont {P.}~\bibnamefont
  {Zhao}}, \bibinfo {author} {\bibfnamefont {K.}~\bibnamefont {Linghu}},
  \bibinfo {author} {\bibfnamefont {Z.}~\bibnamefont {Li}}, \bibinfo {author}
  {\bibfnamefont {P.}~\bibnamefont {Xu}}, \bibinfo {author} {\bibfnamefont
  {R.}~\bibnamefont {Wang}}, \bibinfo {author} {\bibfnamefont {G.}~\bibnamefont
  {Xue}}, \bibinfo {author} {\bibfnamefont {Y.}~\bibnamefont {Jin}},\ and\
  \bibinfo {author} {\bibfnamefont {H.}~\bibnamefont {Yu}},\ }\bibfield
  {title} {\bibinfo {title} {Quantum {C}rosstalk analysis for simultaneous gate
  operations on superconducting qubits},\ }\href
  {https://doi.org/10.1103/PRXQuantum.3.020301} {\bibfield  {journal} {\bibinfo
   {journal} {PRX Quantum}\ }\textbf {\bibinfo {volume} {3}},\ \bibinfo {pages}
  {020301} (\bibinfo {year} {2022})}\BibitemShut {NoStop}%
\bibitem [{\citenamefont {Klauck}(2003)}]{harmut}%
  \BibitemOpen
  \bibfield  {author} {\bibinfo {author} {\bibfnamefont {H.}~\bibnamefont
  {Klauck}},\ }\bibfield  {title} {\bibinfo {title} {Quantum time-space
  tradeoffs for sorting},\ }in\ \href {https://doi.org/10.1145/780542.780553}
  {\emph {\bibinfo {booktitle} {Proceedings of the Thirty-Fifth Annual ACM
  Symposium on Theory of Computing}}},\ \bibinfo {series and number} {STOC
  '03}\ (\bibinfo  {publisher} {Association for Computing Machinery},\ \bibinfo
  {address} {New York, NY, USA},\ \bibinfo {year} {2003})\ p.\ \bibinfo {pages}
  {69–76}\BibitemShut {NoStop}%
\bibitem [{\citenamefont {Sinha}\ and\ \citenamefont
  {Russer}(2010)}]{sinha2010quantum}%
  \BibitemOpen
  \bibfield  {author} {\bibinfo {author} {\bibfnamefont {S.}~\bibnamefont
  {Sinha}}\ and\ \bibinfo {author} {\bibfnamefont {P.}~\bibnamefont {Russer}},\
  }\bibfield  {title} {\bibinfo {title} {Quantum computing algorithm for
  electromagnetic field simulation},\ }\href@noop {} {\bibfield  {journal}
  {\bibinfo  {journal} {Quantum Information Processing}\ }\textbf {\bibinfo
  {volume} {9}},\ \bibinfo {pages} {385} (\bibinfo {year} {2010})}\BibitemShut
  {NoStop}%
\bibitem [{\citenamefont {Barenco}\ \emph {et~al.}(1995)\citenamefont
  {Barenco}, \citenamefont {Bennett}, \citenamefont {Cleve}, \citenamefont
  {DiVincenzo}, \citenamefont {Margolus}, \citenamefont {Shor}, \citenamefont
  {Sleator}, \citenamefont {Smolin},\ and\ \citenamefont
  {Weinfurter}}]{barenco1995}%
  \BibitemOpen
  \bibfield  {author} {\bibinfo {author} {\bibfnamefont {A.}~\bibnamefont
  {Barenco}}, \bibinfo {author} {\bibfnamefont {C.~H.}\ \bibnamefont
  {Bennett}}, \bibinfo {author} {\bibfnamefont {R.}~\bibnamefont {Cleve}},
  \bibinfo {author} {\bibfnamefont {D.~P.}\ \bibnamefont {DiVincenzo}},
  \bibinfo {author} {\bibfnamefont {N.}~\bibnamefont {Margolus}}, \bibinfo
  {author} {\bibfnamefont {P.}~\bibnamefont {Shor}}, \bibinfo {author}
  {\bibfnamefont {T.}~\bibnamefont {Sleator}}, \bibinfo {author} {\bibfnamefont
  {J.~A.}\ \bibnamefont {Smolin}},\ and\ \bibinfo {author} {\bibfnamefont
  {H.}~\bibnamefont {Weinfurter}},\ }\bibfield  {title} {\bibinfo {title}
  {Elementary gates for quantum computation},\ }\href@noop {} {\bibfield
  {journal} {\bibinfo  {journal} {Physical review A}\ }\textbf {\bibinfo
  {volume} {52}},\ \bibinfo {pages} {3457} (\bibinfo {year}
  {1995})}\BibitemShut {NoStop}%
\bibitem [{\citenamefont {Craik}\ \emph {et~al.}(2017)\citenamefont {Craik},
  \citenamefont {Linke}, \citenamefont {Sepiol}, \citenamefont {Harty},
  \citenamefont {Goodwin}, \citenamefont {Ballance}, \citenamefont {Stacey},
  \citenamefont {Steane}, \citenamefont {Lucas},\ and\ \citenamefont
  {Allcock}}]{craik2017}%
  \BibitemOpen
  \bibfield  {author} {\bibinfo {author} {\bibfnamefont {D.~A.}\ \bibnamefont
  {Craik}}, \bibinfo {author} {\bibfnamefont {N.}~\bibnamefont {Linke}},
  \bibinfo {author} {\bibfnamefont {M.}~\bibnamefont {Sepiol}}, \bibinfo
  {author} {\bibfnamefont {T.}~\bibnamefont {Harty}}, \bibinfo {author}
  {\bibfnamefont {J.}~\bibnamefont {Goodwin}}, \bibinfo {author} {\bibfnamefont
  {C.}~\bibnamefont {Ballance}}, \bibinfo {author} {\bibfnamefont
  {D.}~\bibnamefont {Stacey}}, \bibinfo {author} {\bibfnamefont
  {A.}~\bibnamefont {Steane}}, \bibinfo {author} {\bibfnamefont
  {D.}~\bibnamefont {Lucas}},\ and\ \bibinfo {author} {\bibfnamefont
  {D.}~\bibnamefont {Allcock}},\ }\bibfield  {title} {\bibinfo {title}
  {High-fidelity spatial and polarization addressing of ca+ 43 qubits using
  near-field microwave control},\ }\href@noop {} {\bibfield  {journal}
  {\bibinfo  {journal} {Physical Review A}\ }\textbf {\bibinfo {volume} {95}},\
  \bibinfo {pages} {022337} (\bibinfo {year} {2017})}\BibitemShut {NoStop}%
\bibitem [{\citenamefont {Erhard}\ \emph {et~al.}(2019)\citenamefont {Erhard},
  \citenamefont {Wallman}, \citenamefont {Postler}, \citenamefont {Meth},
  \citenamefont {Stricker}, \citenamefont {Martinez}, \citenamefont
  {Schindler}, \citenamefont {Monz}, \citenamefont {Emerson},\ and\
  \citenamefont {Blatt}}]{erhard2019}%
  \BibitemOpen
  \bibfield  {author} {\bibinfo {author} {\bibfnamefont {A.}~\bibnamefont
  {Erhard}}, \bibinfo {author} {\bibfnamefont {J.~J.}\ \bibnamefont {Wallman}},
  \bibinfo {author} {\bibfnamefont {L.}~\bibnamefont {Postler}}, \bibinfo
  {author} {\bibfnamefont {M.}~\bibnamefont {Meth}}, \bibinfo {author}
  {\bibfnamefont {R.}~\bibnamefont {Stricker}}, \bibinfo {author}
  {\bibfnamefont {E.~A.}\ \bibnamefont {Martinez}}, \bibinfo {author}
  {\bibfnamefont {P.}~\bibnamefont {Schindler}}, \bibinfo {author}
  {\bibfnamefont {T.}~\bibnamefont {Monz}}, \bibinfo {author} {\bibfnamefont
  {J.}~\bibnamefont {Emerson}},\ and\ \bibinfo {author} {\bibfnamefont
  {R.}~\bibnamefont {Blatt}},\ }\bibfield  {title} {\bibinfo {title}
  {Characterizing large-scale quantum computers via cycle benchmarking},\
  }\href@noop {} {\bibfield  {journal} {\bibinfo  {journal} {Nature
  communications}\ }\textbf {\bibinfo {volume} {10}},\ \bibinfo {pages} {1}
  (\bibinfo {year} {2019})}\BibitemShut {NoStop}%
\bibitem [{\citenamefont {Lienhard}\ \emph {et~al.}(2019)\citenamefont
  {Lienhard}, \citenamefont {Braum{\"u}ller}, \citenamefont {Woods},
  \citenamefont {Rosenberg}, \citenamefont {Calusine}, \citenamefont {Weber},
  \citenamefont {Veps{\"a}l{\"a}inen}, \citenamefont {O'Brien}, \citenamefont
  {Orlando}, \citenamefont {Gustavsson} \emph {et~al.}}]{lienhard2019}%
  \BibitemOpen
  \bibfield  {author} {\bibinfo {author} {\bibfnamefont {B.}~\bibnamefont
  {Lienhard}}, \bibinfo {author} {\bibfnamefont {J.}~\bibnamefont
  {Braum{\"u}ller}}, \bibinfo {author} {\bibfnamefont {W.}~\bibnamefont
  {Woods}}, \bibinfo {author} {\bibfnamefont {D.}~\bibnamefont {Rosenberg}},
  \bibinfo {author} {\bibfnamefont {G.}~\bibnamefont {Calusine}}, \bibinfo
  {author} {\bibfnamefont {S.}~\bibnamefont {Weber}}, \bibinfo {author}
  {\bibfnamefont {A.}~\bibnamefont {Veps{\"a}l{\"a}inen}}, \bibinfo {author}
  {\bibfnamefont {K.}~\bibnamefont {O'Brien}}, \bibinfo {author} {\bibfnamefont
  {T.~P.}\ \bibnamefont {Orlando}}, \bibinfo {author} {\bibfnamefont
  {S.}~\bibnamefont {Gustavsson}}, \emph {et~al.},\ }\bibfield  {title}
  {\bibinfo {title} {Microwave packaging for superconducting qubits},\ }in\
  \href@noop {} {\emph {\bibinfo {booktitle} {2019 IEEE MTT-S International
  Microwave Symposium (IMS)}}}\ (\bibinfo {organization} {IEEE},\ \bibinfo
  {year} {2019})\ pp.\ \bibinfo {pages} {275--278}\BibitemShut {NoStop}%
\bibitem [{\citenamefont {Murali}\ \emph
  {et~al.}(2019{\natexlab{b}})\citenamefont {Murali}, \citenamefont {Baker},
  \citenamefont {Javadi-Abhari}, \citenamefont {Chong},\ and\ \citenamefont
  {Martonosi}}]{BAKER_murali}%
  \BibitemOpen
  \bibfield  {author} {\bibinfo {author} {\bibfnamefont {P.}~\bibnamefont
  {Murali}}, \bibinfo {author} {\bibfnamefont {J.~M.}\ \bibnamefont {Baker}},
  \bibinfo {author} {\bibfnamefont {A.}~\bibnamefont {Javadi-Abhari}}, \bibinfo
  {author} {\bibfnamefont {F.~T.}\ \bibnamefont {Chong}},\ and\ \bibinfo
  {author} {\bibfnamefont {M.}~\bibnamefont {Martonosi}},\ }\bibfield  {title}
  {\bibinfo {title} {Noise-adaptive compiler mappings for noisy
  intermediate-scale quantum computers},\ }in\ \href
  {https://doi.org/10.1145/3297858.3304075} {\emph {\bibinfo {booktitle}
  {Proceedings of the Twenty-Fourth International Conference on Architectural
  Support for Programming Languages and Operating Systems}}},\ \bibinfo {series
  and number} {ASPLOS '19}\ (\bibinfo  {publisher} {Association for Computing
  Machinery},\ \bibinfo {address} {New York, NY, USA},\ \bibinfo {year}
  {2019})\ p.\ \bibinfo {pages} {1015–1029}\BibitemShut {NoStop}%
\bibitem [{\citenamefont {Molavi}\ \emph {et~al.}(2022)\citenamefont {Molavi},
  \citenamefont {Xu}, \citenamefont {Diges}, \citenamefont {Pick},
  \citenamefont {Tannu},\ and\ \citenamefont {Albarghouthi}}]{molavi2022}%
  \BibitemOpen
  \bibfield  {author} {\bibinfo {author} {\bibfnamefont {A.}~\bibnamefont
  {Molavi}}, \bibinfo {author} {\bibfnamefont {A.}~\bibnamefont {Xu}}, \bibinfo
  {author} {\bibfnamefont {M.}~\bibnamefont {Diges}}, \bibinfo {author}
  {\bibfnamefont {L.}~\bibnamefont {Pick}}, \bibinfo {author} {\bibfnamefont
  {S.}~\bibnamefont {Tannu}},\ and\ \bibinfo {author} {\bibfnamefont
  {A.}~\bibnamefont {Albarghouthi}},\ }\bibfield  {title} {\bibinfo {title}
  {Qubit mapping and routing via maxsat},\ }in\ \href@noop {} {\emph {\bibinfo
  {booktitle} {2022 55th IEEE/ACM International Symposium on Microarchitecture
  (MICRO)}}}\ (\bibinfo {organization} {IEEE},\ \bibinfo {year} {2022})\ pp.\
  \bibinfo {pages} {1078--1091}\BibitemShut {NoStop}%
\bibitem [{\citenamefont {Chhangte}\ and\ \citenamefont
  {Chakrabarty}(2022)}]{chhangte}%
  \BibitemOpen
  \bibfield  {author} {\bibinfo {author} {\bibfnamefont {L.}~\bibnamefont
  {Chhangte}}\ and\ \bibinfo {author} {\bibfnamefont {A.}~\bibnamefont
  {Chakrabarty}},\ }\bibfield  {title} {\bibinfo {title} {Near-optimal circuit
  mapping with reduced search paths on {IBM} quantum architectures},\
  }\href@noop {} {\bibfield  {journal} {\bibinfo  {journal} {Microprocessors
  and Microsystems}\ }\textbf {\bibinfo {volume} {94}},\ \bibinfo {pages}
  {104637} (\bibinfo {year} {2022})}\BibitemShut {NoStop}%
\bibitem [{\citenamefont {Burgholzer}\ \emph {et~al.}(2022)\citenamefont
  {Burgholzer}, \citenamefont {Schneider},\ and\ \citenamefont
  {Wille}}]{burgholzer2022}%
  \BibitemOpen
  \bibfield  {author} {\bibinfo {author} {\bibfnamefont {L.}~\bibnamefont
  {Burgholzer}}, \bibinfo {author} {\bibfnamefont {S.}~\bibnamefont
  {Schneider}},\ and\ \bibinfo {author} {\bibfnamefont {R.}~\bibnamefont
  {Wille}},\ }\bibfield  {title} {\bibinfo {title} {Limiting the search space
  in optimal quantum circuit mapping},\ }in\ \href@noop {} {\emph {\bibinfo
  {booktitle} {2022 27th Asia and South Pacific Design Automation Conference
  (ASP-DAC)}}}\ (\bibinfo {organization} {IEEE},\ \bibinfo {year} {2022})\ pp.\
  \bibinfo {pages} {466--471}\BibitemShut {NoStop}%
\bibitem [{\citenamefont {Onorati}\ \emph {et~al.}(2017)\citenamefont
  {Onorati}, \citenamefont {Buerschaper}, \citenamefont {Kliesch},
  \citenamefont {Brown}, \citenamefont {Werner},\ and\ \citenamefont
  {Eisert}}]{onorati2017}%
  \BibitemOpen
  \bibfield  {author} {\bibinfo {author} {\bibfnamefont {E.}~\bibnamefont
  {Onorati}}, \bibinfo {author} {\bibfnamefont {O.}~\bibnamefont
  {Buerschaper}}, \bibinfo {author} {\bibfnamefont {M.}~\bibnamefont
  {Kliesch}}, \bibinfo {author} {\bibfnamefont {W.}~\bibnamefont {Brown}},
  \bibinfo {author} {\bibfnamefont {A.~H.}\ \bibnamefont {Werner}},\ and\
  \bibinfo {author} {\bibfnamefont {J.}~\bibnamefont {Eisert}},\ }\bibfield
  {title} {\bibinfo {title} {Mixing properties of stochastic quantum
  hamiltonians},\ }\href@noop {} {\bibfield  {journal} {\bibinfo  {journal}
  {Communications in Mathematical Physics}\ }\textbf {\bibinfo {volume}
  {355}},\ \bibinfo {pages} {905} (\bibinfo {year} {2017})}\BibitemShut
  {NoStop}%
\bibitem [{\citenamefont {Blais}\ \emph {et~al.}(2004)\citenamefont {Blais},
  \citenamefont {Huang}, \citenamefont {Wallraff}, \citenamefont {Girvin},\
  and\ \citenamefont {Schoelkopf}}]{wallraff}%
  \BibitemOpen
  \bibfield  {author} {\bibinfo {author} {\bibfnamefont {A.}~\bibnamefont
  {Blais}}, \bibinfo {author} {\bibfnamefont {R.-S.}\ \bibnamefont {Huang}},
  \bibinfo {author} {\bibfnamefont {A.}~\bibnamefont {Wallraff}}, \bibinfo
  {author} {\bibfnamefont {S.~M.}\ \bibnamefont {Girvin}},\ and\ \bibinfo
  {author} {\bibfnamefont {R.~J.}\ \bibnamefont {Schoelkopf}},\ }\bibfield
  {title} {\bibinfo {title} {Cavity quantum electrodynamics for superconducting
  electrical circuits: An architecture for quantum computation},\ }\href
  {https://doi.org/10.1103/PhysRevA.69.062320} {\bibfield  {journal} {\bibinfo
  {journal} {Phys. Rev. A}\ }\textbf {\bibinfo {volume} {69}},\ \bibinfo
  {pages} {062320} (\bibinfo {year} {2004})}\BibitemShut {NoStop}%
\bibitem [{\citenamefont {Bradley}\ \emph {et~al.}(2019)\citenamefont
  {Bradley}, \citenamefont {Randall}, \citenamefont {Abobeih}, \citenamefont
  {Berrevoets}, \citenamefont {Degen}, \citenamefont {Bakker}, \citenamefont
  {Markham}, \citenamefont {Twitchen},\ and\ \citenamefont
  {Taminiau}}]{bradley}%
  \BibitemOpen
  \bibfield  {author} {\bibinfo {author} {\bibfnamefont {C.~E.}\ \bibnamefont
  {Bradley}}, \bibinfo {author} {\bibfnamefont {J.}~\bibnamefont {Randall}},
  \bibinfo {author} {\bibfnamefont {M.~H.}\ \bibnamefont {Abobeih}}, \bibinfo
  {author} {\bibfnamefont {R.~C.}\ \bibnamefont {Berrevoets}}, \bibinfo
  {author} {\bibfnamefont {M.~J.}\ \bibnamefont {Degen}}, \bibinfo {author}
  {\bibfnamefont {M.~A.}\ \bibnamefont {Bakker}}, \bibinfo {author}
  {\bibfnamefont {M.}~\bibnamefont {Markham}}, \bibinfo {author} {\bibfnamefont
  {D.~J.}\ \bibnamefont {Twitchen}},\ and\ \bibinfo {author} {\bibfnamefont
  {T.~H.}\ \bibnamefont {Taminiau}},\ }\bibfield  {title} {\bibinfo {title} {A
  ten-qubit solid-state spin register with quantum memory up to one minute},\
  }\href {https://doi.org/10.1103/PhysRevX.9.031045} {\bibfield  {journal}
  {\bibinfo  {journal} {Phys. Rev. X}\ }\textbf {\bibinfo {volume} {9}},\
  \bibinfo {pages} {031045} (\bibinfo {year} {2019})}\BibitemShut {NoStop}%
\bibitem [{\citenamefont {Bennett}(1973)}]{Uncomp_bennett}%
  \BibitemOpen
  \bibfield  {author} {\bibinfo {author} {\bibfnamefont {C.~H.}\ \bibnamefont
  {Bennett}},\ }\bibfield  {title} {\bibinfo {title} {Logical reversibility of
  computation},\ }\href {https://doi.org/10.1147/rd.176.0525} {\bibfield
  {journal} {\bibinfo  {journal} {IBM Journal of Research and Development}\
  }\textbf {\bibinfo {volume} {17}},\ \bibinfo {pages} {525} (\bibinfo {year}
  {1973})}\BibitemShut {NoStop}%
\bibitem [{\citenamefont {Aaronson}(2003)}]{unco_aaron}%
  \BibitemOpen
  \bibfield  {author} {\bibinfo {author} {\bibfnamefont {S.}~\bibnamefont
  {Aaronson}},\ }\bibfield  {title} {\bibinfo {title} {Quantum lower bound for
  recursive fourier sampling},\ }\href@noop {} {\bibfield  {journal} {\bibinfo
  {journal} {Quantum Info. Comput.}\ }\textbf {\bibinfo {volume} {3}},\
  \bibinfo {pages} {165–174} (\bibinfo {year} {2003})}\BibitemShut {NoStop}%
\bibitem [{\citenamefont {Amy}\ and\ \citenamefont {Ross}(2021)}]{amy_ross}%
  \BibitemOpen
  \bibfield  {author} {\bibinfo {author} {\bibfnamefont {M.}~\bibnamefont
  {Amy}}\ and\ \bibinfo {author} {\bibfnamefont {N.~J.}\ \bibnamefont {Ross}},\
  }\bibfield  {title} {\bibinfo {title} {Phase-state duality in reversible
  circuit design},\ }\href {https://doi.org/10.1103/PhysRevA.104.052602}
  {\bibfield  {journal} {\bibinfo  {journal} {Phys. Rev. A}\ }\textbf {\bibinfo
  {volume} {104}},\ \bibinfo {pages} {052602} (\bibinfo {year}
  {2021})}\BibitemShut {NoStop}%
\bibitem [{\citenamefont {Paradis}\ \emph {et~al.}(2021)\citenamefont
  {Paradis}, \citenamefont {Bichsel}, \citenamefont {Steffen},\ and\
  \citenamefont {Vechev}}]{paradis}%
  \BibitemOpen
  \bibfield  {author} {\bibinfo {author} {\bibfnamefont {A.}~\bibnamefont
  {Paradis}}, \bibinfo {author} {\bibfnamefont {B.}~\bibnamefont {Bichsel}},
  \bibinfo {author} {\bibfnamefont {S.}~\bibnamefont {Steffen}},\ and\ \bibinfo
  {author} {\bibfnamefont {M.}~\bibnamefont {Vechev}},\ }\bibfield  {title}
  {\bibinfo {title} {Unqomp: Synthesizing uncomputation in quantum circuits},\
  }in\ \href {https://doi.org/10.1145/3453483.3454040} {\emph {\bibinfo
  {booktitle} {Proceedings of the 42nd ACM SIGPLAN International Conference on
  Programming Language Design and Implementation}}},\ \bibinfo {series and
  number} {PLDI 2021}\ (\bibinfo  {publisher} {Association for Computing
  Machinery},\ \bibinfo {address} {New York, NY, USA},\ \bibinfo {year}
  {2021})\ p.\ \bibinfo {pages} {222–236}\BibitemShut {NoStop}%
\bibitem [{\citenamefont {Doob}(1953)}]{doob1953stochastic}%
  \BibitemOpen
  \bibfield  {author} {\bibinfo {author} {\bibfnamefont {J.}~\bibnamefont
  {Doob}},\ }\href {https://books.google.co.in/books?id=KvJQAAAAMAAJ} {\emph
  {\bibinfo {title} {Stochastic Processes}}},\ Probability and Statistics
  Series\ (\bibinfo  {publisher} {Wiley},\ \bibinfo {year} {1953})\BibitemShut
  {NoStop}%
\bibitem [{\citenamefont {Kemeny}\ and\ \citenamefont
  {Snell}(1960)}]{kemeny1960finite}%
  \BibitemOpen
  \bibfield  {author} {\bibinfo {author} {\bibfnamefont {J.~G.}\ \bibnamefont
  {Kemeny}}\ and\ \bibinfo {author} {\bibfnamefont {J.~L.}\ \bibnamefont
  {Snell}},\ }\bibfield  {title} {\bibinfo {title} {Finite markov chains. d van
  nostad co},\ }\href@noop {} {\bibfield  {journal} {\bibinfo  {journal} {Inc.,
  Princeton, NJ}\ } (\bibinfo {year} {1960})}\BibitemShut {NoStop}%
\bibitem [{\citenamefont {Grimmett}\ and\ \citenamefont
  {Stirzaker}(2020)}]{grimmett2020probability}%
  \BibitemOpen
  \bibfield  {author} {\bibinfo {author} {\bibfnamefont {G.}~\bibnamefont
  {Grimmett}}\ and\ \bibinfo {author} {\bibfnamefont {D.}~\bibnamefont
  {Stirzaker}},\ }\href@noop {} {\emph {\bibinfo {title} {Probability and
  random processes}}}\ (\bibinfo  {publisher} {Oxford university press},\
  \bibinfo {year} {2020})\BibitemShut {NoStop}%
\bibitem [{\citenamefont {Bena}(2006)}]{bena}%
  \BibitemOpen
  \bibfield  {author} {\bibinfo {author} {\bibfnamefont {I.}~\bibnamefont
  {Bena}},\ }\bibfield  {title} {\bibinfo {title} {Dichotomous markov noise:
  Exact results for out-of-equilibrium systems},\ }\href
  {https://doi.org/10.1142/S0217979206034881} {\bibfield  {journal} {\bibinfo
  {journal} {International Journal of Modern Physics B}\ }\textbf {\bibinfo
  {volume} {20}},\ \bibinfo {pages} {2825} (\bibinfo {year} {2006})},\ \Eprint
  {https://arxiv.org/abs/https://doi.org/10.1142/S0217979206034881}
  {https://doi.org/10.1142/S0217979206034881} \BibitemShut {NoStop}%
\bibitem [{\citenamefont {Yu}\ and\ \citenamefont
  {Eberly}(2006)}]{yu2006sudden}%
  \BibitemOpen
  \bibfield  {author} {\bibinfo {author} {\bibfnamefont {T.}~\bibnamefont
  {Yu}}\ and\ \bibinfo {author} {\bibfnamefont {J.}~\bibnamefont {Eberly}},\
  }\bibfield  {title} {\bibinfo {title} {Sudden death of entanglement:
  classical noise effects},\ }\href@noop {} {\bibfield  {journal} {\bibinfo
  {journal} {Optics Communications}\ }\textbf {\bibinfo {volume} {264}},\
  \bibinfo {pages} {393} (\bibinfo {year} {2006})}\BibitemShut {NoStop}%
\bibitem [{\citenamefont {Sza{\'n}kowski}\ \emph {et~al.}(2015)\citenamefont
  {Sza{\'n}kowski}, \citenamefont {Trippenbach}, \citenamefont {Cywi{\'n}ski},\
  and\ \citenamefont {Band}}]{szankowski2015}%
  \BibitemOpen
  \bibfield  {author} {\bibinfo {author} {\bibfnamefont {P.}~\bibnamefont
  {Sza{\'n}kowski}}, \bibinfo {author} {\bibfnamefont {M.}~\bibnamefont
  {Trippenbach}}, \bibinfo {author} {\bibfnamefont {{\L}.}~\bibnamefont
  {Cywi{\'n}ski}},\ and\ \bibinfo {author} {\bibfnamefont {Y.~B.}\ \bibnamefont
  {Band}},\ }\bibfield  {title} {\bibinfo {title} {The dynamics of two
  entangled qubits exposed to classical noise: role of spatial and temporal
  noise correlations},\ }\href@noop {} {\bibfield  {journal} {\bibinfo
  {journal} {Quantum Information Processing}\ }\textbf {\bibinfo {volume}
  {14}},\ \bibinfo {pages} {3367} (\bibinfo {year} {2015})}\BibitemShut
  {NoStop}%
\bibitem [{\citenamefont {Aolita}\ \emph {et~al.}(2015)\citenamefont {Aolita},
  \citenamefont {De~Melo},\ and\ \citenamefont {Davidovich}}]{aolita2015open}%
  \BibitemOpen
  \bibfield  {author} {\bibinfo {author} {\bibfnamefont {L.}~\bibnamefont
  {Aolita}}, \bibinfo {author} {\bibfnamefont {F.}~\bibnamefont {De~Melo}},\
  and\ \bibinfo {author} {\bibfnamefont {L.}~\bibnamefont {Davidovich}},\
  }\bibfield  {title} {\bibinfo {title} {Open-system dynamics of entanglement:
  a key issues review},\ }\href@noop {} {\bibfield  {journal} {\bibinfo
  {journal} {Reports on Progress in Physics}\ }\textbf {\bibinfo {volume}
  {78}},\ \bibinfo {pages} {042001} (\bibinfo {year} {2015})}\BibitemShut
  {NoStop}%
\bibitem [{\citenamefont {Jurcevic}\ and\ \citenamefont
  {Govia}(2022)}]{Jurcevic_2022}%
  \BibitemOpen
  \bibfield  {author} {\bibinfo {author} {\bibfnamefont {P.}~\bibnamefont
  {Jurcevic}}\ and\ \bibinfo {author} {\bibfnamefont {L.~C.~G.}\ \bibnamefont
  {Govia}},\ }\bibfield  {title} {\bibinfo {title} {Effective qubit dephasing
  induced by spectator-qubit relaxation},\ }\href
  {https://doi.org/10.1088/2058-9565/ac8cad} {\bibfield  {journal} {\bibinfo
  {journal} {Quantum Science and Technology}\ }\textbf {\bibinfo {volume}
  {7}},\ \bibinfo {pages} {045033} (\bibinfo {year} {2022})}\BibitemShut
  {NoStop}%
\bibitem [{\citenamefont {Rudinger}\ \emph {et~al.}(2019)\citenamefont
  {Rudinger}, \citenamefont {Proctor}, \citenamefont {Langharst}, \citenamefont
  {Sarovar}, \citenamefont {Young},\ and\ \citenamefont
  {Blume-Kohout}}]{rudinger}%
  \BibitemOpen
  \bibfield  {author} {\bibinfo {author} {\bibfnamefont {K.}~\bibnamefont
  {Rudinger}}, \bibinfo {author} {\bibfnamefont {T.}~\bibnamefont {Proctor}},
  \bibinfo {author} {\bibfnamefont {D.}~\bibnamefont {Langharst}}, \bibinfo
  {author} {\bibfnamefont {M.}~\bibnamefont {Sarovar}}, \bibinfo {author}
  {\bibfnamefont {K.}~\bibnamefont {Young}},\ and\ \bibinfo {author}
  {\bibfnamefont {R.}~\bibnamefont {Blume-Kohout}},\ }\bibfield  {title}
  {\bibinfo {title} {Probing context-dependent errors in quantum processors},\
  }\href {https://doi.org/10.1103/PhysRevX.9.021045} {\bibfield  {journal}
  {\bibinfo  {journal} {Phys. Rev. X}\ }\textbf {\bibinfo {volume} {9}},\
  \bibinfo {pages} {021045} (\bibinfo {year} {2019})}\BibitemShut {NoStop}%
\bibitem [{\citenamefont {Bowen}\ and\ \citenamefont
  {Mancini}(2004)}]{Bowen_2004}%
  \BibitemOpen
  \bibfield  {author} {\bibinfo {author} {\bibfnamefont {G.}~\bibnamefont
  {Bowen}}\ and\ \bibinfo {author} {\bibfnamefont {S.}~\bibnamefont
  {Mancini}},\ }\bibfield  {title} {\bibinfo {title} {Quantum channels with a
  finite memory},\ }\bibfield  {journal} {\bibinfo  {journal} {Physical Review
  A}\ }\textbf {\bibinfo {volume} {69}},\ \href
  {https://doi.org/10.1103/physreva.69.012306} {10.1103/physreva.69.012306}
  (\bibinfo {year} {2004})\BibitemShut {NoStop}%
\bibitem [{\citenamefont {Specht}\ \emph {et~al.}(2011)\citenamefont {Specht},
  \citenamefont {Nölleke}, \citenamefont {Reiserer}, \citenamefont {Uphoff},
  \citenamefont {Figueroa}, \citenamefont {Ritter},\ and\ \citenamefont
  {Rempe}}]{Specht}%
  \BibitemOpen
  \bibfield  {author} {\bibinfo {author} {\bibfnamefont {H.~P.}\ \bibnamefont
  {Specht}}, \bibinfo {author} {\bibfnamefont {C.}~\bibnamefont {Nölleke}},
  \bibinfo {author} {\bibfnamefont {A.}~\bibnamefont {Reiserer}}, \bibinfo
  {author} {\bibfnamefont {M.}~\bibnamefont {Uphoff}}, \bibinfo {author}
  {\bibfnamefont {E.}~\bibnamefont {Figueroa}}, \bibinfo {author}
  {\bibfnamefont {S.}~\bibnamefont {Ritter}},\ and\ \bibinfo {author}
  {\bibfnamefont {G.}~\bibnamefont {Rempe}},\ }\bibfield  {title} {\bibinfo
  {title} {A single-atom quantum memory},\ }\href
  {https://doi.org/10.1038/nature09997} {\bibfield  {journal} {\bibinfo
  {journal} {Nature}\ }\textbf {\bibinfo {volume} {473}},\ \bibinfo {pages}
  {190} (\bibinfo {year} {2011})}\BibitemShut {NoStop}%
\bibitem [{\citenamefont {Kumar}\ \emph {et~al.}(2018)\citenamefont {Kumar},
  \citenamefont {Banerjee}, \citenamefont {Srikanth}, \citenamefont
  {Jagadish},\ and\ \citenamefont {Petruccione}}]{kumar}%
  \BibitemOpen
  \bibfield  {author} {\bibinfo {author} {\bibfnamefont {N.~P.}\ \bibnamefont
  {Kumar}}, \bibinfo {author} {\bibfnamefont {S.}~\bibnamefont {Banerjee}},
  \bibinfo {author} {\bibfnamefont {R.}~\bibnamefont {Srikanth}}, \bibinfo
  {author} {\bibfnamefont {V.}~\bibnamefont {Jagadish}},\ and\ \bibinfo
  {author} {\bibfnamefont {F.}~\bibnamefont {Petruccione}},\ }\bibfield
  {title} {\bibinfo {title} {Non-markovian evolution: a quantum walk
  perspective},\ }\href@noop {} {\bibfield  {journal} {\bibinfo  {journal}
  {Open Systems \& Information Dynamics}\ }\textbf {\bibinfo {volume} {25}},\
  \bibinfo {pages} {1850014} (\bibinfo {year} {2018})}\BibitemShut {NoStop}%
\bibitem [{\citenamefont {Kretschmann}\ and\ \citenamefont
  {Werner}(2005)}]{Kre}%
  \BibitemOpen
  \bibfield  {author} {\bibinfo {author} {\bibfnamefont {D.}~\bibnamefont
  {Kretschmann}}\ and\ \bibinfo {author} {\bibfnamefont {R.~F.}\ \bibnamefont
  {Werner}},\ }\bibfield  {title} {\bibinfo {title} {Quantum channels with
  memory},\ }\href {https://doi.org/10.1103/PhysRevA.72.062323} {\bibfield
  {journal} {\bibinfo  {journal} {Phys. Rev. A}\ }\textbf {\bibinfo {volume}
  {72}},\ \bibinfo {pages} {062323} (\bibinfo {year} {2005})}\BibitemShut
  {NoStop}%
\bibitem [{\citenamefont {Caruso}\ \emph {et~al.}(2014)\citenamefont {Caruso},
  \citenamefont {Giovannetti}, \citenamefont {Lupo},\ and\ \citenamefont
  {Mancini}}]{Man_rev}%
  \BibitemOpen
  \bibfield  {author} {\bibinfo {author} {\bibfnamefont {F.}~\bibnamefont
  {Caruso}}, \bibinfo {author} {\bibfnamefont {V.}~\bibnamefont {Giovannetti}},
  \bibinfo {author} {\bibfnamefont {C.}~\bibnamefont {Lupo}},\ and\ \bibinfo
  {author} {\bibfnamefont {S.}~\bibnamefont {Mancini}},\ }\bibfield  {title}
  {\bibinfo {title} {Quantum channels and memory effects},\ }\href
  {https://link.aps.org/doi/10.1103/RevModPhys.86.1203} {\bibfield  {journal}
  {\bibinfo  {journal} {Rev. Mod. Phys.}\ }\textbf {\bibinfo {volume} {86}},\
  \bibinfo {pages} {1203} (\bibinfo {year} {2014})}\BibitemShut {NoStop}%
\bibitem [{\citenamefont {Sudarshan}\ \emph {et~al.}(1961)\citenamefont
  {Sudarshan}, \citenamefont {Mathews},\ and\ \citenamefont
  {Rau}}]{sudarshan1961}%
  \BibitemOpen
  \bibfield  {author} {\bibinfo {author} {\bibfnamefont {E.}~\bibnamefont
  {Sudarshan}}, \bibinfo {author} {\bibfnamefont {P.}~\bibnamefont {Mathews}},\
  and\ \bibinfo {author} {\bibfnamefont {J.}~\bibnamefont {Rau}},\ }\bibfield
  {title} {\bibinfo {title} {Stochastic dynamics of quantum-mechanical
  systems},\ }\href@noop {} {\bibfield  {journal} {\bibinfo  {journal}
  {Physical Review}\ }\textbf {\bibinfo {volume} {121}},\ \bibinfo {pages}
  {920} (\bibinfo {year} {1961})}\BibitemShut {NoStop}%
\bibitem [{\citenamefont {Choi}(1975)}]{choi1975}%
  \BibitemOpen
  \bibfield  {author} {\bibinfo {author} {\bibfnamefont {M.-D.}\ \bibnamefont
  {Choi}},\ }\bibfield  {title} {\bibinfo {title} {Completely positive linear
  maps on complex matrices},\ }\href@noop {} {\bibfield  {journal} {\bibinfo
  {journal} {Linear algebra and its applications}\ }\textbf {\bibinfo {volume}
  {10}},\ \bibinfo {pages} {285} (\bibinfo {year} {1975})}\BibitemShut
  {NoStop}%
\bibitem [{\citenamefont {Lindblad}(1976)}]{Lind}%
  \BibitemOpen
  \bibfield  {author} {\bibinfo {author} {\bibfnamefont {G.}~\bibnamefont
  {Lindblad}},\ }\bibfield  {title} {\bibinfo {title} {On the generators of
  quantum dynamical semigroups},\ }\href
  {https://doi.org/https://doi.org/10.1007/BF01608499} {\bibfield  {journal}
  {\bibinfo  {journal} {Commun.Math. Phys.}\ }\textbf {\bibinfo {volume}
  {48}},\ \bibinfo {pages} {119} (\bibinfo {year} {1976})}\BibitemShut
  {NoStop}%
\bibitem [{\citenamefont {Kraus}\ \emph {et~al.}(1983)\citenamefont {Kraus},
  \citenamefont {B{\"o}hm}, \citenamefont {Dollard},\ and\ \citenamefont
  {Wootters}}]{kraus1983}%
  \BibitemOpen
  \bibfield  {author} {\bibinfo {author} {\bibfnamefont {K.}~\bibnamefont
  {Kraus}}, \bibinfo {author} {\bibfnamefont {A.}~\bibnamefont {B{\"o}hm}},
  \bibinfo {author} {\bibfnamefont {J.~D.}\ \bibnamefont {Dollard}},\ and\
  \bibinfo {author} {\bibfnamefont {W.}~\bibnamefont {Wootters}},\ }\href@noop
  {} {\emph {\bibinfo {title} {States, Effects, and Operations Fundamental
  Notions of Quantum Theory: Lectures in Mathematical Physics at the University
  of Texas at Austin}}}\ (\bibinfo  {publisher} {Springer},\ \bibinfo {year}
  {1983})\BibitemShut {NoStop}%
\bibitem [{\citenamefont {Mendl}\ and\ \citenamefont {Wolf}(2009)}]{Mendl}%
  \BibitemOpen
  \bibfield  {author} {\bibinfo {author} {\bibfnamefont {C.~B.}\ \bibnamefont
  {Mendl}}\ and\ \bibinfo {author} {\bibfnamefont {M.~M.}\ \bibnamefont
  {Wolf}},\ }\bibfield  {title} {\bibinfo {title} {Unital quantum channels
  {\textendash} convex structure and revivals of {B}irkhoff's {T}heorem},\
  }\href {https://doi.org/10.1007/s00220-009-0824-2} {\bibfield  {journal}
  {\bibinfo  {journal} {Communications in Mathematical Physics}\ }\textbf
  {\bibinfo {volume} {289}},\ \bibinfo {pages} {1057} (\bibinfo {year}
  {2009})}\BibitemShut {NoStop}%
\end{thebibliography}%

\end{document}